\documentclass[12pt,epsf,qsymbols]{article}
\pdfoutput=1
\usepackage[utf8]{inputenc}
\usepackage[normalem]{ulem}
\usepackage[english]{babel}
\usepackage{tabularx}
\usepackage{array}
\usepackage{graphics}
\usepackage[pdftex]{graphicx}
\usepackage{psfrag}
\usepackage{epsfig}
\usepackage{subfig}
\usepackage{mathtools}
\usepackage{amssymb}
\usepackage{pifont}
\usepackage{bbold}
\usepackage{setspace}
\usepackage{rotating}
\usepackage{colortbl}
\usepackage{longtable}
\usepackage{slashed}
\usepackage{braket}
\usepackage{lineno}
\makeatletter
\usepackage{textcomp}
\usepackage[usenames,dvipsnames]{xcolor}
\usepackage{relsize}
\usepackage{cite}

\usepackage{arydshln}

\usepackage{float}
\usepackage{multirow}
\usepackage{tikz}
\usetikzlibrary{decorations.pathmorphing,shapes}

\usepackage{lipsum}

\usepackage{bbm}
\usepackage{bm}
\usepackage{enumitem}
\usepackage{amsmath}
\usepackage{verbatim}

\usepackage{hyperref}
\hypersetup{colorlinks,citecolor=nicegreen,linkcolor=niceblue}
\hypersetup{colorlinks=true}

\usepackage{pifont}

\usepackage{calligra}

\usepackage{geometry}

\allowdisplaybreaks

\setlongtables

\newcommand{\bea}{\begin{eqnarray}}
\newcommand{\eea}{\end{eqnarray}}
\newcommand{\beq}{\begin{equation}}

\newcommand{\eeq}{\end{equation}}
\newcommand{\ec}{\end{center}}
\newcommand{\bc}{\begin{center}}

\newcommand{\pdir}{p\kern -5.2pt\raise 0.2ex\hbox {/}}

\newcommand{\vdir}{v\kern -5.75pt\raise 0.15ex\hbox {/}}
\newcommand{\kdir}{k\kern -5.75pt\raise 0.15ex\hbox {/}}
\newcommand{\epsdir}{\epsilon\kern -5.0pt\raise 0.15ex\hbox {/}}
\newcommand{\bvdir}{\bar{v}\kern -5.75pt\raise 0.15ex\hbox {/}}
\newcommand{\Ddir}{D\kern -7.75pt\raise 0.20ex\hbox {/}}
\newcommand{\Adir}{A\kern -7.75pt\raise 0.20ex\hbox {/}}
\newcommand{\ldir}{l\kern -5.0pt\raise 0.2ex\hbox{/}}
\newcommand{\varepsdir}{\varepsilon\kern -5.5pt\raise 0.15ex\hbox{/}}

\makeatother

\DeclareMathAlphabet{\mathcalligra}{T1}{calligra}{m}{n}
\DeclareFontShape{T1}{calligra}{m}{n}{<->s*[2.2]callig15}{}

\definecolor{lightgray}{rgb}{0.9,0.9,0.9}
\definecolor{niceblue}{rgb}{0.15,0.15,0.6}
\definecolor{nicegreen}{rgb}{0.1,0.5,0.1}
\definecolor{Red}{rgb}{1.,0.,0.}

\definecolor{Green}{rgb}{0.2,.7,0.2}

\begin{document}
\unitlength = 1mm

\thispagestyle{empty} 
\begin{center}
\vskip 1.8cm\par
{\par\centering \textbf{\LARGE  
\Large \bf Lepton Flavor Violation in Semileptonic Observables}
\vskip 1.2cm\par
{\scalebox{.85}{\par\centering \large  
\sc I.~Plakias$^{\,a}$,  O.~Sumensari$^{\,a}$ }
{\par\centering \vskip 0.7 cm\par}
{\sl 
$^a$~
{Universit\'e Paris-Saclay, CNRS/IN2P3, IJCLab, 91405 Orsay, France}
}}

{\vskip 1.65cm\par}}
\end{center}

\vskip 0.85cm

\begin{abstract}
In this paper, we perform a comprehensive study of Lepton Flavor Violation (LFV) in semileptonic transitions in the framework of an Effective Field Theory with general flavor structure. We account for the Renormalization Group Equations (RGEs), which induce non-trivial correlations between the different types of LFV processes. In particular, we show that these loop effects are needed to improve the bounds on several coefficients that are not efficiently constrained at tree level. For illustration, we consider a few concrete scenarios, with predominant couplings to third-generation quarks, and we explore the correlations between the various tree- and loop-level constraints on these models. As a by-product, we also provide expressions for several semileptonic LFV meson decays by using the latest determinations of the relevant hadronic form factors on the lattice.

\end{abstract}

\setcounter{page}{1}
\setcounter{footnote}{0}
\setcounter{equation}{0}
\noindent

\renewcommand{\thefootnote}{\arabic{footnote}}
 
\setcounter{footnote}{0}

\clearpage

{\small\tableofcontents}

\newpage

\section{Introduction}
\label{sec:intro}

Lepton Flavor Violation (LFV) is strictly forbidden by the accidental symmetries in the Standard Model (SM). However, these symmetries are not necessarily respected by higher-dimensional operators appearing beyond the SM, which are suppressed by inverse powers of a heavy scale $\Lambda$. Indeed, the observation of neutrino masses and oscillation is the first indication that Lepton Number might be broken, which can be described by dimension-five operators (suppressed by $1/\Lambda$)~\cite{Weinberg:1979sa}. The smallness of neutrino masses, however, guarantees that LFV in charged processes remains heavily suppressed unless new dynamics beyond light neutrino masses are present~\cite{Petcov:1976ff}. Therefore, these processes are clean probes of New Physics, as their observation would be unambiguous evidence of new phenomena, typically arising through dimension-six operators (thus suppressed by $1/\Lambda^2$).

Significant progress is expected in the next years from experiments searching for LFV in both leptonic and semileptonic processes, which will improve the current sensitivity in several channels by at least one order of magnitude~\cite{Calibbi:2017uvl}. These include the various experiments targeting the $\mu \to e$ transition, namely MEG-II~\cite{MEGII:2018kmf}, Mu2E~\cite{Mu2e:2014fns}, Mu3E~\cite{Blondel:2013ia} and COMET~\cite{COMET:2018auw}, as well as Belle-II that will improve our sensitivity to both leptonic and hadronic LFV $\tau$-lepton decays~\cite{Belle-II:2018jsg}. There has also been an increasing effort by LHCb and Belle to improve the limits of several LFV decays of $B$-mesons~\cite{LHCb:2017hag,Belle:2023jwr}. Furthermore, experiments with kaons at NA62~\cite{NA62:2017rwk} and KOTO~\cite{Yamanaka:2012yma}, and charmed hadrons at BES-III~\cite{BESIII:2020nme}, are also expected to improve their sensitivity to LFV decays in the near future.

New Physics contributions to LFV processes from heavy mediators can be characterized by the Standard Model Effective Field Theory (SM EFT)~\cite{Grzadkowski:2010es}, defined above the electroweak scale, which is invariant under the SM gauge symmetry $SU(3)_c\times SU(2)_L \times U(1)_Y$. In the SM EFT, the leading effective operators that induce LFV at low energies are the leptonic dipoles ($\psi^2 H X$), as well as four-fermion leptonic and semileptonic operators  ($\psi^4$)~\cite{Kuno:1999jp}. 
The remarkable experimental sensitivity of LFV searches in specific channels implies that some of these operators can be better probed through radiative corrections, instead of their direct tree-level contributions to low-energy processes. For instance, the operator $(\bar{\mu}\gamma^\mu e)(\bar{q}\gamma_\mu q)$, where $q$ denotes a weak doublet with heavy quarks, can be more efficiently probed through its contributions at one-loop to $\mu \to 3e$ rather than by studying semileptonic processes at tree level~\cite{Crivellin:2017rmk,EliasMiro:2021jgu}. Therefore, it is clear that radiative corrections must be systematically included to assess the full potential of LFV processes to discover New Physics. This program has been carried out to a large extent in the leptonic sector~\cite{Crivellin:2017rmk,Pruna:2014asa,EliasMiro:2021jgu}. However, analogous systematic study for semileptonic transitions is still missing, although the one-loop anomalous dimensions are known both in the low-energy EFT~\cite{Crivellin:2017rmk,Jenkins:2017dyc} and in the SM EFT~\cite{Jenkins:2013zja} (see Ref.~\cite{Davidson:2018rqt,Feruglio:2016gvd} for the first steps in this direction).

In this paper, we will perform a comprehensive analysis of LFV in semileptonic observables, accounting for the one-loop Renormalization Group Equations (RGEs) in the low-energy EFT and the SM EFT. We will demonstrate that such loop effects induce non-trivial correlations between the different types of processes. Moreover, we will provide general expressions for the relevant (semi)leptonic meson decays that can be studied experimentally, using the latest inputs for the needed hadronic matrix elements. We will also revisit purely leptonic processes, since they can provide valuable constraints on semileptonic operators through operator mixing, including the known two-loop contributions to the RGEs~\cite{Crivellin:2017rmk,Pruna:2014asa,EliasMiro:2021jgu}. To compare the sensitivity of different observables, we will consider a specific EFT scenario with operators containing only third-generation quarks, which contributes to various transitions through RGE evolution above and below the electroweak scale. Finally, we will also briefly discuss a few concrete New Physics scenarios to illustrate the relevance of the loop effects that we compute.

The remainder of this paper is organized as follows. In Sec.~\ref{sec:eft}, we define our EFT Lagrangian and we discuss the relevant RGE effects in the low-energy EFT and the SM EFT. In Sec.~\ref{sec:LEFT-LFV}, we study the low-energy probes of LFV in meson and lepton decays. In Sec.~\ref{ssec:SMEFT-LFV}, we discuss the high-energy probes of LFV in Higgs, top-quark and $Z$-boson decays, as well as indirect constraints obtained from the high-energy tails of $pp\to\ell_i\ell_j$, with $i\neq j$. In Sec.~\ref{sec:pheno}, we perform a numerical study with a specific EFT scenario and discuss the relevance of our loop constraints to concrete New Physics models. Our findings are briefly summarized in Sec.~\ref{sec:conclusion}.

\section{EFT approach for LFV}
\label{sec:eft}

In this Section, we define our effective approach and describe the RGE effects that are relevant for LFV observables. In Sec.~\ref{ssec:left}, we formulate the low-energy EFT, invariant under $SU(3)_c\times U(1)_\mathrm{em}$, which will be used to compute the low-energy observables in Sec.~\ref{sec:pheno}. In Sec.~\ref{ssec:smeft}, we consider the SM EFT~\cite{Grzadkowski:2010es}, which gives the appropriate description of New Physics arising well above the electroweak scale, as we assume in this paper. The RGE contributions that we consider in our phenomenological analysis are summarized in Sec.~\ref{ssec:summary} (see also Figs.~\ref{fig:gauge-left-rge} and~\ref{fig:gauge-smeft-rge}).

\subsection{Low-Energy EFT}
\label{ssec:left}

We start by defining our low-energy effective Lagrangian,
\begin{equation}
    \label{eq:left-lagrangian}
    \mathcal{L}_\mathrm{LEFT} = \sum_I \dfrac{{C}_I}{v^2} \,{O}_I\,,
\end{equation}
where $v=(\sqrt{2}G_F)^{-1/2}$, $G_F$ is the Fermi constant, and $C_I$ denotes the low-energy effective coefficients of the operators invariant under $SU(3)_c\times U(1)_\mathrm{em}$\,. Below the electroweak scale, the lowest-dimension operators that can violate lepton flavor are~\cite{Kuno:1999jp}
\begin{itemize}
    \item[i)] Operators of semileptonic type ($\psi^4$):
\begin{align}
\begin{split}    {O}_{\substack{V_{XY}\\ ijkl}}^{(q)} =  \big{(} \bar{\ell}_i \gamma_\mu P_X\ell_j\big{)}&\big{(}\bar{q}_k \gamma^\mu P_Y q_l\big{)}\,, \qquad\quad {O}_{\substack{S_{XY}\\ ijkl}}^{(q)} =\big{(} \bar{\ell}_i  P_X\ell_j\big{)}\big{(}\bar{q}_k  P_Y q_l\big{)}\,, \\[0.4em]
 {O}_{\substack{T_{X}\\ ijkl}}^{(q)} &=\big{(} \bar{\ell}_i  \sigma_{\mu\nu} P_X\ell_j\big{)}\big{(}\bar{q}_k \sigma^{\mu\nu} P_X q_l\big{)}\,,
\end{split}
\end{align}
where $X,Y \in \lbrace L,R \rbrace$, $q \in \lbrace u,d\rbrace$, and $\lbrace i,j,k,l\rbrace$ denote flavor indices.
    \item[ii)] Four-fermion leptonic operators ($\psi^4$):
    \begin{align}
    {O}_{\substack{V_{XY}\\{ijkk}}}^{(\ell)} &=  \big{(} \bar{\ell}_i \gamma_\mu P_X\ell_j\big{)}\big{(}\bar{\ell}_k \gamma^\mu P_Y \ell_k\big{)}\,,  \qquad\quad {O}_{\substack{S_{XX}\\ ijkk}}^{(\ell)} =\big{(} \bar{\ell}_i  P_X\ell_j\big{)}\big{(}\bar{\ell}_k  P_X \ell_k\big{)}\,, 
    \end{align}
where $X,Y \in \lbrace L,R \rbrace$, as above, and $k$ can take any value. For $i \neq k\neq j$, there is an additional type of tensor operator that can be written down,
    \begin{align}
 \qquad {O}_{\substack{T_{X}\\ ijkk}}^{(\ell)} &=\big{(} \bar{\ell}_i  \sigma_{\mu\nu} P_X\ell_j\big{)}\big{(}\bar{\ell}_k \sigma^{\mu\nu} P_X \ell_k\big{)}\,,~\qquad\quad (i\neq k\neq j)\,,
    \end{align}
and, for $X\neq Y$, there are additional scalar operators,
    \begin{align}
 \quad {O}_{\substack{S_{XY}\\ ijkk}}^{(\ell)} &\overset{X\neq Y}{=}\big{(} \bar{\ell}_i   P_X\ell_j\big{)}\big{(}\bar{\ell}_k  P_Y \ell_k\big{)}\,,~\qquad\qquad\quad (i\neq k\neq j)\,,
    \end{align}
which would be redundant via Fierz relations if $k=i$ or $k=j$.
\item[iii)] The electromagnetic dipole operators ($\psi^2 X$),
    \begin{align}
    {O}_{\substack{D_X\\ ij}}^{(\ell)} = m_{\ell}\,\big{(}\bar{\ell}_i \sigma_{\mu\nu}P_X\ell_j\big{)}\, F^{\mu\nu}\,.
    \end{align}
where $X \in \lbrace L,R \rbrace$ and $m_{\ell}$ is the mass of the heaviest lepton appearing in the operator, which we assume to be $m_{\ell_j}$ in the following.
\item[iv)] The gluonic operators ($\psi^2 X^2$),~\footnote{Note that similar operators can be defined with the $U(1)_\mathrm{em}$ field-strength tensors, which would be relevant to processes such as $\mu\to e \gamma\gamma$ that are not considered in this paper, see e.g.~Ref.~\cite{Davidson:2016edt,Fortuna:2022sxt}.}
\begin{align}
\label{eq:left-ope-gg}
    {O}_{\substack{{G}_X\\ ij}}^{(\ell)} &= \dfrac{\alpha_s}{4\pi}
\dfrac{m_{\ell}}{v^2}\,\big{(}\bar{\ell}_i P_X\ell_j\big{)} G_{a\,\mu\nu}{G}^{a\,\mu\nu}\,,\qquad\quad
    {O}_{\substack{\widetilde{G}_X \\ ij}}^{(\ell)} = \dfrac{\alpha_s}{4\pi}
\dfrac{m_{\ell}}{v^2}\,\big{(}\bar{\ell}_i P_X\ell_j\big{)} G_{a\,\mu\nu}\widetilde{G}^{a\,\mu\nu}\,,
\end{align}
where $X \in \lbrace L,R \rbrace$, $m_{\ell}\equiv m_{\ell_j}$ is the mass of the heaviest lepton in the operator, as before, and we define the dual field-strength as $\widetilde{G}^{a\,\mu\nu} = 1/2 \,\varepsilon^{\alpha\beta\mu\nu}{G}^{a}_{\alpha\beta}$.
\end{itemize} 

\noindent Our notation for the SM fields is specified in Appendix~\ref{app:conventions}. Flavor indices of semileptonic operators are assigned with the first two indices  corresponding to the lepton flavor and the last two to the quark ones. Furthermore, the  vector Wilson coefficients must satisfy the following Hermiticy condition, 
\begin{align}
C_{\substack{V_{XY}\\ ijkl}}^{(q)} &=C_{\substack{V_{XY}\\ jilk}}^{(q)\,\ast}\,,
\end{align}
whereas the scalar and tensor coefficients satisfy
\begin{align}
C_{\substack{S_{LL}\\ ijkl}}^{(q)} &=C_{\substack{S_{RR}\\ jilk}}^{(q)\,\ast}\,, \qquad\qquad\quad C_{\substack{S_{LR}\\ ijkl}}^{(q)} =C_{\substack{S_{RL}\\ jilk}}^{(q)\,\ast}\,,  \\[0.5em]
 C_{\substack{T_{LL}\\ ijkl}}^{(q)} &=C_{\substack{T_{RR}\\ jilk}}^{(q)\,\ast}\,, \qquad\qquad\quad C_{\substack{T_{LR}\\ ijkl}}^{(q)} =C_{\substack{T_{RL}\\ jilk}}^{(q)\,\ast}\,.
\end{align}
To avoid redundancies in the operator basis, we will express the observables only in terms of Wilson coefficients with the leptonic indices $i,j$ ordered as $i<j$. From the four-lepton coefficients $C_{V_{XX}}^{(\ell)}$, we choose only to write the ones with indices $ijkk$, where $i<j$ and $k$ can be any flavor index, as we are only interested in operators that violate lepton flavors by one unit, i.e.~with $\Delta L_i = -\Delta L_j = -1$.

\begin{figure}[t!]
\centering
\includegraphics[width=0.67\linewidth]{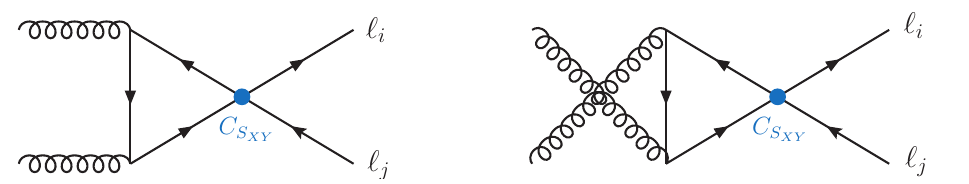}
\caption{\small \sl Finite one-loop contribution from scalar coefficients $C_{S_{XY}}$ to the gluonic operators $C_{G_X}^{(\ell)}$ defined in Eq.~\eqref{eq:left-ope-gg}, which is relevant e.g.~for the phenomenology of $\mu \to e$ conversion in nuclei. }
\label{fig:ggtilde} 
\end{figure}

Note, in particular, that the CP-odd and CP-even gluonic operators $\smash{{O}_{G_X}^{(\ell)}}$ and $\smash{{O}_{\widetilde{G}_X}^{(\ell)}}$ defined in Eq.~\eqref{eq:left-ope-gg} are of higher dimension. We list them among the other operators since they can be produced via an anomaly triangle when heavy quarks are integrated out from the low-energy EFT at one-loop, cf.~Fig.~\ref{fig:ggtilde}. The shifts of the gluonic coefficients induced by the heavy quarks read~\cite{Shifman:1978zn}
\begin{align}
 \label{eq:GGtilde1}
  C_{\substack{G_X\\ij}}^{(\ell)} &\to C_{\substack{G_X\\ij}}^{(\ell)} -\frac{v}{3m_{\ell_j}} \sum_{q_k=c,b,t} \frac{v}{m_{q_k}} \big{(} C_{\substack{S_{XL}\\ijkk}}^{(q)} + C_{\substack{S_{XR}\\ijkk}}^{(q)} \big{)}\,,\\[0.35em]
C_{\substack{\widetilde{G}_X\\ij}}^{(\ell)} &\to C_{\substack{\widetilde{G}_X\\ij}}^{(\ell)} -\frac{iv}{2m_{\ell_j}} \sum_{q_k=c,b,t} \frac{v}{m_{q_k}} \big{(} C_{\substack{S_{XL}\\ijkk}}^{(q)} - C_{\substack{S_{XR}\\ijkk}}^{(q)} \big{)}\,, 
\end{align}

\noindent where $X\in \lbrace L,R\rbrace$, we take $j>i$, as before, and we have kept the first term in the $1/m_q$ expansion. These finite matching contributions will be considered in our phenomenological analysis since they are the leading contributions to gluonic operators in our setup, having implications to processes such as $\mu N\to e N$ (see also Ref.~\cite{Bhattacharya:2018ryy}).

\subsubsection*{Operator mixing}

\begin{figure}[t!]
\centering
\includegraphics[width=0.95\linewidth]{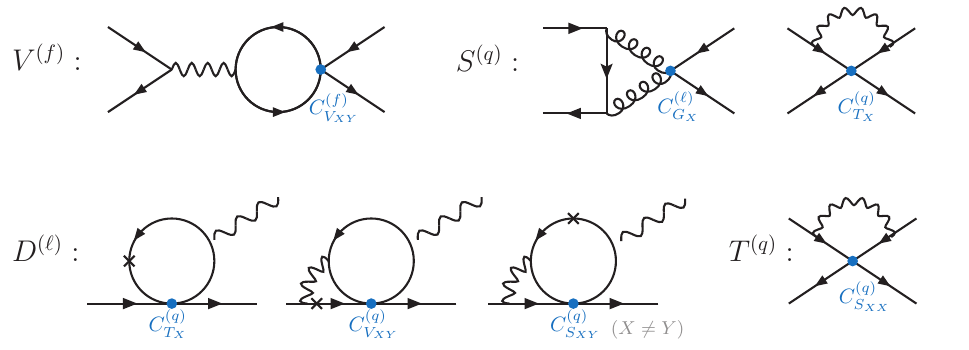}
\caption{\small \sl Schematic representation of one- and two-loop diagrams that can induce non-diagonal operator-mixing via QED and QCD below the electroweak scale~\cite{Crivellin:2017rmk}, cf.~Eq.~\eqref{eq:adm-left} and Eqs.~\eqref{eq:vector-into-dipole}--\eqref{eq:scalar-into-dipole}. The blue dots represent insertions of specific $d=6$ operators appearing in Eq.~\eqref{eq:adm-left}.}
\label{fig:gauge-left-rge} 
\end{figure}

The one-loop running of the LFV low-energy EFT has been computed in full generality Ref.~\cite{Crivellin:2017rmk} (see also Ref.~\cite{Jenkins:2017dyc}). Besides the known renormalization of the scalar and tensor semileptonic operators by QCD~\cite{Gonzalez-Alonso:2017iyc}, it has been shown that the QED running induces a nontrivial mixing of LFV operators, cf.~Fig.~\ref{fig:gauge-left-rge}. For instance, the operator $\smash{(\bar{\ell}_i \gamma^\mu P_X \ell_j)(\bar{q} \gamma_\mu q)}$ mixes into $\smash{(\bar{\ell}_i \gamma^\mu P_X \ell_j)(\bar{\ell}_k \gamma_\mu \ell_k)}$ via a penguin diagram depicted in Fig.~\ref{fig:gauge-left-rge}. Such effects allow us to constrain semileptonic operators with same-flavor quarks by using the experimental limits on $\ell_j^- \to \ell_i^- \ell_k^+\ell_k^-$, with $i \leq k <j$.  Another important example is the one-loop mixing of $\smash{(\bar{\ell}_i \sigma^{\mu\nu} P_X \ell_j)(\bar{q} \sigma_{\mu\nu} P_X q)}$ into the dipoles $\smash{(\bar{\ell}_i \sigma^{\mu\nu} P_X \ell_j)F_{\mu\nu}}$, which induce chirality-enhanced contributions to $\ell_j\to \ell_i\gamma$, as shown in Fig.~\ref{fig:gauge-left-rge}~\cite{Crivellin:2017rmk,Crivellin:2013hpa,Feruglio:2018fxo}.

In the following, we consider LFV operators with fixed lepton flavor indices $i<j$. The \emph{leading logarithm}  solution of the RGEs from the electroweak scale ($\mu_\mathrm{ew}$) to the relevant low-energy scale ($\mu_\mathrm{low}$) is given by
\begin{equation}
\label{eq:leading-log}
C_I(\mu_\mathrm{low}) \simeq C_I(\mu_\mathrm{ew})+\dfrac{\gamma_{JI}}{16\pi^2} \log(\mu_\mathrm{low}/\mu_\mathrm{ew})\,C_J(\mu_\mathrm{ew})\,,
\end{equation}

\noindent where $\gamma_{JI}$ are the elements of the one-loop anomalous-dimension matrix, which is defined via $16 \pi^2\, \mu \mathrm{d} C_I /\mathrm{d}\mu = \gamma_{JI} \,C_J$. These \emph{leading logarithm}  effects can be schematically written in our case as follows~\cite{Crivellin:2017rmk},

\begin{align}
\label{eq:adm-left}
\begin{pmatrix}
\vec{\bm{C}}_{V} \\[0.3em]
\vec{\bm{C}}_{S} \\[0.3em] 
\vec{\bm{C}}_{T} \\[0.3em] 
\vec{\bm{C}}_{D} \\[0.3em] 
\vec{\bm{C}}_{G}  
\end{pmatrix}_{\mu_\mathrm{low}}
\hspace{-1.2em}\overset{\tiny 1-\mathrm{loop}}{\simeq}\begin{pmatrix}
U_{VV} &  0&  0&  0 & 0\\[0.3em] 
0 & U_{SS} & U_{ST}  & 0 & U_{SG} \\[0.3em] 
0 & U_{TS} & U_{TT} & 0 & 0 \\[0.3em] 
0 & 0 & U_{DT} & U_{DD} & 0 \\[0.3em] 
0 & 0 & 0 & 0  & U_{GG}
\end{pmatrix}
\cdot 
\begin{pmatrix}
\vec{\bm{C}}_{V} \\[0.3em]
\vec{\bm{C}}_{S} \\[0.3em] 
\vec{\bm{C}}_{T} \\[0.3em] 
\vec{\bm{C}}_{D} \\[0.3em] 
\vec{\bm{C}}_{G}  
\end{pmatrix}_{\mu_\mathrm{ew}}\,,
\end{align}
\noindent where we collect the $\Delta L_{i}=-\Delta L_{j}$ Wilson coefficients in the vectors
\begin{align}
    \vec{\bm{C}}_{V} &= \Big{(} \vec{{C}}_{\substack{V_{LL}\\ij }}^{(d)}\,,~\vec{{C}}_{\substack{V_{LR}\\ij  }}^{(d)}\,,~\vec{{C}}_{\substack{V_{LL}\\ij }}^{(u)}\,,~\vec{{C}}_{\substack{V_{LR}\\ij  }}^{(u)}\,,~\vec{{C}}_{\substack{V_{LL}\\ij }}^{(\ell)}\,,~\vec{{C}}_{\substack{V_{LR}\\ij }}^{(\ell)}\Big{)}\,,\\[0.35em]
    \vec{\bm{C}}_{S} &= \Big{(} \vec{{C}}_{\substack{S_{LL}\\ij }}^{(d)}\,,~\vec{{C}}_{\substack{S_{LR}\\ij  }}^{(d)}\,,~\vec{C}_{\substack{S_{LL}\\ij }}^{(u)}\,,~\vec{{C}}_{\substack{S_{LR}\\ij  }}^{(u)}\,,~\vec{C}_{\substack{S_{LL}\\ij }}^{(\ell)}\,,~\vec{{C}}_{\substack{S_{LR}\\ij  }}^{(\ell)} \Big{)}\,, \\[0.35em] 
    \vec{\bm{C}}_{T} &= \Big{(} \vec{{C}}_{\substack{T_L\\ij }}^{(d)}\,,~\vec{{C}}_{\substack{T_L\\ij  }}^{(u)}\,,~\vec{{C}}_{\substack{T_L\\ij  }}^{(\ell)}\Big{)}\,, \\[0.35em] 
    \vec{\bm{C}}_{D} &= \Big{(}C_{\substack{D_L\\ ij}}\Big{)} \,, \\[0.35em] 
    \vec{\bm{C}}_{G} &= \Big{(}C_{\substack{G_L\\ ij}}\,,~\widetilde C_{\substack{G_L\\ij}}\Big{)}\,,
\end{align}

\noindent where lepton flavor indices $i,j$ are fixed, and the sub-vectors with all possible quark flavor indices are defined e.g.~at the $\mu=\mu_\mathrm{EW}$ scale as~\footnote{Note, in particular, that the up-type quark coefficients should not contain the top-quark as it has been integrated out. Furthermore, one should use the basis defined in Sec.~\ref{ssec:left} to avoid redundancies related to Fierz identities.}
\begin{equation}
\vec{{C}}_{\substack{V_{LL}\\ ij}}^{(d)} = \Big{(} C_{\substack{V_{LL}\\ij  11}}^{(d)}\,,~C_{\substack{V_{LL}\\ij  12}}^{(d)}\,,~C_{\substack{V_{LL}\\ij  13}}^{(d)}\,,~C_{\substack{V_{LL}\\ij  21}}^{(d)}\,,~C_{\substack{V_{LL}\\ij  22}}^{(d)}\,,~C_{\substack{V_{LL}\\ij  23}}^{(d)}\,,~C_{\substack{V_{LL}\\ij  31}}^{(d)}\,,~C_{\substack{V_{LL}\\ij  32}}^{(d)}\,,~C_{\substack{V_{LL}\\ij  33}}^{(d)} \Big{)}\,.
\end{equation}

\noindent Similar definitions hold for the four-lepton operators, where we only consider operators that violate lepton flavor by one unit, e.g., 
\begin{equation}
\vec{{C}}_{\substack{V_{LL}\\ ij}}^{(\ell)} = \Big{(} C_{\substack{V_{LL}\\ij  11}}^{(\ell)}\,,~C_{\substack{V_{LL}\\ij22}}^{(\ell)}\,,~C_{\substack{V_{LL}\\ij  33}}^{(\ell)} \Big{)}\,.
\end{equation}

\noindent Expressions for the operators with a flipped chirality can be obtained from Eq.~\eqref{eq:adm-left} via the trivial replacement $L\leftrightarrow R$.

Going beyond the leading-logarithm approximation can be important since coefficients such as the dipoles $C_{D_{L(R)}}$ are subject to very stringent experimental constraints from $\mu\to e\gamma$ and $\mu N \to e N$~\cite{Zyla:2020zbs}, which are far more constraining than the ones on LFV mesons decays, and which will be considerably improved in the coming years. These constraints can supersede the direct ones on operators that are not efficiently constrained at the leading order, provided that they contribute at higher loop orders to dipoles~\cite{Ardu:2021koz}. A noticeable example is the vector four-fermion operators, which mix into the dipoles at the two-loop level, yielding~\cite{Crivellin:2017rmk}~\footnote{This is a well-known effect in the quark sector for the $b\to s\ell\ell$ transition, where the effective coefficient $C_7$ mixes into $C_9$ at two-loop order in QCD, cf.~e.g.~Ref.~\cite{Ciuchini:1993fk}. See Ref.~\cite{Crivellin:2017rmk} for a brief discussion on the renormalization-scheme dependence of these two-loop contributions.}
\begin{align}
    \label{eq:vector-into-dipole}
    C_{\substack{D_L\\ ij}} (\mu_\mathrm{low}) \propto \dfrac{e^3 Q_f^2 Q_l N_c^{(f)}}{(16\pi^2)^2}\log \bigg{(}\dfrac{\mu_\mathrm{low}}{\mu_\mathrm{ew}}\bigg{)} C_{\substack{V_{XY}\\ij kk}}^{(f)} (\mu_\mathrm{ew}) \,,
\end{align}
where $f \in \lbrace u,d,\ell \rbrace$ denotes a fermion, with electric charge $Q_f$ and color number $N_c^{(f)}$. For instance, for the $\mu\to e$ transition and for a heavy fermion $f\in \lbrace c,\tau,b,t \rbrace$, the two-loop constraint from $\mu\to e\gamma$ obtained in this way can be competitive with the one-loop constraint from $\mu\to eee$~\cite{Crivellin:2017rmk}. Another relevant example is the scalar $C_{S_{LR}}^{(q)}$, which can be Fierzed into a vector operator that also mixes into $C_{D_L}$ at two-loop order, thus being constrained by $\mu\to e\gamma$~\cite{Crivellin:2017rmk}.

Besides the above-mentioned (single logarithm) two-loop effects, there are also double-logarithm contributions that are relevant to phenomenology. The most relevant one is the product of the one-loop matrices $U_{DT} \times U_{TS}$, which induces the one-loop mixing of a scalar into a tensor operator, which then mixes into the dipole at one-loop again~\cite{Davidson:2016edt}, 
\begin{align}
    \label{eq:scalar-into-dipole}
    C_{\substack{D_L\\ ij}}  (\mu_\mathrm{low})\propto  \frac{e^3 Q_q^2 Q_{\ell} N_c^{(q)} }{(16\pi^2)^2} \dfrac{m_q}{m_\ell}\Big{[}\log\Big{(}\frac{\mu_{\text{low}}}{\mu_{\text{ew}}}\Big{)} \Big{]}^2 C^{(q)}_{\substack{S_{LL}\\ijkk}}(\mu_\mathrm{ew}) \,.
\end{align}
The loop-level constraints on $C_{S_{LL}}^{(q)}$ from $\ell_j \to \ell_i \gamma$ can also supersede the ones from tree-level processes despite the additional suppression by $1/(16\pi^2)$. In this case, the chirality enhancement ($\propto m_q/m_\ell$) plays an important role in making this contribution sizable when $q$ is a heavy quark.

Even though the two-loop anomalous-dimension matrix in the low-energy EFT is not yet fully known for a consistent analysis of these processes at order $1/(16\pi^2)^2$, we opt to account for these contributions, by using the expressions from Ref.~\cite{Crivellin:2017rmk}, since they can be the dominant effects in some cases. However, the constraints depending on these effects should be more carefully reassessed when the full two-loop anomalous-dimensions calculations will become available. Finally, we remind the reader that the RGE contributions that we consider below the electroweak scale are summarized in Fig.~\ref{fig:gauge-left-rge}.

\subsection{SM EFT}
\label{ssec:smeft}

The dimension $d=6$ SMEFT Lagrangian can be written as
\begin{equation}
\label{eq:SMEFT-Lag}
    \mathcal{L}_\mathrm{SMEFT}^{(6)} = \sum_i \dfrac{\mathcal{C}_I}{\Lambda^2} \,\mathcal{O}_I\,,
\end{equation}
where $\mathcal{C}_I$ denotes the relevant Wilson coefficients and $\Lambda$ stands for the EFT cutoff. There are only a few types of  $d=6$ operators defined above the electroweak scale that can violate lepton flavors by one unit:

\begin{itemize}
\item[i)] The semileptonic operators ($\psi^4$) collected in Table~\ref{tab:SMEFT-ope}, which contribute at tree level to the various semileptonic transitions;
\item[ii)] The four-fermion leptonic operators ($\psi^4$):
\begin{align}
\begin{split}
\hspace{-1.2em}\mathcal{O}_{\substack{ll\\ijkk}} &= \big{(}\bar{l}_i \gamma^\mu l_j\big{)}\big{(}\bar{l}_k \gamma_\mu l_k\big{)}\,,\quad\,\,
\mathcal{O}_{\substack{le\\ijkl}} = \big{(}\bar{l}_i \gamma^\mu l_j\big{)}\big{(}\bar{e}_k \gamma_\mu e_k\big{)}\,,\quad\,\,
\mathcal{O}_{\substack{ee\\ijkl}} = \big{(}\bar{e}_i \gamma^\mu e_j\big{)}\big{(}\bar{e}_k \gamma_\mu e_k\big{)}\,,
\end{split}
\end{align}

\noindent and the leptonic dipole ($\psi^2 H X$) operators:
\begin{align}
\qquad \mathcal{O}_{\substack{eB\\ij}} &=\big{(}\bar{l}_i \sigma^{\mu\nu} e_j \big{)} H B_{\mu\nu}\,, \qquad\qquad
\mathcal{O}_{\substack{eW\\ij}} =\big{(}\bar{l}_i \sigma^{\mu\nu} \tau^I e_j \big{)} H W^I_{\mu\nu}\,,
\end{align}

\noindent where $\tau^I$ ($I=1,2,3$) are the Pauli matrices. These operators become relevant when including RGE effects which mix semileptonic operators into purely leptonic ones.

\item[iii)] The Yukawa operators ($\psi^2 H^3$):
\begin{align}
\mathcal{O}_{\substack{eH\\ij}} = \big{(}\bar{l}_i e_j H \big{)}\big{(}H^\dagger H\big{)}\,,
\end{align}
which induces LFV Higgs decays after electroweak-symmetry breaking.
\item[iv)] The Higgs-current operators ($\psi^2 H^2 D$):~\footnote{We define $H^\dagger\overleftrightarrow{D}_\mu H \equiv H^\dagger \big{(}{D}_\mu H\big{)}- \big{(}{D}_\mu H^\dagger\big{)}  H$ and $H^\dagger\overleftrightarrow{D}_\mu^I H \equiv H^\dagger \tau^I\big{(}{D}_\mu H\big{)}- \big{(}{D}_\mu H^\dagger\big{)} \tau^I  H$.}
\begin{align}
\mathcal{O}_{\substack{Hl\\ ij}}^{(1)} = \big{(}H^\dagger  \overleftrightarrow{D}_\mu H  \big{)}&\big{(}\bar{l}_i \gamma^\mu l_j\big{)}\,,\qquad\quad
\mathcal{O}_{\substack{Hl\\ ij}}^{(3)} = \big{(}H^\dagger  \overleftrightarrow{D}_\mu^I H  \big{)}\big{(}\bar{l}_i \tau^I \gamma^\mu l_j\big{)}\,,\\[0.45em]
&\mathcal{O}_{\substack{He}} = \big{(}H^\dagger  \overleftrightarrow{D}_\mu H  \big{)}\big{(}\bar{e}_i \gamma^\mu e_j\big{)}\,,
\end{align}
which induce LFV $Z$-boson decays, in addition to LFV in semileptonic amplitudes mediated by the $Z$-boson, which are quark-flavor conserving.
\end{itemize}

\begin{table}[!t]
\renewcommand{\arraystretch}{2.}
\centering
\begin{tabular}{cc}
\centering
\begin{tabular}{c|c}
$\psi^4$ &   Operator\\ \hline\hline
$\mathcal{O}_{lq}^{(1)}$ & $\big{(}\bar{l}_i \gamma^\mu l_j\big{)}\big{(}\bar{q}_k \gamma_\mu q_l\big{)}$ \\ 
$\mathcal{O}_{lq}^{(3)}$ & $\big{(}\bar{l}_i \gamma^\mu \tau^I l_j\big{)}\big{(}\bar{q}_k \gamma_\mu \tau^I q_l\big{)}$\\ 
$\mathcal{O}_{lu}$ & $\big{(}\bar{l}_i \gamma^\mu l_j\big{)}\big{(}\bar{u}_k \gamma_\mu u_l\big{)}$\\ 
$\mathcal{O}_{ld}$ & $\big{(}\bar{l}_i \gamma^\mu l_j\big{)}\big{(}\bar{d}_k \gamma_\mu d_l\big{)}$\\ 
$\mathcal{O}_{eq}$ & $\big{(}\bar{e}_i \gamma^\mu e_j\big{)}\big{(}\bar{q}_k \gamma_\mu q_l\big{)}$\\ 
$\mathcal{O}_{eu}$ & $\big{(}\bar{e}_i \gamma^\mu e_j\big{)}\big{(}\bar{u}_k \gamma_\mu u_l\big{)}$\\ 
$\mathcal{O}_{ed}$ & $\big{(}\bar{e}_i \gamma^\mu e_j\big{)}\big{(}\bar{d}_k \gamma_\mu d_l\big{)}$\\ 
\end{tabular}
&
\hspace{1.2cm}
\begin{tabular}{c|c}
$\psi^4$ &   Operator $+\mathrm{h.c.}$\\ \hline\hline
$\mathcal{O}_{ledq}$ & $\big{(}\bar{l}^a_i e_j\big{)}\big{(}\bar{d}_k q^a_l\big{)}$ \\ 
$\mathcal{O}_{lequ}^{(1)}$ & $\big{(}\bar{l}^a_i e_j \big{)}\varepsilon_{ab}\big{(}\bar{q}^b_k u_l\big{)}$\\ 
$\mathcal{O}_{lequ}^{(3)}$ & $\big{(}\bar{l}^a_i \sigma^{\mu\nu}e_j \big{)}\varepsilon_{ab}\big{(}\bar{q}^b_k \sigma_{\mu\nu}u_l\big{)}$\\
\end{tabular}
\end{tabular}
\vspace{0.2cm}
\caption{\small \sl Hermitian (left) and non-Hermitian (right) $d=6$ semileptonic operators in the SMEFT. We consider the Warsaw basis~\cite{Grzadkowski:2010es}, and we renamed the operator $\mathcal{O}_{qe}$ as $\mathcal{O}_{eq}$ to have lepton flavor indices before the quark ones. The $SU(2)_L$ indices are denoted by $a,b$, with $\varepsilon_{12}=-\varepsilon_{21}=+1$, $\tau^I$ ($I=1,2,3$) are the Pauli matrices, and $SU(3)_c$ indices are omitted. Flavor indices are denoted by $i,j,k,l$ and should be assigned as $\mathcal{O} \equiv \mathcal{O}_{ijkl}$ in the left columns of the table. See Appendix~\ref{app:conventions} for our notation for the SM fields.}
\label{tab:SMEFT-ope} 
\end{table}

\noindent Similarly to Sec.~\ref{ssec:left}, flavor indices are denoted by Latin symbols. For semileptonic operators, our convention is that the first two indices always correspond to lepton flavors. For Hermitian operators such as $\mathcal{O}_{ld}$, we further impose that lepton indices satisfy $i<j$ to avoid redundancy in the operator basis. Moreover, we assume that down-quark Yukawas are diagonal, i.e.~the Cabibbo–Kobayashi–Maskawa (CKM) matrix $V$ appears in the upper component of $q_i=[(V^\dagger u_{L})_i\,~d_{Li}]^T$, cf.~Appendix~\ref{app:conventions}. 

The SMEFT Lagrangian in Eq.~\eqref{eq:SMEFT-Lag} can be matched to the Low-Energy EFT defined in Eq.~\eqref{eq:left-lagrangian}, as provided at tree level in Appendix~\ref{app:SMEFT-matching} (see also Ref.~\cite{Jenkins:2017jig}). We note, in particular, that the leptonic scalar and tensor coefficients $C_{S_{XY}}^{(\ell)}$ and $C_{T_{X}}^{(\ell)}$ vanish at $d=6$, only appearing at higher dimensions via insertions of the Higgs doublets, which are needed to make the operator invariant under $SU(2)_L\times U(1)_Y$. The same holds for the down-type tensor coefficient $C_{T_{X}}^{(d)}$ and other four-fermion semileptonic operators, cf.~Ref.~\cite{Alonso:2014csa}.

\subsubsection*{Operator Mixing}

\begin{figure}[t!]
\centering
\includegraphics[width=0.92\linewidth]{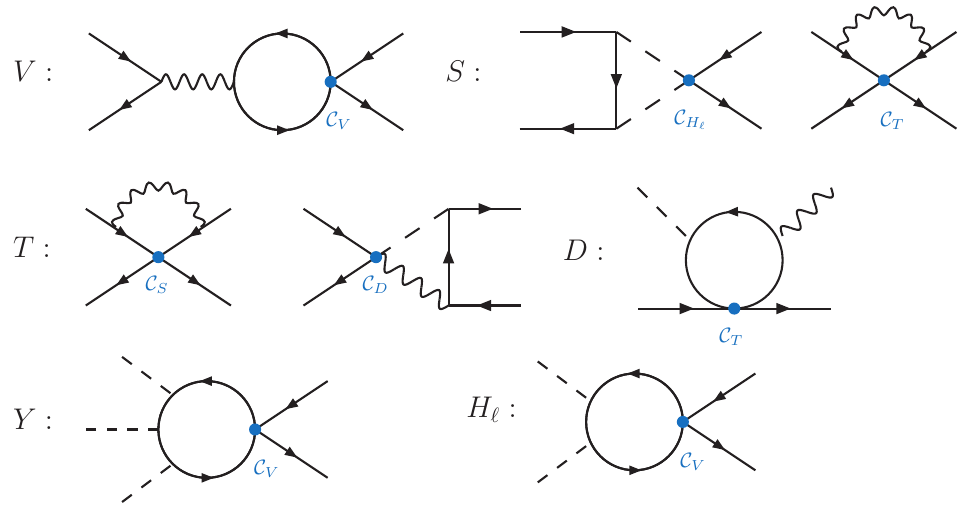}
\caption{\small \sl Schematic representation of one-loop diagrams that can induce non-diagonal operator-mixing via the $SU(2)_L\times U(1)_Y$ gauge and Yukawa interactions above the electroweak scale~\cite{Jenkins:2013zja}. The blue dots represent insertions of specific $d=6$ operators appearing in Eq.~\eqref{eq:adm-smeft}. }
\label{fig:gauge-smeft-rge} 
\end{figure}

The different types of LFV operators listed above can mix through renormalization group evolution~\cite{Jenkins:2013zja}. For the operators that we consider, the relevant RGE effects are induced by the electroweak and the Yukawa interactions, which mix different operators, in addition to the QCD running that can only change the magnitude of scalar and tensor semileptonic Wilson coefficients. The RGE effects generated by the electroweak interactions are necessarily flavor-conserving, whereas the Yukawa ones can induce quark-flavor violation through the CKM matrix, as depicted in Fig.~\ref{fig:gauge-smeft-rge}.

In what follows, we consider the $d=6$ LFV operators with fixed lepton flavor indices $i<j$.~\footnote{To this order in the EFT expansion, these operators do not mix with operators with different combinations of lepton flavors. See Ref.~\cite{Ardu:2022pzk} for the double-insertion of $d=6$ operators and their mixing into $d=8$ operators at one-loop.} The leading-logarithm solution to the RGEs can then be schematically written as follows~\cite{Jenkins:2013zja},

\begin{align}
\label{eq:adm-smeft}
\begin{pmatrix}
\vec{\bm{\mathcal{C}}}_{V} \\[0.3em]
\vec{\bm{\mathcal{C}}}_{S} \\[0.3em] 
\vec{\bm{\mathcal{C}}}_{T} \\[0.3em] 
\vec{\bm{\mathcal{C}}}_{D} \\[0.3em] 
\vec{\bm{\mathcal{C}}}_{\mathrm{Yuk}}  \\[0.3em] 
\vec{\bm{\mathcal{C}}}_{H_\ell}  
\end{pmatrix}_{\mu_\mathrm{ew}}
\hspace{-1.2em}\overset{\tiny 1-\mathrm{loop}}{\simeq}\begin{pmatrix}
\mathcal{U}_{VV} &  0&  0&  0 & 0 & \mathcal{U}_{VH_\ell}\\[0.4em] 
0 & \mathcal{U}_{SS} & \mathcal{U}_{ST}  & 0 & 0 &0 \\[0.4em] 
0 & \mathcal{U}_{TS} & \mathcal{U}_{TT} & \mathcal{U}_{TD} & 0 &0\\[0.4em] 
0 & 0 & \mathcal{U}_{DT} & \mathcal{U}_{DD} & 0  &0\\[0.4em] 
0 & \mathcal{U}_{\mathrm{Y}S} & 0 & 0  & \mathcal{U}_{\mathrm{Y}\mathrm{Y}} &0 \\[0.4em] 
\mathcal{U}_{H_\ell V} & 0 & 0 & 0  &0  &\mathcal{U}_{H_\ell H_\ell}
\end{pmatrix}
\cdot 
\begin{pmatrix}
\vec{\bm{\mathcal{C}}}_{V} \\[0.3em]
\vec{\bm{\mathcal{C}}}_{S} \\[0.3em] 
\vec{\bm{\mathcal{C}}}_{T} \\[0.3em] 
\vec{\bm{\mathcal{C}}}_{D} \\[0.3em] 
\vec{\bm{\mathcal{C}}}_{\mathrm{Yuk}}  \\[0.3em] 
\vec{\bm{\mathcal{C}}}_{H_\ell}  
\end{pmatrix}_{\Lambda}\,,
\end{align}

\noindent where we only kept the contributions proportional to the top and bottom quark Yukawas, as well as the SM gauge couplings, which are taken at the scale $\mu\simeq \mu_\mathrm{ew}$. In the leading-logarithm approximation, the matrices $U_{IJ}$ can be obtained from the SMEFT anomalous dimensions $\gamma_{JI}$ given in Ref.~\cite{Jenkins:2013zja} via the analogous expression of Eq.~\eqref{eq:leading-log}. The effective coefficients are combined in the above equation in the following vectors,
\begin{align}
    \bm{\vec{\mathcal{C}}}_{V} &= \Big{(}{\vec{\mathcal{C}}}_{\substack{lq\\ ij}}^{\,(1)}\,,~{\vec{\mathcal{C}}}_{\substack{lq\\ ij}}^{\,(3)}\,,~{\vec{\mathcal{C}}}_{\substack{lu\\ ij}}\,,~{\vec{\mathcal{C}}}_{\substack{ld\\ ij}}\,,~{\vec{\mathcal{C}}}_{\substack{eq\\ ij}}\,,~{\vec{\mathcal{C}}}_{\substack{eu\\ ij}}\,,~{\vec{\mathcal{C}}}_{\substack{ed\\ ij}}\,,~{\vec{\mathcal{C}}}_{\substack{ll\\ ij}}\,,~{\vec{\mathcal{C}}}_{\substack{l e\\ ij}}\,,~{\vec{\mathcal{C}}}_{\substack{ee\\ ij}}\Big{)}\,,\\[0.35em]
    \bm{\vec{\mathcal{C}}}_{S} &= \Big{(}\vec{\mathcal{C}}_{\substack{ledq\\ ij}}\,,~\vec{\mathcal{C}}_{\substack{lequ\\ ij}}^{\,(1)}\Big{)}\,,\\[0.35em]
    \bm{\vec{\mathcal{C}}}_{T} &= \Big{(} \vec{\mathcal{C}}_{\substack{lequ\\ ij}}^{\,(3)} \Big{)}\,,\\[0.35em] 
    \bm{\vec{\mathcal{C}}}_{D} &= \Big{(} \mathcal{C}_{\substack{eB\\ ij}}\,,~\mathcal{C}_{\substack{eW\\ ij}}\Big{)}\,,\\[0.35em]
    \bm{\vec{\mathcal{C}}}_{\mathrm{Yuk}} &= \Big{(} \mathcal{C}_{\substack{eH\\ ij}}\Big{)}\,,\\[0.35em]
    \bm{\vec{\mathcal{C}}}_{H_\ell} &= \Big{(} \mathcal{C}_{\substack{Hl\\ ij}}^{(1)}\,,~\mathcal{C}_{\substack{Hl\\ ij}}^{(3)}\,,~\mathcal{C}_{\substack{He\\ ij}}\Big{)}\,,
\end{align}
where the four-fermion subvectors are defined in such a way as to comprise all possible quark flavor indices, for fixed lepton flavor indices $\lbrace i,j\rbrace$, e.g.,
\begin{equation}
    {\vec{\mathcal{C}}}_{\substack{lq\\ ij}}^{\,(1)} = \Big{(} \mathcal{C}_{\substack{lq\\ij  11}}^{(1)}\,,~\mathcal{C}_{\substack{lq\\ij  12}}^{(1)}\,,~\mathcal{C}_{\substack{lq\\ij  13}}^{(1)}\,,~\mathcal{C}_{\substack{lq\\ij  21}}^{(1)}\,,~\mathcal{C}_{\substack{lq\\ij  22}}^{(1)}\,,~\mathcal{C}_{\substack{lq\\ij  23}}^{(1)}\,,~\mathcal{C}_{\substack{lq\\ij  31}}^{(1)}\,,~\mathcal{C}_{\substack{lq\\ij  32}}^{(1)}\,,~\mathcal{C}_{\substack{lq\\ij  33}}^{(1)} \Big{)}\,,
\end{equation}

\noindent with similar definitions for the other operators.

Besides the leading-logarithm contributions discussed above, we stress once again that two-loop RGE effects can be relevant if they introduce a new source of mixing that does not appear at the one-loop level. A relevant example in the SMEFT is the double-logarithm mixing of the scalar operator $\mathcal{C}_{lequ}^{(1)}$ into the tensor $\mathcal{C}_{lequ}^{(3)}$, which then mixes into the dipoles $\mathcal{C}_{eW}$ and $\mathcal{C}_{eB}$, via the product $\mathcal{U}_{DT}\times \mathcal{U}_{TS}$ in Eq.~\eqref{eq:adm-smeft}. This mixing induces a chirality-enhancement similar to Eq.~\eqref{eq:scalar-into-dipole}, which provides the most stringent constraint on the scalar coefficient $\smash{\mathcal{C}_{lequ}^{(1)}}$ with third-generation quarks, as will be shown in Sec.~\ref{sec:pheno}.

Finally, we note that Eq.~\eqref{eq:adm-smeft} was obtained by neglecting the lepton Yukawas ($y_e$). However, we found that one particular type of contribution proportional to the lepton Yukawas is relevant to phenomenology, namely the Yukawa-induced mixing of the vector coefficients into tensors~\cite{Jenkins:2017dyc},
\begin{align}
\begin{split}
\label{eq:smeft-double-log}
\mu\, \dfrac{\mathrm{d}}{\mathrm{d}\mu} \mathcal{C}_{\substack{lequ\\ijkl}}^{(3)} \overset{\mathrm{yuk.}}{=} &-\dfrac{[Y_u^\dagger]_{kw} [Y_e^\dagger]_{iv}}{32\pi^2}  \,\Big{(} \mathcal{C}_{\substack{eu\\vj wl}}
+\mathcal{C}_{\substack{lu\\vj wl}}\Big{)}-\dfrac{[Y_u^\dagger]_{wl} [Y_e^\dagger]_{vj}}{32\pi^2}  \,\Big{(} \mathcal{C}_{\substack{lq\\iv kw}}^{(1)}-3\,\mathcal{C}_{\substack{lq\\iv kw
}}^{(3)}\Big{)}\\
&-\dfrac{[Y_u^\dagger]_{vl} [Y_e^\dagger]_{iw}}{32\pi^2}\mathcal{C}_{\substack{eq\\wj kv}}+\dots\,,
\end{split}
\end{align}

\noindent where $Y_u=\mathrm{diag}(y_u,y_c,y_t)\cdot V$ and $Y_e=\mathrm{diag}(y_e,y_\mu,y_\tau)$. This effect is sizable for coefficients third-generation quarks since their contribution is proportional to $y_t$. The tensor coefficient then mixes into dipoles~\cite{Jenkins:2017dyc}, inducing contributions scaling such as
\begin{align}
\label{eq:smeft-double-log-2}
 \mathcal{C}_{\substack{eW\\ij}} (\mu_\mathrm{ew}) \propto \dfrac{g_2\,N_c\,y_t^2\,y_\ell}{(16\pi^2)^2} \Big{[}\log\Big{(}\frac{\mu_{\text{ew}}}{\Lambda}\Big{)} \Big{]}^2 \mathcal{C}_{\substack{V\\ij33}}(\Lambda)\,,
\end{align}

\noindent where $\mathcal{C}_V$ denotes one of the coefficients appearing in the right-hand-side of Eq.~\eqref{eq:smeft-double-log}, and $y_\ell$ can be either $y_{\ell_i}$ or $y_{\ell_j}$. Similar expressions apply for $\mathcal{C}_{eB}$ by replacing $g_2\to g_1$. We will include these effects in our analysis in Sec.~\ref{sec:pheno}, which allow us to derive stringent limits on vector semileptonic coefficients, with third-generation quarks, by using $\ell_j\to\ell_i\gamma$ (see also Ref.~\cite{EliasMiro:2021jgu}).

\subsection{Summary}
\label{ssec:summary}

Finally, we briefly summarize the loop effects that will be considered in our phenomenological analysis in Sec.~\ref{sec:pheno}, which are classified in terms of a logarithmic expansion. Firstly, for the low-energy EFT, we will consider
\begin{itemize}
    \item The one-loop mixing of the low-energy LFV operators from Ref.~\cite{Crivellin:2017rmk}, which are proportional to
    \begin{equation}
         \dfrac{1}{16\pi^2} \log \bigg{(}\dfrac{\mu_\mathrm{low}}{\mu_\mathrm{ew}}\bigg{)}\,,
    \end{equation}
    as described in Eq.~\eqref{eq:adm-left} (cf.~also Ref.~\cite{Jenkins:2017dyc}).
    \item The two-loop single logarithm mixing of semileptonic vector and operators into leptonic dipoles calculated in Ref.~\cite{Crivellin:2017rmk}, 
    \begin{equation}
         \dfrac{1}{(16\pi^2)^2} \log\Big{(}\frac{\mu_{\text{low}}}{\mu_{\text{ew}}}\Big{)} \,,
    \end{equation}
    as illustrated in Eqs.~\eqref{eq:vector-into-dipole} and~\eqref{eq:scalar-into-dipole}.
    \item The two-loop double logarithm mixing $C_{S_{LL}}^{(q)} \overset{\mathrm{1-loop}}{\longrightarrow} C_{T_{LL}}^{(q)}\overset{\mathrm{1-loop}}{\longrightarrow} C_{D_L}^{(\ell)}$, which induces contributions to dipoles proportional to
    \begin{equation}
         \dfrac{1}{(16\pi^2)^2} \bigg{[}\log\Big{(}\frac{\mu_{\text{low}}}{\mu_{\text{ew}}}\Big{)} \bigg{]}^2\,,
    \end{equation}
    which are chirality-enhanced for heavy-quarks.
    \item The finite matching contributions of scalar semileptonic operators to the gluonic ones~\cite{Shifman:1978zn}, as given in Eq.~\eqref{eq:GGtilde1}. 
\end{itemize}

\noindent For the SMEFT, we will consider
\begin{itemize}
    \item The one-loop mixing of the SMEFT operators taken from Ref.~\cite{Jenkins:2013zja}, which are proportional to 
    \begin{equation}
         \dfrac{1}{16\pi^2} \log \Big{(}\frac{\mu_{\text{ew}}}{\Lambda}\Big{)}\,,
    \end{equation}
    as shown in Eq.~\eqref{eq:adm-smeft}.
    
    \item The double-logarithm two-loop mixing $\mathcal{O}_{lequ}^{(1)} \overset{\mathrm{1-loop}}{\longrightarrow} \mathcal{O}_{lequ}^{(3)} \overset{\mathrm{1-loop}}{\longrightarrow}  \mathcal{O}_{eV}$ into dipoles, which is induced by the product of gauge and Yukawa running, and the Yukawa-induced ${\lbrace \mathcal{O}_{lq}^{(1,3)}, \mathcal{O}_{lu}, \mathcal{O}_{eq}, \mathcal{O}_{eu} \rbrace \overset{\mathrm{1-loop}}{\longrightarrow} \mathcal{O}_{lequ}^{(3)} \overset{\mathrm{1-loop}}{\longrightarrow} \mathcal{O}_{eV}}$ (with $V=W,B$), which is suppressed by lepton masses, but which remains relevant for phenomenology. These effects are proportional to
    \begin{equation}
         \dfrac{1}{(16\pi^2)^2} \bigg{[}\log\Big{(}\frac{\mu_{\text{ew}}}{\Lambda}\Big{)}\bigg{]}^2 \,.
    \end{equation}
    as described e.g.~in~Eq.~\eqref{eq:smeft-double-log-2}.
    
\end{itemize}

\noindent The expressions for the relevant anomalous dimensions can be found in the references given above. Finally, we stress once again that our analysis of two-loop contributions is not entirely self-consistent and should be rather understood as a qualitative indication of the dominant effects, which must be refined in the future with a precise two-loop calculation.

\section{Low-energy probes}
\label{sec:LEFT-LFV}

In this Section, we provide the general expressions for the low-energy LFV observables in terms of the EFT defined in Sec.~\ref{ssec:left}. Firstly, we will discuss direct probes of semileptonic operators at tree-level in LFV meson and $\tau$-lepton decays in Sec.~\ref{ssec:qq-LFV} and \ref{ssec:tau-LFV}, respectively.~\footnote{In principle, we could also consider baryon decays such as $\Lambda_b \to \Lambda \ell_i \ell_j$~\cite{Angelescu:2021lln,Bordone:2021usz}, but there are no experimental limits available for these decays yet.} In Sec.~\ref{ssec:llpgamma-LFV}, we will then discuss the purely leptonic LFV decays, which are sensitive to semileptonic operators at one-loop, cf.~Fig.~\ref{fig:gauge-left-rge} and \ref{fig:gauge-smeft-rge}. 

\begin{table}[!p]
\renewcommand{\arraystretch}{1.6}
\centering
\resizebox{\columnwidth}{!}{\begin{tabular}{|c|cc||c|cc||c|cc|}
\hline 
Process & Exp.~limit & Ref. & Process & Exp.~limit & Ref. & Process & Exp.~limit & Ref.\\ \hline\hline
$\pi^0 \to e \mu$   & $4.8 \times 10^{-10}$  & \cite{Zyla:2020zbs}  &  $\tau \to e \pi^0$  & $1.1 \times 10^{-7}$ & \cite{Zyla:2020zbs} & $\tau \to \mu \pi^0$  & $1.5\times 10^{-7}$ & \cite{Zyla:2020zbs}\\
$\eta \to e \mu$   & $8.1 \times 10^{-6}$  & \cite{Zyla:2020zbs}  &  $\tau \to e \eta$  & $1.2\times 10^{-7}$ & \cite{Zyla:2020zbs} & $\tau \to \mu \eta$  & $8.7\times 10^{-8}$ & \cite{Zyla:2020zbs}\\
$\eta^\prime \to e \mu$   & $6.3 \times 10^{-4}$  & \cite{Zyla:2020zbs}  &  $\tau \to e \eta^\prime$  & $2.1 \times 10^{-7}$ & \cite{Zyla:2020zbs} & $\tau \to \mu \eta^\prime$  & $1.8\times 10^{-7}$ & \cite{Zyla:2020zbs}\\ 
 &   &   &  $\tau\to e \rho^0$ & $2.4\times 10^{-8}$ & \cite{Zyla:2020zbs} &  $\tau\to\mu\rho^0$ &  $1.6\times 10^{-8}$ &  \cite{Zyla:2020zbs} \\ \hline
$K_L \to e \mu$   &  $6.3 \times 10^{-12}$ & \cite{Zyla:2020zbs} &  $\tau \to e K_S$   & $3.5\times 10^{-8}$ & \cite{Zyla:2020zbs} &  $\tau \to \mu K_S$ & $3.1\times 10^{-8}$ & \cite{Zyla:2020zbs}\\
$K^+\to \pi^+ \mu^- e^+$  & $1.1\times 10^{-10}$ & \cite{Zyla:2020zbs} &  $\tau \to e K^{\ast 0}$  & $4.3\times 10^{-8}$ & \cite{Zyla:2020zbs} & $\tau \to \mu K^{\ast 0}$ & $7.9\times 10^{-8}$ & \cite{Zyla:2020zbs} \\
$K_L \to \pi^0 e \mu$ & $1.0\times 10^{-10}$  & \cite{Zyla:2020zbs} &    &  &  &  & & \\
\hline
$D^0 \to e \mu$ & $1.7\times 10^{-8}$ & \cite{Zyla:2020zbs} &    $\tau \to e D^0$ & -- &  &  &  &  \\
$D^+ \to \pi^+ e \mu $ & $4.5 \times 10^{-7}$ & \cite{LHCb:2020car}   &   &  &  &  &  &  \\ 
$D_s \to K^+ e \mu $ & $1.5 \times 10^{-6}$ & \cite{LHCb:2020car}   &   &  &  &  &  &  \\ \hline
$B^0 \to  e\mu$ & $1.3\times 10^{-9}$ & \cite{Zyla:2020zbs} &    $B^0 \to  e\tau$ & $2.1\times 10^{-5}$ & \cite{Zyla:2020zbs} & $B^0 \to  \mu\tau$ & $1.4\times 10^{-5}$ & \cite{Zyla:2020zbs} \\
$B^+ \to \pi^+  e\mu$ & $1.2\times 10^{-7}$ & \cite{BaBar:2007xeb} & $B^+ \to \pi^+  e\tau$ & $1.0\times 10^{-4}$   & \cite{Zyla:2020zbs} & $B^+ \to \pi^+  \mu\tau$ & $9.7\times 10^{-5}$ & \cite{Zyla:2020zbs} \\ 
$B^0\to \rho^0 e \mu$ & -- & & $B^0\to \rho^0 e \tau$ & -- & & $B^0\to \rho^0 \mu\tau$ & -- & \\ 
$B_s\to K^0 e \mu$ & -- & & $B_s\to K^0 e \tau$ & -- & & $B_s\to K^0 \mu\tau$ & -- & \\\hline
$B_s \to  e \mu$ & $7.2\times 10^{-9}$ & \cite{Zyla:2020zbs} &    $B_s \to  e \tau$ & $1.9\times 10^{-3}$ & \cite{Belle:2023jwr} & $B_s \to  \mu\tau$ & $4.2\times 10^{-5}$ & \cite{Zyla:2020zbs} \\
$B^+ \to K^+  e\mu$ & $1.8 \times 10^{-8}$ & \cite{LHCb:2017hag} &  $B^+ \to K^+  e\tau$  & $4.1\times 10^{-5}$ & \cite{Belle:2023jwr} & $B^+ \to K^+  \mu\tau$ & $4.1\times 10^{-5}$ & \cite{Belle:2023jwr} \\
$B^0 \to K^\ast  e \mu$ & $1.2 \times 10^{-8}$ & \cite{LHCb:2017hag} &   $B^0 \to K^\ast  e \tau$ & -- & &$B^0 \to K^\ast \mu\tau$ & $2.2 \times 10^{-5}$ & \cite{LHCb:2017hag} \\
$B_s\to \phi e \mu$ & $2.0 \times 10^{-8}$ & \cite{LHCb:2017hag} & $B_s\to \phi e \tau$ & -- &  & $B_s\to \phi \mu \tau$ & -- &  \\ \hline
$\phi \to  e \mu$  & $2.7\times 10^{-6}$ & \cite{Zyla:2020zbs} &   $\tau \to e \phi$& $5.5\times 10^{-8}$ & \cite{Zyla:2020zbs} & $\tau \to \mu \phi$ & $1.1\times 10^{-7}$ & \cite{Zyla:2020zbs} \\ 
$J/\psi \to  e \mu$  & $6.1 \times 10^{-9}$ & \cite{Li:2023icf} &   $J/\psi \to  e \tau$  & $1.0\times 10^{-7}$ & \cite{Li:2023icf} & $J/\psi \to  \mu \tau$  & $2.7\times 10^{-6}$ & \cite{Zyla:2020zbs} \\ 
$\Upsilon \to  e \mu$ & $5.2 \times 10^{-7}$ & \cite{Zyla:2020zbs} & $\Upsilon \to  e \tau$   & $3.6\times 10^{-6}$ & \cite{Zyla:2020zbs} & $\Upsilon \to  \mu \tau $ & $3.6\times 10^{-6}$ & \cite{Zyla:2020zbs}\\ 
\hline
\end{tabular}}
\caption{\small \sl Experimental limits on LFV decays considered in our analysis (95\% CL). For the decays with two leptons in the final state, we quote the limits on the sum of the branching fractions with different lepton charges, i.e.~$\ell_i\ell_j\equiv \ell_i^-\ell_j^+ + \ell_i^+\ell_j^-$\,. The dash symbol represent decays that are kinematically allowed, but for which there is not an experimental limit yet.  Experimental results provided at (90\% CL) are converted to (95\% CL) following Ref.~\cite{Calibbi:2017uvl}.}
\label{tab:exp-lfv-had} 
\end{table}

\subsection{$q_l\to q_k \ell_i \ell_j$}
\label{ssec:qq-LFV}

Firstly, we consider the leptonic and semileptonic decays of pseudoscalar mesons. For convenience, we define the following combinations of (axial-)vector coefficients that will allow us to express the branching fractions in a more compact form,
\begin{align}
\label{eq:left-redef-1}
C_{VV}^{(q)} &= {C_{V_{RR}}^{(q)}+C_{V_{RL}}^{(q)}+(L\leftrightarrow R)}\,, & C_{AA}^{(q)} &= {C_{V_{RR}}^{(q)}-C_{V_{RL}}^{(q)}+(L\leftrightarrow R)}\,,\\[0.4em]
\label{eq:left-redef-2}
C_{AV}^{(q)} &= {C_{V_{RR}}^{(q)}+C_{V_{RL}}^{(q)}-(L\leftrightarrow R)}\,, & C_{VA}^{(q)} &= {C_{V_{RR}}^{(q)}-C_{V_{RL}}^{(q)}-(L\leftrightarrow R)}\,,
\end{align}

\noindent and similarly, for the (pseudo)scalar coefficients,
\begin{align}
\label{eq:left-redef-3}
C_{SS}^{(q)} &= {C_{S_{RR}}^{(q)}+C_{S_{RL}}^{(q)}+(L\leftrightarrow R)}\,, & C_{PP}^{(q)} &= {C_{S_{RR}}^{(q)}-C_{S_{RL}}^{(q)}+(L\leftrightarrow R)}\,,\\*[0.4em]
\label{eq:left-redef-4}
C_{PS}^{(q)} &= {C_{S_{RR}}^{(q)}+C_{S_{RL}}^{(q)}-(L\leftrightarrow R)}\,, & C_{SP}^{(q)} &= {C_{S_{RR}}^{(q)}-C_{S_{RL}}^{(q)}-(L\leftrightarrow R)}\,,
\end{align}

\noindent and for the tensor ones,
\begin{align}
C_{T}^{(q)} &= C_{T_R}^{(q)}+C_{T_L}^{(q)}\,, \qquad \quad \;\; C_{T_5}^{(q)} = C_{T_R}^{(q)}-C_{T_L}^{(q)}\,,
\end{align}
where, for simplicity, flavor indices are omitted. The current experimental limits on the most relevant processes of this type are collected in Table~\ref{tab:exp-lfv-had}.

\subsubsection{Leptonic decays: $P\to \ell_i\ell_j$}

One of the simplest probes of LFV at low-energies are leptonic decays of pseudoscalar mesons. Firstly, we consider decays of flavored mesons of the type $P=\bar{q}_kq_l$, with $k\neq l$, such as $K$-, $D$- and $B$-mesons. The hadronic matrix element needed to compute the decay rates reads
\begin{align}
\label{eq:P-decayconst}
\langle 0 \vert \bar{q}_k \gamma^\mu \gamma_5 q_l \vert P(P) \rangle = i f_P p^\mu\,,
\end{align}

\noindent where $f_P$ denotes the $P$-meson decay constant, which has been computed for the most relevant transition by means of numerical simulations of QCD on the lattice, cf.~Table~\ref{tab:decay-cte}~\cite{FlavourLatticeAveragingGroupFLAG:2021npn}. In terms of the low-energy EFT defined in Eq.~\eqref{eq:left-lagrangian}, the $P\to \ell_i\ell_j$ branching fraction then reads
\begin{align}
\begin{split}
   \mathcal{B}(P\to \ell_i^- \ell_j^+)&= \tau_P \, \dfrac{f_P^2 m_P m_{\ell_j}^2}{128 \pi v^4} \left(1-\dfrac{m_{\ell_j}^2}{m_P^2}\right)^2\\
   &\times\bigg{\lbrace} \bigg{|}C_{\substack{VA\\ijkl}}^{(q)} +\dfrac{m_P^2}{m_{\ell_j}(m_{q_k}+m_{q_l})}\,C_{\substack{SP\\ijkl}}^{(q)}\bigg{|}^2+ \bigg{|}C_{\substack{AA\\ijkl}}^{(q)} - \dfrac{m_P^2}{m_{\ell_j}(m_{q_k}+m_{q_l})}\,C_{\substack{PP\\ijkl}}^{(q)}\bigg{|}^2\bigg{\rbrace} \,,
\end{split}
\end{align}

\noindent where $m_P$ denotes the $P$-meson mass, and we have assumed $i<j$ and neglected the mass of the lightest lepton ($m_{\ell_i}$).~\footnote{See e.g.~Ref.~\cite{Becirevic:2016zri} for the expression without this approximation.} Note, also, that the sign of the interference term between ${C_{VA}^{(q)}}$ and ${C_{SP}^{(q)}}$ changes for the decay $P\to \ell_i^+\ell_j^-$, with opposite electric charges for the leptons, as it is proportional to the difference of lepton masses, see e.g.~Ref.~\cite{Becirevic:2016zri}. The above expressions can be used for the decays of $D$- and $B_{(s)}$-mesons with the appropriate replacement of the flavor indices. Moreover, for the kaon decays $K_{L(S)} \to e\mu$, we have to take into account that $\vert K_{L(S)} \rangle \simeq (\vert K^0\rangle \pm \vert \overline{K^0}\rangle)/\sqrt{2}$, which implies that these decays probe both $K_0 \to e\mu$ and $\overline{K}_0 \to e\mu$. In this case, the above expression can be used after making the following substitution~\cite{Borsato:2018tcz},
\begin{equation}
    C_{\substack{I\\ ijkl}}^{(q)} \to \frac{1}{\sqrt{2}}\Big{(} C^{(d)}_{\substack{I \\ 1212}} \pm C^{(d)}_{\substack{I \\ 1221}}  \Big{)} \;,
\end{equation}

\noindent where $I \in \lbrace VA,AA,SP,PP\rbrace$, and the upper (lower) sign corresponds to the $K_{L}$ ($K_S$) decay.

\subsubsection{Leptonic decays: $V\to\ell_i\ell_j$}

Next, we provide the expressions for the leptonic decays $V\to \ell_i^-\ell_j^+$ of vector mesons of the type $V=\bar{q}_k q_k$, such as $V\in \lbrace \rho, \phi,  J/\psi, \Upsilon\rbrace$. These processes are typically much less constraining than the other probes discussed in this paper, since vector mesons typically have a large total-width ($\tau_V$) that suppresses the LFV branching fractions~\cite{Angelescu:2020uug,Descotes-Genon:2023pen}, but we also discuss them for completeness. The relevant hadronic matrix elements can be defined as,
\begin{align}
\label{eq:V-decayconst}
\begin{split}\langle 0 \vert \bar{q}_k \gamma^\mu  q_k \vert V(p,\lambda) \rangle &= f_V m_V \varepsilon_\lambda^\mu(p)\,,\\*[0.45em]
\langle 0 \vert \bar{q_k} \sigma^{\mu\nu}  q_k \vert V(p,\lambda) \rangle &= i f_V^T(\mu) \big{[}\varepsilon_\lambda^\mu(p) p^\nu-\varepsilon_\lambda^\nu(p) p^\mu \big{]}\,,
\end{split}
\end{align}

\noindent where $m_V$ is $V$-meson mass,  $\varepsilon_\lambda^\mu$ denotes its polarization-vector, and $f_V$ and $f_V^T$ stand for the vector and tensor decay constants, respectively. See Ref.~\cite{Donald:2013pea,Becirevic:2012dc,Becirevic:2013bsa,Lewis:2012ir,Colquhoun:2014ica,RBC-UKQCD:2008mhs,Becirevic:2003pn} for lattice QCD calculations of the relevant vector-meson decay constants. The branching fraction can then be written in terms of Eq.~\eqref{eq:left-lagrangian} as follows,
\begin{align}
   \mathcal{B}(V\to &\ell_i^- \ell_j^+)= \tau_V \, \dfrac{f_V^2 m_V^3}{192 \pi v^4}  \left(1-\dfrac{m_{\ell_j}^2}{m_V^2}\right)^2\\*
&\times\bigg{\lbrace}\Big{[}|C_{\substack{VV\\ijkk}}^{(q)}|^2+|C_{\substack{AV\\ijkk}}^{(q)}|^2\Big{]} \bigg{(}1+\dfrac{m_{\ell_j}^2}{2m_V^2}\bigg{)}+8 \bigg{(}\dfrac{f_V^T}{f_V}\bigg{)}^2\Big{[}|{C}_{\substack{T\\ ijkk}}^{(q)}|^2+|{C}_{\substack{T_5\\ijkk}}^{(q)}|^2\Big{]}\bigg{(}1+\dfrac{2m_{\ell_j}^2}{m_V^2}\bigg{)}   \nonumber\\*[0.35em]
& +12 \dfrac{m_{\ell_j}}{m_V}\dfrac{f_V^T}{f_V} \, \mathrm{Re}\big{[}{C}_{\substack{T\\ ijkk}}^{(q)}\,C_{\substack{VV\\ijkk}}^{(q)\ast}-{C}_{\substack{T_{5}\\ ijkk}}^{(q)}\,C_{\substack{AV\\ijkk}}^{(q)\ast}\big{]} \bigg{\rbrace} \,, \nonumber
\end{align}
where we have once again assumed $i<j$ and neglected the light-lepton mass, $m_{\ell_i}$. Note, in particular, that scalar operators do not contribute to these decays, since the hadronic-matrix element with a scalar density vanishes in this case. Furthermore, we have not included the contributions from the leptonic dipoles, as they are already tightly constrained by $\ell_j\to\ell_i\gamma$.

\begin{table}[!t]
\renewcommand{\arraystretch}{1.6}
\centering
\begin{tabular}{|c|cc||c|cc|}
\hline 
$P$  &  $f_P~[\mathrm{MeV}]$ & Ref. & $V$  &  $f_V~[\mathrm{MeV}]$ & Ref.\\ \hline\hline
$f_\pi$ & $130.2(8)$ & \cite{FlavourLatticeAveragingGroupFLAG:2021npn} & $f_\rho$ &$209.4(1.5)$ &  \cite{Zyla:2020zbs} \\ 
$f_K$ & $155.7(3)$ & \cite{FlavourLatticeAveragingGroupFLAG:2021npn} & $f_\phi$ & $241(18)$ & \cite{Donald:2013pea} \\ 
$f_D$ & $212.0(7)$ & \cite{FlavourLatticeAveragingGroupFLAG:2021npn} & $f_{J/\psi}$ & $418(10)$ & \cite{Becirevic:2012dc} \\ 
$f_B$ & $190.0(1.3)$ & \cite{FlavourLatticeAveragingGroupFLAG:2021npn} & $f_{\Upsilon}$ & $649(31)$ & \cite{Colquhoun:2014ica} \\ 
$f_{B_s}$ & $230.3(1.3)$ & \cite{FlavourLatticeAveragingGroupFLAG:2021npn} &  & & \\ \hline
\end{tabular}
\caption{\small \sl Pseudoscalar $(f_P)$ and vector meson $(f_V)$ decay constants considered in our analysis.}
\label{tab:decay-cte} 
\end{table}

\subsubsection{Semileptonic decays: $P\to P^\prime\ell_i \ell_j$ }

The $P\to P^\prime\ell_i^- \ell_j^+$  decays (with $i<j$) can be generally written in terms of the effective coefficients defined in Eq.~\eqref{eq:left-lagrangian} as
\begin{align}
    \label{eq:semilep-numeric}
   \mathcal{B}(P\to P^\prime\ell_i^- \ell_j^+) &= a_{VV}\,|C_{VV}^{(q)}|^2 + a_{AV}\,|C_{AV}^{(q)}|^2+ a_{TT}\,(|C_{T}^{(q)}|^2+|C_{T_5}^{(q)}|^2)\\*[0.4em]
   & + a_{SS}\,|C_{SS}^{(q)}|^2+ a_{PS}\,|C_{PS}^{(q)}|^2 +  a_{VS} \,\mathrm{Re}[C_{VV}^{(q)} C_{SS}^{(q)\,\ast}]+ a_{AP} \,\mathrm{Re}[C_{AV}^{(q)} C_{PS}^{(q)\,\ast}]\nonumber\\*[0.4em]
   &+  a_{VT} \,\mathrm{Re}[C_{VV}^{(q)} \,C_{T}^{(q)\,\ast}]+  a_{AT_5}\, \mathrm{Re}[C_{AV}^{(q)} \,C_{T_5}^{(q)\,\ast}]\,,\nonumber
\end{align}

\noindent where $C_I$ are the effective coefficients, evaluated at the relevant low-energy scale, and $a_I$ stand for numeric coefficients that depend on the form factors. Flavor indices are omitted in the above equation, for simplicity, and should be replaced for the $q_k\to q_l \ell_i^- \ell_j^+$ transition as follows,
\begin{align}
\label{eq:wc-semilep}
C_{I}^{(q)} \to C_{\substack{I\\ ij kl}}^{(q)}\,,
\end{align}

\noindent where $I$ labels the Lorentz structure of the Wilson coefficients. In the limit where the light-lepton mass ($m_{\ell_i}$) is neglected, we find that
\begin{align}
    \label{eq:semilep-massless}
    a_{VV} &\simeq a_{AV}\,, & a_{SS} &\simeq a_{PS}\,, & a_{VS} &\simeq-a_{AP}\,, & a_{VT} &\simeq-a_{AT_5}\,,
\end{align}

\noindent which are valid up to $\mathcal{O}(m_{\ell_i}/m_{\ell_j})$ corrections. In the following, we compute these numerical coefficients for the most relevant decays based on down- and up-type quark transitions, using the analytical expressions from Appendix~\ref{app:diff-semileptonic} and the $P\to P^\prime$ form factors described below:

\begin{itemize}
    \item[$\bullet$] {$d_l\to d_k \ell_i^- \ell_j^+$}: Our numerical coefficients for the down-type quark transitions are collected in Table~\ref{tab:weights-PPpll}, where flavor indices are omitted for simplicity and should be replaced following Eq.~\eqref{eq:wc-semilep}. For these decays, the tensor coefficients $C_{T_{X}}$ are highly suppressed since they only appear at $d=8$ once the $SU(2)_L\times U(1)_Y$ symmetry is imposed~\cite{Alonso:2014csa}, cf.~also~Appendix~\ref{app:SMEFT-matching}. For this reason, we will not quote the numerical values for the coefficients involving tensor operators. The numeric coefficients are computed using lattice QCD form-factors for the $K\to \pi$~\cite{Carrasco:2016kpy}, $B\to K$~\cite{Bailey:2015dka} (cf.~Ref.~\cite{Becirevic:2023aov}) and $B_s\to K$~\cite{FlavourLatticeAveragingGroupFLAG:2021npn} transitions. For $B\to \pi$ decays, we have used the combined fit of experimental~\cite{HFLAV:2022esi} and Lattice QCD~\cite{FermilabLattice:2015mwy} data made in Ref.~\cite{FlavourLatticeAveragingGroupFLAG:2021npn} to have better control of the uncertainties associated with the extrapolation of the lattice form-factors to the entire physical region.~\footnote{Note, also, that we have not included effects from $B_s-\overline{B}_s$ mixing in the above expressions, but these could be easily implemented following Ref.~\cite{Descotes-Genon:2015hea}.}

We remind the reader that the $K_{L(S)}\to \pi^0 e^- \mu^+$ decays have to be treated separately, since $\vert K_{L(S)} \rangle \simeq (\vert K^0\rangle \pm \vert \overline{K^0}\rangle)/\sqrt{2}$~\cite{Borsato:2018tcz}. Therefore, we have to amend Eq.~\eqref{eq:semilep-numeric} via the replacements,
\begin{align}
\label{eq:repl-K0}
    C_{V_{XY}}^{(q)} \rightarrow \frac{1}{2}\Big{(}C_{\substack{V_{XY}\\12 21}}^{(q)} \mp C_{\substack{V_{XY}\\ 1212}}^{(q)}\Big{)}\,, \qquad\quad  C_{S_{XY}}^{(q)} \rightarrow  \frac{1}{2}\Big{(} C_{\substack{S_{XY}\\12 21}}^{(q)} \pm C_{\substack{S_{XY}\\ 1212}}^{(q)}\Big{)} \,, 
\end{align}

\noindent for $X,Y \in \lbrace L,R\rbrace$, where the upper (lower) sign corresponds  to the $K_L$ ($K_S$) decays. Similar replacements must be made for the coefficients with $L\leftrightarrow R$.

\item[$\bullet$] {$u_l\to u_k \ell_i^- \ell_j^+$}: The only up-type transition relevant for our study is $c\to u e \mu$, which can induce e.g.~the decays $D\to \pi e \mu$ and $D_s\to K e\mu$. Notice that processes with the $\tau$-lepton are phase-space forbidden. For the $c\to u e\mu$ decays, we consider the $D\to \pi$ lattice QCD form-factors from~Ref.~\cite{Lubicz:2017syv} and the $D_s\to K$ ones from Ref.~\cite{FermilabLattice:2022gku},~\footnote{See also the recent lattice QCD results for the $D\to \pi$ from Ref.~\cite{FermilabLattice:2022gku}.} which allow us to determine the numerical coefficients collected in Table~\ref{tab:weights-PPpll-up} by neglecting the electron mass.

\end{itemize}

\begin{table}[!p]
\renewcommand{\arraystretch}{1.5}
\centering
\scalebox{0.93}{
\begin{tabular}{|c|cccccc|}
\hline
$P\to P^\prime \ell_i^-\ell_j^+$ &  $a_{VV}$ & $a_{AV}$ & $a_{SS}$ & $a_{PS}$ & $a_{VS}$ & $a_{AP}$ \\ \hline\hline
$K^+ \to \pi^+ e^- \mu^+$  &$0.1570(11) $ &$0.1578(11)$ &$2.71(3)$ &$2.72(3)$ &$-0.723(7)$ &$0.735(7)$ \\ 
$K_L \to \pi^0 e^- \mu^+$ & $0.690(5)$ & $0.693(5)$ & $12.18(14)$ & $12.23(14)$ & $-3.19(3)$ & $3.24(3)$ \\ 
$K_S \to \pi^0 e^- \mu^+$ & $0.001208(9)$ & $0.001213(9)$ & $0.0213(2)$ & $0.0214(2)$ & $-0.00558(5)$ & $0.00567(5)$ \\ \hline
$B^+ \to \pi^+ e^- \mu^+$  & $1.46(12)$ &$1.46(12)$ &$2.37(14)$ &$2.37(14)$ &$-0.154(10)$ &$0.156(10)$ \\
$B^+ \to \pi^+ e^- \tau^+$  &$1.01(7)$ &$1.01(7)$ &$1.49(9)$ &$1.49(9)$ &$-1.29(8)$ &$1.29(8)$ \\
$B^+ \to \pi^+ \mu^- \tau^+$  &$1.00(7)$ &$1.03(7)$ &$1.45(8)$ &$1.53(9)$ &$-1.17(7)$ &$1.42(8)$ \\ \hline
$B_s \to K_S e^- \mu^+$  &$0.42(9)$ &$0.42(9)$ &$0.69(8)$ &$0.69(8)$ &$-0.043(8)$ &$0.043(8)$ \\ 
$B_s \to K_S e^- \tau^+$  &$0.31(4)$ &$0.31(4)$ &$0.43(4)$ &$0.43(4)$ &$-0.39(4)$ &$0.39(5)$ \\ 
$B_s \to K_S \mu^- \tau^+$  &$0.31(4) $ &$0.32(4)$ &$0.42(4)$ &$0.44(4)$ &$-0.35(4)$ &$0.42(5)$ \\ \hline
$B^+ \to K^+ e^- \mu^+$  & $1.92(6)$ & $1.92(6)$ & $2.72(7)$ & $2.72(7)$ & $-0.209(6)$ & $0.211(6)$ \\ 
$B^+ \to K^+ e^- \tau^+$  & $1.20(3)$ &$1.20(3)$ &$1.55(3)$ &$1.55(3)$ &$-1.53(4)$ &$1.53(4)$ \\ 
$B^+ \to K^+ \mu^- \tau^+$  &$1.18(3)$ &$1.22(3)$ &$1.49(3)$ &$1.60(4)$ &$-1.37(3)$ &$1.68(4)$ \\ \hline
\end{tabular}
}
\caption{\sl \small Numerical coefficients defined in Eq.~\eqref{eq:semilep-numeric} for the decays $P\to P^\prime \ell_i^- \ell_j^+$ based on the down-type transition $\smash{d_k \to d_l \ell_i^- \ell_j^+}$ (cf.~Eq.~\eqref{eq:wc-semilep}). Decays with opposite lepton electric charges in the final state can be obtained via the replacement $ a_{VS} \to - a_{VS}$. Moreover, the expressions for the $\smash{K_{L(S)} \to \pi^0 e^+ \mu^-}$ decays are obtained through the replacement specified in Eq.~\eqref{eq:repl-K0}. See text for details on the hadronic inputs considered.}
\label{tab:weights-PPpll} 
\vspace{5em}
\renewcommand{\arraystretch}{1.5}
\centering
\begin{tabular}{|c|ccccc|}
\hline
$P\to P^\prime \ell_i^-\ell_j^+$ &  $a_{VV}=a_{AV}$ & $a_{SS}=a_{PS}$ & $a_{VS}=-a_{AP}$ & $a_{TT}$ & $a_{VT} = a_{AT_5}$\\ \hline\hline
$D^+ \to \pi^+ e^- \mu^+$  & $0.0166(10) $ & $0.0337(11)$ &$-0.0066(3)$  & $0.019(3)$ & $0.0081(7)$  \\ 
$D^0 \to \pi^0 e^- \mu^+$  &$0.00326(19) $ &  $0.0066(2)$ & $-0.00130(5)$ & $0.0038(6)$ & $0.00160(14)$  \\ 
$D_s \to K^+ e^- \mu^+$  & $0.00622(6)$  & $0.00893(3)$ & $-0.002364(8)$ & $0.00377(4)$ & $0.00250(2)$ \\ \hline
\end{tabular}
\caption{\sl \small  Numerical coefficients defined in Eq.~\eqref{eq:semilep-numeric} for the decays $P\to P^\prime e^- \mu^+$ based on the up-type transition $\smash{c \to u e^- \mu^+}$. These expressions have been obtained by neglecting the electron mass, cf.~Eq.~\eqref{eq:semilep-massless}. See caption of Fig.~\ref{tab:weights-PPpll} for details. }
\vspace{3em}
\label{tab:weights-PPpll-up} 
\end{table}

\begin{table}[!t]
\renewcommand{\arraystretch}{1.7}
\centering
\resizebox{\columnwidth}{!}{\begin{tabular}{|c|cccccccc|}
\hline
$P\to V \ell\ell$ &  $a_{VV}$ & $a_{VA}$ & $a_{AV}$ & $a_{AA}$ & $a_{PP}$ & $a_{SP}$ & $a_{AP}$ & $a_{AS}$ \\ \hline\hline
$B^0 \to \rho e^- \mu^+$  &$0.62(10) $ &$2.9(6) $ &$0.62(10) $ &$2.9(6) $ &$1.1(2) $ &$1.1(2) $ &$0.12(2) $ &$-0.12(2) $ \\ 
$B^0 \to \rho e^- \tau^+$  &$0.32(5) $ &$1.6(3) $ &$0.32(5) $ &$1.6(3) $ &$0.49(9) $ &$0.49(9) $ &$0.63(12) $ &$-0.63(12) $ \\ 
$B^0 \to \rho \mu^- \tau^+$  &$0.35(6) $ &$1.6(3) $ &$0.31(5) $ &$1.5(3) $ &$0.52(10) $ &$0.47(9) $ &$0.56(11) $ &$-0.71(13) $ \\ \hline
$B^0 \to K^{\ast\,0}e^- \mu^+$  & $0.68(11)$ & $3.4(5)$ & $0.68(11)$ &$3.4(5)$ & $1.06(15)$ & $1.06(15)$ &$0.127(17)$ & $-0.128(18)$ \\ 
$B^0 \to K^{\ast\,0} e^- \tau^+$  & $0.34(5)$ & $1.8(3)$ &$0.33(5)$ &$1.8(3)$ &$0.46(7)$ &$0.46(7)$ &$0.62(9)$ & $-0.62(9)$ \\ 
$B^0 \to K^{\ast\,0} \mu^- \tau^+$  & $0.35(6)$ &$1.9(3)$ &$0.31(5)$ &$1.8(3)$ &$0.48(7)$ &$0.43(7)$ &$0.54(8)$ & $-0.69(10)$ \\ \hline
$B_s \to \phi e^- \mu^+$ &$0.58(5)$ &$3.7(5)$ &$0.58(5)$ &$3.7(5)$ &$1.21(17)$ &$1.21(17)$ &$0.15(2)$&$-0.15(2)$ \\ 
$B_s \to \phi  e^- \tau^+$  &$0.29(2)$ &$1.9(2)$ &$0.29(2)$ & $1.9(2)$ &$0.51(7)$ &$0.51(7)$ &$0.71(10)$ & $-0.71(10)$\\ 
$B_s \to \phi  \mu^- \tau^+$ &$0.30(3)$ &$1.9(3)$ &$0.27(2)$ &$1.8(2)$ &$0.53(8)$ &$0.48(7)$ &$0.62(9)$&$-0.79(11)$ \\ \hline
\end{tabular}}
\caption{\sl \small Numerical coefficients defined in Eq.~\eqref{eq:semilep-numeric} for the decays $P\to V \ell_i^- \ell_j^+$ based on the down-type transition $\smash{d_k \to d_l \ell_i^- \ell_j^+}$. Flavor indices are omitted and should be replace as in Eq.~\eqref{eq:wc-semilep}. Decays with opposite lepton electric charges in the final state can be obtained via the replacement $a_{AP} \to - a_{AP}$. See text for details on the hadronic inputs considered.}
\label{tab:weights-PVll} 
\end{table}

\subsubsection{Semileptonic decays: $P\to V\ell_i \ell_j$}

We turn now our attention to the semileptonic decays $P\to V\ell_i^- \ell_j^+$ with a vector meson $V$ in the final state  decays. In this case, the branching fraction can be expressed in terms of Eq.~\eqref{eq:left-lagrangian} as follows,
\begin{align}
    \mathcal{B}(P\to V &\ell_i^- \ell_j^+)=  a_{VV}\,|C_{VV}|^2 + a_{VA}\,|C_{VA}|^2 + a_{AV}\,|C_{AV}|^2 + a_{AA}\,|C_{AA}|^2 \nonumber\\*[0.35em]
   &+a_{PP}\,|C_{PP}|^2+ a_{SP}\,|C_{SP}|^2+ a_{AP} \,\mathrm{Re}[C_{VA} C_{SP}^\ast]+ a_{AS}\, \mathrm{Re}[C_{AA} C_{PP}^\ast] \nonumber\\*[0.35em]
   &+a_{TT}\, \big{(} |C_T|^2 + |C_{T_5}|^2 \big{)} + a_{ST} \,\text{Re}[C_{SP} C_T^\ast]+ a_{PT}\, \text{Re}[C_{PP} C_T^\ast]  \nonumber\\*[0.35em]
   &+ a_{ST_5} \,\text{Re}[C_{SP} C_{T_5}^\ast]  + a_{PT_5}\, \text{Re}[C_{PP} C_{T_5}^\ast]   \,,
\end{align}
with $C_{I}$ are the relevant Wilson coefficients, defined at the relevant low-energy scale, and $a_I$ are numerical coefficients related to the hadronic matrix elements. Flavor indices are omitted and should be replaced as in Eq.~\eqref{eq:semilep-numeric}. In the limit where the light-lepton mass ($m_{\ell_i}$) is neglected, we find the relations
\begin{align}
    \label{eq:semilep-massless}
    a_{VV} &\simeq a_{AV}\,, & a_{AP} &\simeq a_{AS}\,, & a_{PT} &\simeq - a_{ST_5}\,, \\[0.35em]
    a_{VA} &\simeq a_{AA}\,, & a_{PP} &\simeq a_{SP}\,, & a_{ST} &\simeq - a_{PT_5}\,, \nonumber 
\end{align}

\noindent which are once again valid up to corrections of the order $m_{\ell_i}/m_{\ell_j}$. 

For the $P\to V$ based on the $d_k\to d_l \ell_i \ell_j$ transition, we use the general analytical expressions provided in Ref.~\cite{Becirevic:2016zri}, which are summarized in the Appendix~\ref{app:diff-semileptonic-V}, and the Light-Cone-Sum-Rules form-factors from Ref.~\cite{Bharucha:2015bzk} to obtain the $a_i$ values collected in Table~\ref{tab:weights-PVll} for the $B\to \rho$, $B\to K^\ast$ and $B_s\to\phi$ decays (see also Ref.~\cite{Gubernari:2018wyi}). Once again, we do not quote the numerical coefficients corresponding to tensor coefficients since the corresponding effective coefficients are necessarily suppressed once $SU(2)_L \times U(1)_Y$ gauge symmetry is imposed~\cite{Alonso:2014csa}. The only up-type decays that would appear in this case are $D^+\to \rho^+ \mu^- e^+$ and $D_s\to K^{\ast +} \mu^- e^+$, which are experimentally challenging, and for which the form-factors have not yet been determined on the lattice. For these reasons, we only report numerical results for the $P\to V$ based on down-type quark transitions.

\subsection{$\tau \to \ell M$}
\label{ssec:tau-LFV}

The simplest LFV decays of the $\tau$-lepton into hadrons are $\tau\to \ell M$, where $M=\bar{q}_k q_l$ can be a light pseudoscalar or vector meson. Here, it is important to distinguish light pseudoscalar mesons $P$ with open flavor ($k\neq l$), such as $K_S$, from the unflavored ones ($k=l$), such as $P\in \lbrace \pi^0,\eta,\eta^{\prime}\rbrace$~\cite{Black:2002wh}. This is because the latter decays are also sensitive to the CP-odd gluonic operators defined in Eq.~\eqref{eq:left-ope-gg} in addition to the semileptonic ones. We will also briefly discuss the $\tau$ decays into vector mesons $V$, such as $K^\ast$, and $\rho$ and $\omega$. 

\subsubsection{Pseudoscalar mesons: $\tau\to \ell P$}

Firstly, we consider the $\tau \to \ell P$ process, where $P=\bar{q}_k q_l$  denotes a generic pseudoscalar meson. We discuss separately the case where $k\neq l$ from the one where $k=l$:

\paragraph{Flavored mesons} In the first case, it is sufficient to consider the decay constant $f_P$ defined in Eq.~\eqref{eq:P-decayconst}, which allows us to write e.g.~for $P=K_S$,
\begin{align}
\begin{split}
   \mathcal{B}(\tau \to \ell_i K_S)= \tau_{\tau} \, &\dfrac{f_P^2 m_{\tau}^3}{256 \pi v^4} \left(1-\dfrac{m_P^2}{m_{\tau}^2}\right)^2\\
   &\times\bigg{\lbrace} \bigg{|}C_{VA}^{(q)} +  \dfrac{m_P^2\,{C}_{SP}^{(q)}}{m_{\tau}(m_Q+m_q)}\bigg{|}^2+ \bigg{|}C_{AA}^{(q)} - \dfrac{m_P^2\,{C}_{PP}^{(q)}}{m_{\tau}(m_Q+m_q)}\bigg{|}^2\bigg{\rbrace} \,,
\end{split}
\end{align}

\noindent where we have used that $|K_S\rangle \simeq (|K^0\rangle - |\overline{K^0}\rangle)/\sqrt{2} $ and we have neglected the light-lepton mass ($m_{\ell_i}$), as before. Flavor indices in the above expressions are to be replaced as follows,
\begin{equation}
 C_I^{(q)} \to \frac{1}{\sqrt{2}}\Big{(}C_{\substack{I\\ i312}}^{(d)} - C_{\substack{I\\ i321}}^{(d)} \Big{)}   \,,
\end{equation}
for $I \in \{VA,AA,SP,PP\}$. Similar expressions can be obtained for $\tau\to\ell_i K_L$ by changing the relative sign of the coefficients in the above equation. However, these decay modes are experimentally challenging due to the $K_L$ lifetime. 

\paragraph{Unflavored mesons} The case of light unflavored mesons is rather different since there is not a simple relation between the axial and pseudoscalar matrix elements, and since the matrix element of the $G\widetilde{G}$ operator does not vanish in this case. Following Ref.~\cite{Beneke:2002jn}, we define the axial and pseudoscalar densities as follows
\begin{align}
\langle P(p) |\bar{q}\gamma^\mu\gamma_5 q|0 \rangle &\equiv  - \dfrac{i  f_P^{(q)}}{\sqrt{2}} p^\mu\,,\qquad\quad 2 m_q \langle P(p) |\bar{q}\gamma_5 q|0 \rangle \equiv - \dfrac{i h_P^{(q)}}{\sqrt{2}}\,,\\*[0.35em]
\langle P(p) |\bar{s}\gamma^\mu\gamma_5 s|0 \rangle &\equiv  - {i  f_P^{(s)}} p^\mu\,,\qquad\quad 2 m_s \langle P(p) |\bar{s}\gamma_5 s|0 \rangle \equiv - i h_P^{(s)}\,,
\end{align}
and
\begin{align}
 \langle P(p) \vert \dfrac{\alpha_s}{4\pi}G_{\mu\nu}\widetilde{G}^{\mu\nu}\vert 0\rangle \equiv a_P \,,
\end{align}
where $q=u$ or $d$, and exact isospin symmetry is assumed, with $m_q \equiv(m_u+m_d)/2$. The $\pi^0$ pseudoscalar density is given in this limit by~$\smash{h_\pi^{(u)}=-h_\pi^{(d)}=f_\pi\,m_{\pi}^2}$, and the anomaly contribution can be computed by taking the divergence of the axial current~\cite{Gross:1979ur}, 
\begin{align}
a_\pi = - \dfrac{1-z}{1+z}\dfrac{f_\pi m_\pi^2}{\sqrt{2}}\,,
\end{align}
where $z=m_u/m_d$. For $\eta^{(\prime)}$, we rely on the computation of $a_{\eta^{(\prime)}}$, $f_{\eta^{(\prime)}}^{(q)}$ and $h_{\eta^{(\prime)}}^{(q)}$ from the so-called Feldmann-Kroll-Stech (FKS) mixing scheme~\cite{Feldmann:1998vh,Beneke:2002jn}, which leads to the phenomenological estimations collected in Table~\ref{tab:FSK}. The final branching fraction for $P\in \lbrace \pi^0, \eta, \eta^\prime \rbrace$ thus reads 
\begin{table}[!t]
\renewcommand{\arraystretch}{1.6}
\centering
\begin{tabular}{|c|ccccc|}
\hline 
$P$  &  $f_P^{(q)}~[\mathrm{MeV}]$ &   $f_P^{(s)}~[\mathrm{MeV}]$  &  $h_P^{(q)}~[\mathrm{GeV}^3]$   & $h_P^{(s)}~[\mathrm{GeV}^3]$ & $a_P~[\mathrm{GeV}^3]$  \\ \hline\hline
$\eta$  &  $108(3)$ & $-111(6)$  & $0.001(3)$  & $-0.055(3)$  & $-0.022(2)$  \\ 
$\eta^\prime$  &  $89(3)$ &  $136(6)$ & $0.001(2)$  & $0.068(5)$  &   $-0.057(2)$ \\  \hline
\end{tabular}
\caption{\small \sl Hadronic inputs for $\eta^{(\prime)}$ obtained in Ref.~\cite{Feldmann:1998vh,Beneke:2002jn} by using the FKS scheme. }
\label{tab:FSK} 
\end{table}
\begin{align}
\begin{split}
   \mathcal{B}(\tau \to \ell_i P)&= \tau_{\tau} \, \dfrac{ m_{\tau}^3}{256 \pi v^4} \left(1-\dfrac{m_P^2}{m_{\tau}^2}\right)^2  \Big{[} \big{|}\mathcal{A}_V^{\tau\to\ell_i P} \big{|}^2 + \big{|}  \mathcal{A}_A^{\tau\to\ell_i P}\big{|}^2 \Big{]}\,,
\end{split}
\end{align}
with
\begin{align}
\mathcal{A}_V^{\tau\to\ell_i P} &= \sum_{q_k=u,d,s}b_{q_k}\bigg{[} f_P^{(q_k)}C_{\substack{VA\\i3kk}}^{(q)} + \dfrac{ h_P^{(q_k)}}{ 2m_{\tau}m_{q_k}}\,{C}_{\substack{SP\\i3kk}}^{(q)} \bigg{]}-i \dfrac{2 a_P}{v^2}(\widetilde{C}_{G_R}-\widetilde{C}_{G_L})\,,\\[0.35em]
\mathcal{A}_A^{\tau\to\ell_i P} &= \sum_{q_k=u,d,s}b_{q_k}\bigg{[}  f_P^{(q_k)} C_{\substack{AA\\i3kk}}^{(q)}- \dfrac{ h_P^{(q_k)}}{2m_{\tau}m_{q_k}} \,{C}_{\substack{PP\\i3kk}}^{(q)}\bigg{]}  +i \dfrac{2 a_P}{v^2}(\widetilde{C}_{G_R}+\widetilde{C}_{G_L})\,,
\end{align}
where $k$ spans the light-quark flavors (i.e., $q_k \in \lbrace u,d,s\rbrace$) and we have neglected the light-lepton mass, $m_{\ell_i}$. The pre-factors $b_q$ are given by $b_{u(d)} = {1}/{\sqrt{2}}$ and $b_s = 1$ for $P=\eta^{(\prime)}$, and $b_u=-b_d={1}/{\sqrt{2}}$ and $b_s=0$ for $P=\pi^0$. The $\smash{C_{{G}_L}^{(\ell)}}$ and $\smash{C_{\Tilde{G}_L}^{(\ell)}}$ coefficients are generated by heavy quarks at one loop, cf.~Eq.~\eqref{eq:GGtilde1}.

\subsubsection{Vector mesons: $\tau\to \ell V$} Lastly, we consider the decays $\tau \to \ell V$, where $V=\bar{q}_k q_l$ denotes a generic vector meson such as $V\in \lbrace K^\ast; \omega, \rho, \phi \rbrace$. By using the decay constants defined in Eq.~\eqref{eq:V-decayconst}, we can show that
\begin{align}
   \mathcal{B}(\tau \to \ell_i V)&= \tau_\tau \, \dfrac{f_V^2 m_\tau^3}{256 \pi v^4}  \left(1-\dfrac{m_{V}^2}{m_{\tau}^2}\right)^2\\*
&\times\bigg{\lbrace}\Big{[}|C_{\substack{VV\\i3kl}}^{(q)}|^2+|C_{\substack{AV\\i3kl}}^{(q)}|^2\Big{]} \bigg{(}1+\dfrac{2 m_V^2}{m_{\tau}^2}\bigg{)}+32 \bigg{(}\dfrac{f_V^T}{f_V}\bigg{)}^2\Big{[}|\Bar{C}_{\substack{T\\i3kl}}^{(q)}|^2+|\Bar{C}_{\substack{T_5\\i3kl}}^{(q)}|^2\Big{]}\bigg{(}1+\dfrac{m_V^2}{2m_{\tau}^2}\bigg{)}  \nonumber\\*
& +24 \dfrac{m_V}{m_{\tau}}\dfrac{f_V^T}{f_V} \, \mathrm{Re}\Big{[}\Bar{C}_{\substack{T\\i3kl}}^{(q)}\,C_{\substack{VV\\i3kl}}^{(q)\ast}-\Bar{C}_{\substack{T_{5}\\i3kl}}^{(q)}\,C_{\substack{AV\\i3kl}}^{(q)\ast}\Big{]} \bigg{\rbrace}\,, \nonumber
\end{align}
with
\begin{align}
\Bar{C}_{\substack{T\\i3kl}}^{(q)} &\equiv C_{\substack{T\\i3kl}}^{(q)} - \delta_{kl} e Q_q \frac{m_{\tau} f_V}{m_V f_V^T}  \Big{(} C_{\substack{D_R\\i3}}^{(\ell)} + C_{\substack{D_L\\i3}}^{(\ell)}  \Big{)} \;,  \\[0.35em]
\Bar{C}_{\substack{T_5\\i3kl}}^{(q)} &\equiv C_{T_5}^{(q)} - \delta_{kl} e Q_q  \frac{m_{\tau} f_V}{m_V f_V^T} \Big{(} C_{\substack{D_R\\i3}}^{(\ell)} - C_{\substack{D_L\\i3}}^{(\ell)}  \Big{)} 
\;,
\end{align}
\noindent where $Q_q$ denotes the $q_{k,l}$ electric charges, and we have neglected again the light-lepton mass, $m_{\ell_i}$. For $V \in \lbrace\rho,\omega\rbrace$, in the isospin limit, we have to perform the following trivial replacements in the above expression,
\begin{align}
    C_{I}^{(q)} \overset{V=\rho}{\to} \frac{1}{\sqrt{2}}\Big{(}C_{\substack{I\\i311}}^{(u)} - C_{\substack{I\\i311}}^{(d)}\Big{)} \,, \qquad\quad C_{I}^{(q)}  \overset{V=\omega}{\to} \frac{1}{\sqrt{2}}\Big{(}C_{\substack{I\\i311}}^{(u)} + C_{\substack{I\\i311}}^{(d)}\Big{)} \,,
\end{align}
where $I\in\lbrace VV,AV,T,T_5\rbrace$.

\subsection{$\ell_j\to \ell_i \gamma$ and $\ell_j\to \ell_i \ell_k \ell_k$}
\label{ssec:llpgamma-LFV}

We now turn our discussion to purely leptonic observables (see e.g.~Ref.~\cite{Kuno:1999jp,Bruser:2015yka,Crivellin:2017rmk} for previous EFT studies). These processes receive contributions from semileptonic operators at loop level, as illustrated in Fig.~\ref{fig:gauge-smeft-rge}. The experimental limits considered in our analysis are collected in Table~\ref{tab:exp-leptonic}.

\paragraph{Radiative decays} The simplest of these processes is the radiative decay $\ell_j\to \ell_i \gamma$ with $i<j$, where the photon is on-shell. This process is described by
\begin{align}
   \mathcal{B}(\ell_j\to \ell_i \gamma) = \frac{\tau_{\ell_j} m_{\ell_j}^5}{4\pi v^4 } \Big{(} |C_{\substack{D_L\\ij}}^{(\ell)}|^2 + |C_{\substack{D_R\\ij}}^{(\ell)}|^2 \Big{)} \,.
\end{align}
\noindent where we have neglected the lightest lepton mass, as before.

\paragraph{Three-body decays} The $\ell_j^-\to \ell_i^- \ell_k^+ \ell_k^-$ decays receive contributions from the dipole operators, in addition to operators with four leptons. We distinguish the case where $j>i=k$ from $j>i\neq k$ which have slightly different expressions. In the first case (i.e., $\tau^- \to e^-e^+e^+$ and $\tau^- \to \mu^-\mu^+\mu^+$)~\cite{Kuno:1999jp}, 
\begin{align}
   \mathcal{B}(\ell_j^-\to \ell_i^-\ell_i^+ \ell_i^-)  = \frac{\tau_{\ell_j} m_{\ell_j}^5}{1536 \pi^3 v^4} &\bigg{\lbrace} 2|C_{\substack{V_{LL}\\ijii}}^{(\ell)}|^2  +|C_{\substack{V_{LR}\\ijii}}^{(\ell)}|^2  + 64 e^2 \Big{(} \log \frac{m_{\ell_j}}{m_{\ell_i}} - \frac{11}{8} \Big{)} |C_{\substack{D_L\\ij}}^{(\ell)}|^2\\*[0.3em]
   &    + 8 e \text{Re}\Big{[} C_{\substack{D_R\\ij}}^{(\ell)}(2 C_{\substack{V_{LL}\\ijii}}^{(\ell)} + C_{\substack{V_{LR}\\ijii}}^{(\ell)} )^\ast\Big{]} +(L\leftrightarrow R)\bigg{\rbrace}   \,, \nonumber
\end{align}

\noindent where the light lepton masses are neglected in the above expression, except in the dipole term, which would otherwise be infrared divergent. Note, also, that we have not included the contributions from the scalar coefficients $\smash{C_{{S_{XY}}}}$  (with $X,L\in \lbrace L,R\rbrace$), as they are not induced by $d=6$ operators in the SMEFT, cf.~Appendix~\ref{app:SMEFT-matching}.

A similar expression can be obtained for the decays with $j >i \neq k$ (i.e., $\tau^- \to \mu^- e^+e^-$ and $\tau^- \to e^- \mu^+\mu^-$), which has no identical particles in the final state~\cite{Kuno:1999jp},
\begin{align}
   \mathcal{B}(\ell_j^-\to \ell_i^-\ell_k^+ \ell_k^-)  = \frac{\tau_{\ell_j} m_{\ell_j}^5}{1536 \pi^3 v^4} &\bigg{\lbrace} |C_{\substack{V_{LL}\\ijkk}}^{(\ell)}|^2  +|C_{\substack{V_{LR}\\ijkk}}^{(\ell)}|^2  + 64 e^2 \Big{(} \log \frac{m_{\ell_j}}{m_{\ell_k}} - \frac{3}{2} \Big{)} |C_{\substack{D_L\\ij}}^{(\ell)}|^2\\*[0.3em]
   &    + 8 e \,\text{Re}\Big{[} C_{\substack{D_R\\ij}}^{(\ell)}( C_{\substack{V_{LL}\\ ijkk}}^{(\ell)} + C_{\substack{V_{LR}\\ijkk}}^{(\ell)} )^\ast\Big{]} +(L\leftrightarrow R)\bigg{\rbrace}   \,, \nonumber
\end{align}
where we have kept the mass $m_{\ell_k}$ in the logarithmic term, which regularizes the infrared divergence coming from the photon propagator in the squared dipole term. Moreover, we neglect $m_{\ell_i}$ in this expression. We do not include the contributions from the scalar $C_{S_{XY}}$ and tensor $C_{T_{X}}$ coefficients (with $X,L\in \lbrace L,R\rbrace$), as before, since they do not appear in the tree-level matching to the SMEFT, as shown in Appendix~\ref{app:SMEFT-matching}.

\begin{table}[!t]
\renewcommand{\arraystretch}{1.6}
\centering
\begin{tabular}{|c|c||c|c||c|c|}
\hline 
Process  &  Exp.~limit  & Process  &  Exp.~limit & Process  &  Exp.~limit  \\ \hline\hline
$\mu\to e \gamma$  & $5.6\times 10^{-13}$   & $\tau\to e \gamma$  & $4.4\times 10^{-8}$    &  $\tau\to \mu\gamma$ & $5.6\times 10^{-8}$   \\
$\mu\to eee$  & $1.3\times 10^{-12}$    &  $\tau\to eee$ & $3.6\times 10^{-8}$    &  $\tau \to \mu \mu\mu$ & $2.8\times 10^{-8}$   \\ 
 &   &   $\tau \to e \mu\mu$  & $3.6\times 10^{-8}$   &  $\tau \to \mu ee$  & $2.4\times 10^{-8}$   \\ \hline
\end{tabular}
\caption{\small \sl Experimental limits on purely leptonic LFV decays at 95$\%$ CL~\cite{Zyla:2020zbs}.} 
\label{tab:exp-leptonic} 
\end{table}

\subsection{$\mu N \to e N$}
\label{ssec:muN-eN}

Lastly, we discuss $\mu \to e$ conversion in nuclei, which has been thoroughly studied within the EFT approach e.g.~in Ref.~\cite{Kuno:1996kv,Kitano:2002mt}. The nucleon effective coefficients are given in terms of our Wilson coefficients as follows~\cite{Kuno:1996kv},
\begin{align}
\begin{split}
 \widetilde{C}^{VX}_{(p)} &= \sum_{q_k=u,d,s} \sum_{Y = L,R} C^{(q)}_{\substack{V_{XY}\\12kk}}\, f^{(q_k)}_{Vp} \,,   \\[0.45em]  
 \widetilde{C}^{SX}_{(p)} &= \sum_{q_k=u,d,s} \sum_{Y=L,R} \dfrac{m_{p}}{m_{q_k}}C^{(q)}_{\substack{S_{XY}\\12kk}}\,f^{(q_k)}_{S\,p} + \dfrac{m_\mu m_{p}}{4\pi v^2}{C}^{(\ell)}_{\substack{G_X\\12}}  \,f_{G\,p}\,,  
\end{split}
\end{align}
where $X = L, R$, with analogous expressions for $p \leftrightarrow n$. The effective coefficients are evaluated at $\mu \approx 1$~GeV, thus including contributions from scalar operators made of heavy quarks~\cite{Shifman:1978zn}, as described in Eq.~\eqref{eq:GGtilde1}. For the vector current, the nucleon form-factors are given by $f^{(u)}_{Vp}=f^{(d)}_{Vn}=2$, $f^{(d)}_{Vp}=f^{(u)}_{Vn}=1$ and $f^{(s)}_{Vp}=f^{(s)}_{Vn}=0$. We consider the numerical results for the scalar form-factors from Ref.~\cite{Crivellin:2017rmk}, which are based on Ref.~\cite{Crivellin:2014cta},
\begin{align}
f_{S\,p}^{(u)}&=(20.8 \pm 1.5)\times 10^{-3}\,, &f_{S\,n}^{(u)}&=(18.9 \pm 1.4)\times 10^{-3}\,,\\*[0.4em]
f_{S\,p}^{(d)}&=(41.1 \pm 2.8)\times 10^{-3}\,, &f_{S\,n}^{(u)}&=(45.1 \pm 2.7)\times 10^{-3}\,,\\*[0.4em]
f_{S\,p}^{(s)}&= f_{Sn}^{(s)}= (53\pm 27)\times 10^{-3}\,,
\end{align}
with the gluonic form-factors given by
\begin{align}
f_{G\,p(n)} = - \dfrac{8\pi}{9}\Big{(}1- \sum_{q=u,d,s} f_{S\,p(n)}^{(q)}\Big{)}\,.
\end{align}
The $\mu\to e$ conversion rate normalized by the muon capture rate $(\Gamma_\mathrm{capt})$ is denoted by $\mathcal{B}^{(N)}_{\mu e}$ and it can be written in terms of nucleon EFT Wilson coefficients as follows~\cite{Kuno:1996kv},
\begin{align}
  \mathcal{B}^{(N)}_{\mu e} = \frac{m_{\mu}^5}{4 \upsilon^4 \Gamma^{(N)}_{\text{capt}}} \Big{|} C^{(\ell)}_{D_L} \,D_N + 2 \Big{[} \widetilde{C}^{VR}_{(p)}\,V_N^{(p)} +\widetilde{C}^{SL}_{(p)}\, S^{(p)}_N + (p \to n) \Big{]} \Big{|}^2  + (L \leftrightarrow R)  \,.
\end{align}

\begin{table}[!t]
\renewcommand{\arraystretch}{1.5}
\centering
\begin{tabular}{|c|ccccc|}
\hline 
Nuclei  &  $D$ &  $V^{(p)}$  & $V^{(n)}$   &  $S^{(p)}$  & $S^{(n)}$  \\ \hline\hline
Au  & $0.189 $  & $0.0974 $  & $0.146 $  & $0.0614 $  & $ 0.0918 $   \\ 
Al  & $ 0.0362 $  & $ 0.0161 $  & $ 0.0173 $  & $ 0.0155 $  & $ 0.0167 $  \\  \hline
\end{tabular}
\caption{\small \sl Overlap integrals for gold and aluminum atoms from Ref.~\cite{Heeck:2022wer}.} 
\label{tab:nuclear-inputs} 
\end{table}

\noindent For the nuclear-physics inputs, we consider the results of Ref.~\cite{Kitano:2002mt} (see also Ref.~\cite{Heeck:2022wer}), which are summarized in Table~\ref{tab:nuclear-inputs}, and the nuclear capture rates determined experimentally \cite{Suzuki:1987jf},
\begin{align}
\Gamma_\mathrm{capt}^{(\mathrm{Al})} &\simeq { 6.99 \times 10^{-7}~\mathrm{ ps}^{-1} }\,,\qquad\qquad 
\Gamma_\mathrm{capt}^{(\mathrm{Au})} \simeq { 1.32 \times 10^{-5}~\mathrm{ ps}^{-1}}\,.
\end{align}
Currently, the most stringent experimental limit is $\mathcal{B}^{(\mathrm{Au})}_{\mu e}<7 \times 10^{-13}$~(90\% CL.), which was set by the SINDRUM-II experiment~\cite{SINDRUMII:2006dvw}. The experimental sensitivity is planned to be considerably improved in the near future by the Mu2e experiment at Fermilab~\cite{Mu2e:2014fns} and COMET~\cite{COMET:2018auw} at J-PARC, which are expected to reach a sensitivity of $\mathcal{O}(10^{-17})$ with aluminum atoms.

\section{High-energy probes}
\label{ssec:SMEFT-LFV}

In this Section, we describe the high-energy probes of LFV in terms of the SM EFT Lagrangian defined in Eq.~\eqref{eq:SMEFT-Lag}. We will consider the decays of the $Z$-boson, the Higgs boson, and the top quark, which are subject to the experimental limits collected in Table~\ref{tab:exp-lfv-H-Z-top}. Furthermore, we will briefly discuss the Drell-Yan processes $pp\to\ell_i\ell_j$ at the LHC that are sensitive to non-resonant contributions from the SM EFT operators.

\subsection{$Z\to \ell_i \ell_j$}
\label{ssec:Z-LFV}

The $Z$-boson LFV decays receive contributions at tree level from the effective coefficients $\smash{\mathcal{C}_{Hl}^{(1)}}$ and $\smash{\mathcal{C}_{Hl}^{(3)}}$, as well as the dipoles $\mathcal{C}_{eB}$ and $\mathcal{C}_{eW}$, which are both sensitive to one-loop contributions from semileptonic operators~(see e.g.~Ref.\cite{Calibbi:2021pyh}),~\footnote{We note that the orthogonal combination $\mathcal{C}_{e \gamma} \equiv -s_W \,\mathcal{C}_{eW} + c_W \,\mathcal{C}_{eB}$ of the dipole coefficients is already tightly constrained by $\ell_i\to\ell_j \gamma$.}
\begin{align}
   \mathcal{B}(Z\to \ell_i^- \ell_j^+)= \dfrac{\tau_Z m_Z^3 v^2}{24\pi \Lambda^4}\Big{[}\big{|}\mathcal{C}_{\substack{Hl\\ ij}}^{(1+3)}\big{|}^2+\big{|}\mathcal{C}_{\substack{He\\ ij}}\big{|}^2+\big{|}\mathcal{C}_{\substack{eZ\\ ij}}\big{|}^2+\big{|}\mathcal{C}_{\substack{eZ\\ ji}}\big{|}^2\Big{]}\,,
\end{align}

\noindent where we define  $\mathcal{C}_{e Z} \equiv -\cos {\theta_W} \,\mathcal{C}_{eW}  - \sin {\theta_W}\, \mathcal{C}_{eB}$, $\theta_W$ is the Weinberg angle, and  $\mathcal{C}_{Hl}^{(1\pm 3)} \equiv \mathcal{C}_{Hl}^{(1)} \pm \mathcal{C}_{Hl}^{(3)}$. The $Z$-boson mass and lifetime are denoted by $m_Z$ and $\tau_Z$, respectively, and
we neglect lepton masses in the above expression. We notice that far more stringent constraints on $\mathcal{C}_{eW}$ and $\mathcal{C}_{eB}$ can be obtained by the $\ell_j\to \ell_i \gamma$ processes. However, these observables can still provide useful constraints on $\smash{\mathcal{C}_{Hl}^{(1,3)}}$ and $\mathcal{C}_{He}$. The latter coefficients are related to $\smash{\mathcal{C}_{lq}^{(1,3)}}$ and $\mathcal{C}_{eu}$, respectively, with third-generation quarks, through the RGE effects depicted in Fig.~\ref{fig:gauge-smeft-rge}, cf.~e.g.~Ref.~\cite{Feruglio:2016gvd}.

\subsection{$h\to \ell_i \ell_j$}
\label{ssec:Higgs-LFV}
The Higgs boson decays are also efficient probes of LFV~\cite{Han:2000jz} (see also Ref.~\cite{Paradisi:2005tk}). The only SM EFT $d=6$ operator that contributes at tree level to this process is $\mathcal{O}_{eH}$, which can receive sizable one-loop contributions from the scalar operator $\mathcal{C}_{ledq}$ through RGEs~\cite{Feruglio:2018fxo}. After electroweak-symmetry breaking, this contributes not only to the Higgs coupling to leptons, but it also induces a non-diagonal contribution to the fermion masses,
\begin{align}
\begin{split}
    \mathcal{L}_\mathrm{SMEFT} &\supset \dfrac{1}{\Lambda^2} \mathcal{C}_{\substack{eH\\ij}} \big{(}\bar{l}_i e_j H\big{)} \big{(}H^\dagger H\big{)} + \mathrm{h.c.}\\[0.35em]
    &\overset{\langle H \rangle \neq 0}{\rightarrow}\dfrac{v^3}{2 \sqrt{2}\Lambda^2} \mathcal{C}_{\substack{eH\\ij}} \big{(}\bar{\ell}_{i} P_R \ell_{j}\big{)} \Big{(}1 + 3  \frac{h}{v} +3 \frac{h^2}{v^2} +\frac{h^3}{v^3}\Big{)} + \mathrm{h.c.}\,,
\end{split}
\end{align}

\begin{table}[!t]
\renewcommand{\arraystretch}{1.6}
\centering
\resizebox{\columnwidth}{!}{\begin{tabular}{|c|cc||c|cc||c|cc|}
\hline 
Process & Exp.~limit & Ref. & Process & Exp.~limit & Ref. & Process & Exp.~limit & Ref.\\ \hline\hline
$Z \to e \mu$   & $2.6\times 10^{-7}$ & \cite{ATLAS:2022uhq}  & $Z \to e \tau $   & $5.0\times 10^{-6}$ & \cite{Zyla:2020zbs}  & $Z \to \mu \tau$   & $6.5\times 10^{-6}$ & \cite{Zyla:2020zbs} \\  
$h \to e \mu $   & $4.4\times 10^{-5}$ & \cite{CMS:2023pte}  & $h \to e \tau $   & $2.0\times 10^{-3}$ & \cite{ATLAS:2023mvd}  & $h \to \mu \tau$   & $1.5\times 10^{-3}$ & \cite{Zyla:2020zbs} \\ 
$t \to c e \mu $   & $2.6\times 10^{-6}$  & \cite{CMS:2022ztx}   & $t  \to c e \tau$   & $1.9\times 10^{-5}$ & \cite{Gottardo:2018ptv} & $t \to c \mu \tau$   & $1.1\times 10^{-6}$  & \cite{ATLAS:2023fcw}\\
\hline
\end{tabular}}
\caption{\small \sl Experimental limits on LFV decays of the $Z$-boson, Higgs and top-quark at 95$\%$ CL. Note that the decays with opposite lepton charges are combined, i.e.~$\ell_i\ell_j \equiv \ell_j^+\ell_i^- + \ell_j^-\ell_i^+$.}
\label{tab:exp-lfv-H-Z-top} 
\end{table}

\noindent which must be re-absorbed by a rotation of the lepton fields. The effective Higgs coupling to leptons is then given by~\footnote{See Appendix~\ref{app:yuk-rge} for a discussion of Yukawa coupling running in the SM.}
\begin{align}
y_{\substack{e\\ij }}^\mathrm{eff} \equiv \sqrt{2}\, \dfrac{m_{\ell_i}}{v} \delta_{ij} - \dfrac{v^2}{\Lambda^2} \mathcal{C}_{\substack{eH\\ ji}}^\ast \,,
\end{align}
\noindent which implies that
\begin{align}
   \mathcal{B}(H\to \ell_i^- \ell_j^+)=  \frac{\tau_h v^4 m_h }{32 \pi \Lambda^4} \Big{[} |\mathcal{C}_{\substack{eH\\ij}}|^2 + |\mathcal{C}_{\substack{eH\\ji}}|^2 \Big{]} \,,
\end{align}

\noindent where the lepton masses have been neglected, and $m_h$ and $\tau_h$ denote the Higgs-boson mass and lifetime, respectively.

\subsection{$t\to c \ell_i\ell_j$}

The top-quark decays $t\to c \ell_i\ell_j$ can also be used to probe LFV at high-energies~\cite{Davidson:2015zza}. The current experimental precision only allows us to consistently probe tree-level contributions, cf.~Table~\ref{tab:exp-lfv-H-Z-top}. The branching fractions for these decays are given by,
\begin{align}
\begin{split}
   \mathcal{B}(t\to c\ell_i^- \ell_j^+) = \frac{\tau_t m_t^5}{ 1536\pi^3\Lambda^4}&\bigg{[} \big{|}\mathcal{C}_{\substack{lq\\ij23}}^{\prime (1-3)}\big{|}^2 + \big{|}\mathcal{C}^{\prime}_{\substack{eq\\ij23}}\big{|}^2 + \big{|}\mathcal{C}_{\substack{eu\\ ij23}}\big{|}^2 + \big{|}\mathcal{C}_{\substack{lu\\ij23}}\big{|}^2  \\*[0.3em]
   & + \frac{1}{4}\Big{(} \big{|}\mathcal{C}_{\substack{lequ\\ij23}}^{\prime (1)} \big{|}^2 + \big{|}\mathcal{C}_{\substack{lequ\\ji32}}^{\prime (1)} \big{|}^2\Big{)} + 12\Big{(} \big{|}\mathcal{C}_{\substack{lequ\\ij23}}^{\prime (3)} \big{|}^2 + \big{|}\mathcal{C}_{\substack{lequ\\ji32}}^{\prime (3) } \big{|}^2\Big{)}\bigg{]} \,, 
\end{split}
\end{align}

\noindent where lepton and charm-quark masses have been neglected, and $m_t$ and $\tau_t$ denote the top-quark mass and lifetime, respectively. We use once again the shorthand notation $\mathcal{C}_{lq}^{(1\pm 3)} = \mathcal{C}_{lq}^{(1)} \pm \mathcal{C}_{lq}^{(3)}$, and we define the primed Wilson coefficients as follows,
\begin{align}
\begin{split}
\mathcal{C}_{lq}^{\prime\,(1-3)}  &\equiv  V   \mathcal{C}_{lq}^{(1-3)} V^\dagger \,,\\[0.4em]
\mathcal{C}_{eq}^{\prime}  &\equiv V   \mathcal{C}_{eq} V^\dagger\,,\\[0.4em]
\mathcal{C}_{lequ}^{\prime\,(1)}  &\equiv V    \mathcal{C}_{lequ}^{\,(1)}\,, \\[0.4em]
\mathcal{C}_{lequ}^{\prime\,(3)}  &\equiv V  \mathcal{C}_{lequ}^{\,(3)}\,,
\end{split}\label{eq:WC2}
\end{align}

\noindent where the CKM matrix ($V$) acts on quark-flavor indices, e.g.~$\smash{\mathcal{C}^{\prime\,(1)}_{\substack{lequ\\ijkl}} \equiv\sum_{k^\prime}V_{kk^\prime}\,\mathcal{C}_{\substack{lequ\\ijk^\prime l}}^{(1)}}$\,. Analogous expressions apply to the $t\to u\ell_i\ell_j$ decays.

\subsection{$p p\to \ell_i \ell_j$}
\label{ssec:LHC-LFV}

The study of $p p\to \ell_i^- \ell_j^+$ at high-$p_T$ can also provide useful probes of LFV in semileptonic operators since these EFT contributions are energy-enhanced at the tails of the distributions~\cite{Farina:2016rws}. The $q_l \bar{q}_k \to \ell_i^- \ell_j^+$ partonic cross-section (with $q=u,d$) can be written as~\cite{Allwicher:2022gkm}
\begin{equation}
  \label{eq:partonic-xsection}
 \hat{\sigma} (q_l \bar{q}_k \to \ell_i^- \ell_j^+) = \dfrac{\hat{s}}{144 \pi v^4} \sum_{IJ} \mathcal{C}_I^{(q)\ast} \,\Omega_{IJ} \,\mathcal{C}_J^{(q)}\,,
\end{equation}
where $\hat{s}$ is the partonic center-of-mass energy, and $\mathcal{C}^{(u)}$ and $\mathcal{C}^{(d)}$ are vectors of effective coefficients~\cite{Angelescu:2020uug,Descotes-Genon:2023pen},~\footnote{The zeros in the $\mathcal{C}^{(d)}$ reflect the fact there are no tensor effective coefficients for the $d_i\to d_j\ell\ell$ transition at $d=6$ in the SM EFT~\cite{Alonso:2014csa}.}
\begin{align}
\vec{\mathcal{C}}^{\,(u)} &= \Big{[}\mathcal{C}_{\substack{lq\\ijkl}}^{\prime \,(1-3)}\,,\;\mathcal{C}_{\substack{lu\\ijkl}}\,,\;\mathcal{C}_{\substack{eq\\ijkl}}^\prime\,,\;\mathcal{C}_{\substack{eu\\ijkl}}\,,\;\mathcal{C}_{\substack{lequ\\ijkl}}^{\prime\,(1)} \,,\;\mathcal{C}_{\substack{lequ\\jilk}}^{\prime\,(1)\,\ast}\,,\;\mathcal{C}_{\substack{lequ\\ijkl}}^{\prime\,(3)} \,,\;\mathcal{C}_{\substack{lequ\\jilk}}^{\prime\,(3)\,\ast} \Big{]}\,,\\[0.4em]
\vec{\mathcal{C}}^{\,(d)} &= \Big{[}\mathcal{C}_{\substack{lq\\ ijkl}}^{(1+3)}\,,\;\mathcal{C}_{\substack{ld\\ijkl}}\,,\;\mathcal{C}_{\substack{eq\\ijkl}}\,,\;\mathcal{C}_{\substack{ed\\ijkl}}\,,\;\mathcal{C}_{\substack{ledq\\ijkl}}\,,\;\mathcal{C}_{\substack{ledq\\jilk}}^\ast \,,\;0\,,\;0\Big{]}\,,
\end{align}

\noindent where the primed effective-coefficients appearing in $\mathcal{\vec{C}}^{(u)}$ are defined in Eq.~\eqref{eq:WC2}. Moreover, $\Omega$ is a $8\times 8$ matrix that takes a diagonal form $\Omega_{IJ}=\Omega_I \, \delta_{IJ}$ for the full partonic cross-section,
\begin{equation}
  \Omega =  \mathrm{diag}\big{(}1\,,\;1\,,\;1\,,\;1\,,\;3/4\,,\;3/4\,,\;4\,,\;4 \big{)}\,.
\end{equation}

\noindent The Drell-Yan cross-section is given by the convolution of~\eqref{eq:partonic-xsection} with  the parton luminosity functions,
\begin{align}
   {\sigma}(p p\to \ell_i^- \ell_j^+)=  \sum_{k,l} \int \dfrac{\mathrm{d}\hat{s}}{s} \mathcal{L}_{q_k \bar q_l}\, \hat{\sigma}(q_k \bar q_l\to \ell_i^- \ell_j^+)  \,,
\end{align}
where
\begin{align}
\mathcal{L}_{ q_k \bar  q_l}(\hat{s}) \equiv \int_{\hat{s}/s}^1 \dfrac{\mathrm{d}x}{x} \Big{[} f_{{q}_k} (x,\mu_F) f_{\bar{q}_l} (\frac{\hat{s}}{s x},\mu_F)+({q}_k \leftrightarrow \bar q_l)\Big{]}\,,
\end{align}
where $f_{{q}_k}$ and $f_{\bar{q}_l}$ are the Parton Distribution Functions (PDFs) of $q_k$ and $\bar{q}_l$, and $\mu_F$ denotes the factorization scale. In this paper, we consider the constraints on SMEFT operators derived in the {\tt HighPT} package~\cite{Allwicher:2022mcg} through an appropriate recast of the latest CMS search (with $140~\mathrm{fb}^{-1}$) for heavy resonances decaying into LFV lepton pairs~\cite{CMS:2021tau}.

\section{Numerical results}
\label{sec:pheno}

\subsection{LFV from top-quark loops}


In this Section, we illustrate our results by considering EFT scenarios with semileptonic operators only involving third-generation quarks at high energies. This choice is motivated by Minimal Flavor Violation~\cite{DAmbrosio:2002vsn} or the $U(2)^5$ flavor symmetry~\cite{Barbieri:2011ci} (see also Ref.~\cite{Faroughy:2020ina}), in which the couplings to quarks are hierarchical. This example is also convenient from a pragmatic point of view since the various quark-level transitions become related through RGE effects induced e.g.~by the top-quark Yukawa, allowing us to compare the sensitivity of different observables with a minimal set of Wilson coefficients.

With the assumption that only the effective operators made of third-generation quarks are present at the scale $\Lambda\gg v$, we have derived constraints on $\Lambda$ (for $\mathcal{C} = 1$) for each semileptonic operator appearing in Table~\ref{tab:SMEFT-ope}, by using the various processes that receive contributions at tree- and loop-level, cf.~Figs.~\ref{fig:gauge-left-rge} and \ref{fig:gauge-smeft-rge}. We collect these results in Table~\ref{tab:lfv-mue}, \ref{tab:lfv-taue} and \ref{tab:lfv-taumu} for the $\mu\to e$, $\tau\to e$ and $\tau \to \mu$ transitions, respectively, in which the superscript symbols are used to distinguish the origin of the leading contribution for each operator. We find that the different low-energy processes are complementary in probing these operators through loop-level contributions, which can appear at one or two loops. The only relevant constraints at tree level for such scenarios are the high-$p_T$ processes $pp\to \ell_i\ell_j$~\cite{Angelescu:2020uug,Allwicher:2022gkm,Allwicher:2022mcg,Descotes-Genon:2023pen}. By inspecting Table~\ref{tab:lfv-mue}--\ref{tab:lfv-taumu}, we arrive at the following conclusions:

\begin{table}[!t]
\renewcommand{\arraystretch}{1.8}
\centering
\resizebox{\columnwidth}{!}{
\begin{tabular}{|c|ccccccccc|}
\hline 
\multicolumn{10}{|c|}{Lower limits on $\Lambda$ (for $\mathcal{C}=1$)}\\ \hline\hline
Coeff.  &  $s\to d e\mu$ & $b\to d e\mu$ & $b\to s e\mu$  & $\mu \to e\gamma$ & $\mu\to eee$ & $\mu N \to e N$ & $Z\to e\mu$ & $h\to e \mu$ & $pp\to e\mu$ \\[0.35em] \hline
$\mathcal{C}_{\substack{lq\\1233}}^{(1+3)}$ & $\ast$ & $\ast$ &  $1.0$~TeV$^{~\color{blue}\bullet}$ & \cellcolor{gray!10} {$69$~TeV}$^{~\color{blue}\bullet\bullet}$ & $14$~TeV$^{~\color{blue}\bullet}$ & $44$~TeV$^{~\color{blue}\bullet}$ & $\ast$ & -- & $2.5$~TeV \\[0.4em]  
$\mathcal{C}_{\substack{lq\\1233}}^{(1-3)}$ & -- & -- & -- & {$100$~TeV}$^{~\color{blue}\bullet\bullet}$ &  $70$~TeV$^{~\color{blue}\bullet}$ &\cellcolor{gray!10}  $200$~TeV$^{~\color{blue}\bullet}$ & $2.0$~TeV$^{~\color{blue}\bullet}$ & -- & --  \\[0.4em] 
$\mathcal{C}_{\substack{eq\\1233}}$ & \color{gray}-- & -- & -- & {$69$~TeV}$^{~\color{blue}\bullet\bullet}$ & $70$~TeV$^{~\color{blue}\bullet}$ & \cellcolor{gray!10}  $210$~TeV$^{~\color{blue}\bullet}$ & $2.1$~TeV$^{~\color{blue}\bullet}$ & -- &  {\color{black} $2.5$~TeV} \\[0.4em] 
$\mathcal{C}_{\substack{eu\\1233}}$ & $\ast$ & $\ast$ &  $0.6$~TeV$^{~\color{blue}\bullet}$ & {$69$~TeV}$^{~\color{blue}\bullet\bullet}$ & $71$~TeV$^{~\color{blue}\bullet}$ & \cellcolor{gray!10} $220$~TeV$^{~\color{blue}\bullet}$ & $2.1$~TeV$^{~\color{blue}\bullet}$ & -- & --  \\[0.4em] 
$\mathcal{C}_{\substack{lu\\1233}}$ & $\ast$ & $\ast$ & $0.6$~TeV$^{~\color{blue}\bullet}$ & {$69$~TeV}$^{~\color{blue}\bullet\bullet}$ & $70$~TeV$^{~\color{blue}\bullet}$ & \cellcolor{gray!10}$220$~TeV$^{~\color{blue}\bullet}$ & $2.1$~TeV$^{~\color{blue}\bullet}$ & -- & -- \\[0.4em]  
$\mathcal{C}_{\substack{ld\\1233}}$ & -- & -- & -- & {$3.7$~TeV}$^{~\color{red}\diamond\diamond}$ & $17$~TeV$^{~\color{blue}\bullet}$ & \cellcolor{gray!10}$43$~TeV$^{~\color{blue}\bullet}$ & $\ast$ & -- & $2.5$~TeV  \\[0.4em] 
$\mathcal{C}_{\substack{ed\\1233}}$ & -- & -- & -- & {$4.6$~TeV}$^{~\color{red}\diamond\diamond}$ & $17$~TeV$^{~\color{blue}\bullet}$ & \cellcolor{gray!10}$43$~TeV$^{~\color{blue}\bullet}$ & $\ast$ & -- & $2.5$~TeV  \\[0.4em]  
$\mathcal{C}_{\substack{lequ\\1233}}^{(1)}$ & $\ast$ & $\ast$ & $\ast$ & \cellcolor{gray!10} {$3.2 \times 10^3$~TeV}$^{~\color{blue}\bullet\bullet}$ & {$600$~TeV}$^{~\color{blue}\bullet\bullet}$ & {$580$~TeV}$^{~\color{blue}\bullet\bullet}$ & $\ast$ & $7.6$~TeV$^{~\color{blue}\bullet}$ & --  \\[0.4em] 
$\mathcal{C}_{\substack{lequ\\1233}}^{(3)}$ & -- & -- & -- & \cellcolor{gray!10}$4.0 \times 10^4$~TeV$^{~\color{blue}\bullet}$ & $8.4\times 10^3$~TeV$^{~\color{blue}\bullet}$ &$8.2 \times 10^3$~TeV$^{~\color{blue}\bullet}$ & $1.0$~TeV$^{~\color{blue}\bullet}$ & -- & --  \\[0.4em] 
$\mathcal{C}_{\substack{ledq\\1233}}$ & $\ast$ & $2.2$~TeV$^{~\color{blue}\bullet}$ & $3.7$~TeV$^{~\color{blue}\bullet}$ & {$23$~TeV}$^{~\color{red}\diamond\diamond}$ & {$5.1$~TeV}$^{~\color{red}\diamond\diamond}$ & \cellcolor{gray!10} {$57$~TeV}$^{~\color{magenta} \square}$ & -- & $\ast$ & $2.4$~TeV \\[0.4em] \hline 
\end{tabular}}
\caption{\small \sl Lower limits on $\Lambda$ in TeV units (for $\mathcal{C}=1$) derived from LFV observables for each semileptonic operator coupled to third-generation couplings that contributes to the $\mu\to e$ transition. The cells highlighted in gray correspond to the most stringent limits on each operator. The dashes denote processes that are not sensitive to a given coefficient within the approximations described in Sec.~\ref{sec:eft}. The asterisks denote upper limits for which the SM EFT is not the valid description, i.e.~with $\Lambda/\sqrt{|\mathcal{C}|} \lesssim v/\sqrt{4\pi}$. The symbols in superscript identify the order of the leading contribution for each operator in the logarithmic expansion described in Sec.~\ref{ssec:summary}: tree-level (no symbol), one-loop single logarithm ($\color{blue}\bullet$), two-loop double logarithm ($\color{blue}\bullet\bullet$) and two-loop single logarithm ($\color{red}\diamond\diamond$); as well as the one-loop matching of gluonic operators ($\color{magenta}\square$).
}
\label{tab:lfv-mue} 
\end{table}

\begin{itemize}
    \item[$\bullet$] $\mu\to e$: For the $\mu\to e$ transition, the dominant bounds come from the experimental limits on the purely leptonic processes $\mu\to e\gamma$ and $\mu\to eee$, as well as on the $\mu\to e$ conversion in nuclei (see Ref.~\cite{Crivellin:2017rmk,EliasMiro:2021jgu}). More specifically, we find that the best constraints on the vector-type operators $\big{\lbrace} \mathcal{C}_{lq}^{(1)},\,\mathcal{C}_{lq}^{(3)},\,\mathcal{C}_{eq},\, \mathcal{C}_{eu},\,\mathcal{C}_{lu},\,\mathcal{C}_{ld}\big{\rbrace}$ come from $\mu N\to e N$, which receives contributions from one-loop penguin diagrams and Yukawa induced mixing, cf.~Figs.~\ref{fig:gauge-left-rge} and \ref{fig:gauge-smeft-rge}. For the scalar operator $\mathcal{C}_{ledq}$, we find that the most stringent constraint comes from the one-loop finite contribution to the gluonic operators ${O}_{G_{X}}^{(\ell)}$ (cf.~Fig.~\ref{fig:ggtilde}), which is again strictly constrained by $\mu N \to e N$. Lastly, the tensor $\mathcal{C}_{lequ}^{(3)}$ and the scalar $\mathcal{C}_{lequ}^{(1)}$ are better constrained by $\mu\to e \gamma$ via the one-loop $(T\to D_\ell)$ and the two-loop double-logarithm mixing $(S\to T \to D_\ell)$, which supersede the constraints from $\mu N\to e N$ and give the most stringent constraint in Table~\ref{tab:lfv-mue}, probing scales as large as $\mathcal{O}(10^4~\mathrm{TeV})$. This can be understood from the chirality enhancement ($\propto m_t/m_\mu$) of the one-loop tensor contribution to the dipole operators entering $\mu\to e \gamma$~\cite{Feruglio:2018fxo}.

    \item[$\bullet$] $\tau\to \ell$ (with $\ell=e,\mu$): For the $\tau \to \ell$ transitions (with $\ell=e,\mu$), there is an even more pronounced complementarity between the various experimental probes that are available. We find that the most stringent constraints come from the $\tau\to \ell P$ and $\tau\to \ell V$ decays, which are induced at loop level, or from the LHC searches for $pp\to \ell\tau$ at high-$p_T$, which receives tree-level contributions in our setup. The $\tau$-lepton decays are particularly useful to probe the vector operators containing the top-quark (i.e., $\mathcal{C}_{lq}^{(1-3)}\,,\mathcal{C}_{eq}\,,\mathcal{C}_{eu}\,,\mathcal{C}_{lu}$), which mixes into operators such as $(\bar{\mu} \gamma^\mu \tau)(\bar{q}\gamma_\mu q)$ (with $q=u,d,s$) via gauge-induced penguins, as well as the Yukawa running depicted in Fig.~\ref{fig:gauge-smeft-rge}. Instead, for the vector operators containing the $b$-quark (i.e.,~$\mathcal{C}_{lq}^{(1+3)},\,\mathcal{C}_{ld},\,\mathcal{C}_{ed}$), these contributions are smaller, giving constraints that are weaker than the tree-level ones arising from Drell-Yan processes. Finally, we find once again that the tensor $\mathcal{C}_{lequ}^{(3)}$ and the scalar $\smash{\mathcal{C}_{lequ}^{(1)}}$ are tightly constrained thanks to their chirality-enhanced contributions to $\tau\to \ell \gamma$ at one- and one-loop squared, respectively.
\end{itemize}

\begin{table}[!p]
\vspace{-3.em}
\renewcommand{\arraystretch}{1.8}
\centering
\resizebox{0.93\columnwidth}{!}{
\begin{tabular}{|c|ccccccccc|}
\hline 
\multicolumn{10}{|c|}{Lower limits on $\Lambda$ (for $\mathcal{C}=1$)}\\ \hline\hline
Coeff.  &  $b\to d e\tau$ & $b\to s e\tau$ & $\tau\to e\gamma$  & $\tau \to e\ell\ell$ & $\tau\to eP$ & $\tau \to e V$ & $Z\to e\tau$ & $h\to e \tau$ & $pp\to e\tau$\\[0.35em] \hline
$\mathcal{C}_{\substack{lq\\1333}}^{(1+3)}$ & $\ast$ & $\ast$ & {$0.9$~TeV}$^{~\color{blue}\bullet\bullet}$ & $0.7$~TeV$^{~\color{blue}\bullet}$ & $\ast$ &  $0.8$~TeV$^{~\color{blue}\bullet}$ & $\ast$ & -- & \cellcolor{gray!10} $1.8$~TeV \\[0.4em]
$\mathcal{C}_{\substack{lq\\1333}}^{(1-3)}$ & -- & -- & {$1.7$~TeV}$^{~\color{blue}\bullet\bullet}$ &  $2.6$~TeV$^{~\color{blue}\bullet}$ & $2.2$~TeV$^{~\color{blue}\bullet}$ &\cellcolor{gray!10}$3.1$~TeV$^{~\color{blue}\bullet}$ & $0.8$~TeV$^{~\color{blue}\bullet}$ & -- & -- \\[0.4em]
$\mathcal{C}_{\substack{eq\\1333}}$ & $\ast$ & $\ast$ & {$0.9$~TeV}$^{~\color{blue}\bullet\bullet}$ & $2.7$~TeV$^{~\color{blue}\bullet}$ & $2.2$~TeV$^{~\color{blue}\bullet}$ & \cellcolor{gray!10} $3.0$~TeV$^{~\color{blue}\bullet}$ & $0.9$~TeV$^{~\color{blue}\bullet}$ & -- & $1.8$~TeV\\[0.4em]
$\mathcal{C}_{\substack{eu\\1333}}$ & $\ast$ & $\ast$ & {$0.9$~TeV}$^{~\color{blue}\bullet\bullet}$ & $2.7$~TeV$^{~\color{blue}\bullet}$ & $2.2$~TeV$^{~\color{blue}\bullet}$ & \cellcolor{gray!10}$2.9$~TeV$^{~\color{blue}\bullet}$ & $0.8$~TeV$^{~\color{blue}\bullet}$ & -- & -- \\[0.4em]
$\mathcal{C}_{\substack{lu\\1333}}$ & $\ast$ & $\ast$ & {$1.0$~TeV}$^{~\color{blue}\bullet\bullet}$ & $2.7$~TeV$^{~\color{blue}\bullet}$ & $2.2$~TeV$^{~\color{blue}\bullet}$ & \cellcolor{gray!10} $2.9$~TeV$^{~\color{blue}\bullet}$ &  $0.8$~TeV$^{~\color{blue}\bullet}$ & -- & -- \\[0.4em]
$\mathcal{C}_{\substack{ld\\1333}}$ & -- & -- & $\ast$ & $0.7$~TeV$^{~\color{blue}\bullet}$ & $\ast$ & $0.8$~TeV$^{~\color{blue}\bullet}$ & $\ast$ & -- & \cellcolor{gray!10} $1.8$~TeV \\[0.4em]
$\mathcal{C}_{\substack{ed\\1333}}$ & -- & -- & $\ast$ & $0.7$~TeV$^{~\color{blue}\bullet}$ & $\ast$ & $0.8$~TeV$^{~\color{blue}\bullet}$ & $\ast$ & -- & \cellcolor{gray!10} $1.8$~TeV \\[0.4em] 
$\mathcal{C}_{\substack{lequ\\1333}}^{(1)}$ & $\ast$ & $\ast$ & \cellcolor{gray!10} {$15$~TeV}$^{~\color{blue}\bullet\bullet}$ & {$3.6$~TeV}$^{~\color{blue}\bullet\bullet}$ & $\ast$ & {$2.4$~TeV}$^{~\color{blue}\bullet\bullet}$ & $\ast$ & $2.5$~TeV$^{~\color{blue}\bullet}$ & -- \\[0.4em]
$\mathcal{C}_{\substack{lequ\\1333}}^{(3)}$ & -- & -- & \cellcolor{gray!10} $300$~TeV$^{~\color{blue}\bullet}$ & $93$~TeV$^{~\color{blue}\bullet}$ & $\ast$ & $69$~TeV$^{~\color{blue}\bullet}$ & $\ast$ & -- & -- \\[0.4em]
$\mathcal{C}_{\substack{ledq\\1333}}$ & $\ast$ & $\ast$ & $\ast$ & $\ast$ & {$0.9$~TeV}$^{~\color{magenta} \square}$ & $\ast$ & -- & $\ast$ & \cellcolor{gray!10} $1.7$~TeV \\[0.4em] \hline
\end{tabular}}
\caption{\small \sl Same as Table~\ref{tab:lfv-mue} for the $\tau\to e$ transition.
}
\vspace{2em}
\label{tab:lfv-taue} 
\renewcommand{\arraystretch}{1.8}
\centering
\resizebox{0.93\columnwidth}{!}{
\begin{tabular}{|c|ccccccccc|}
\hline 
\multicolumn{10}{|c|}{Lower limits on $\Lambda$ (for $\mathcal{C}=1$)}\\ \hline\hline
Coeff.  &  $b\to d \mu\tau$ & $b\to s \mu\tau$ & $\tau\to \mu\gamma$  & $\tau \to \mu\ell\ell$ & $\tau\to \mu P$ & $\tau \to \mu V$ & $Z\to \mu\tau$ & $h\to \mu \tau$ & $pp\to \mu\tau$\\[0.35em] \hline
$\mathcal{C}_{\substack{lq\\2333}}^{(1+3)}$ & $\ast$ & $\ast$ & $0.8$~TeV$^{~\color{blue}\bullet\bullet}$ & $0.7$~TeV$^{~\color{blue}\bullet}$ & $\ast$ & $0.7$~TeV$^{~\color{blue}\bullet}$ & $\ast$ & -- & \cellcolor{gray!10} $2.3$~TeV \\[0.4em]
$\mathcal{C}_{\substack{lq\\2333}}^{(1-3)}$ & -- & -- & $1.5$~TeV$^{~\color{blue}\bullet\bullet}$ & $2.8$~TeV$^{~\color{blue}\bullet}$ &  $2.0$~TeV$^{~\color{blue}\bullet}$ &\cellcolor{gray!10} $3.5$~TeV$^{~\color{blue}\bullet}$ & $0.7$~TeV$^{~\color{blue}\bullet}$ & -- & -- \\[0.4em]
$\mathcal{C}_{\substack{eq\\2333}}$ & $\ast$ & $\ast$ & {$0.8$~TeV}$^{~\color{blue}\bullet\bullet}$ & $2.9$~TeV$^{~\color{blue}\bullet}$ & $1.9$~TeV$^{~\color{blue}\bullet}$ & \cellcolor{gray!10} $3.3$~TeV$^{~\color{blue}\bullet}$ & $0.8$~TeV$^{~\color{blue}\bullet}$ & -- & $2.4$~TeV \\[0.4em]
$\mathcal{C}_{\substack{eu\\2333}}$ & $\ast$ & $\ast$ &  $0.8$~TeV$^{~\color{blue}\bullet\bullet}$ & $2.9$~TeV$^{~\color{blue}\bullet}$ & $1.9$~TeV$^{~\color{blue}\bullet}$ & \cellcolor{gray!10}$3.1$~TeV$^{~\color{blue}\bullet}$ & $0.8$~TeV$^{~\color{blue}\bullet}$ & -- & -- \\[0.4em]
$\mathcal{C}_{\substack{lu\\2333}}$ & $\ast$ & $\ast$ &  $0.9$~TeV$^{~\color{blue}\bullet\bullet}$ & $2.9$~TeV$^{~\color{blue}\bullet}$ & $2.0$~TeV$^{~\color{blue}\bullet}$ & \cellcolor{gray!10} $3.1$~TeV$^{~\color{blue}\bullet}$ &  $0.8$~TeV$^{~\color{blue}\bullet}$ & -- & -- \\[0.4em]
$\mathcal{C}_{\substack{ld\\2333}}$ & -- & -- & $\ast$ & $0.8$~TeV$^{~\color{blue}\bullet}$ & $\ast$ & $0.8$~TeV$^{~\color{blue}\bullet}$ & $\ast$ & -- & \cellcolor{gray!10} $2.4$~TeV \\[0.4em]
$\mathcal{C}_{\substack{ed\\2333}}$ & -- & -- & $\ast$ & $0.8$~TeV$^{~\color{blue}\bullet}$ & $\ast$ & $0.8$~TeV$^{~\color{blue}\bullet}$ & $\ast$ & -- & \cellcolor{gray!10} $2.4$~TeV \\[0.4em] 
$\mathcal{C}_{\substack{lequ\\2333}}^{(1)}$ & $\ast$ & $\ast$ & \cellcolor{gray!10} {$13$~TeV}$^{~\color{blue}\bullet\bullet}$ & {$4.0$~TeV}$^{~\color{blue}\bullet\bullet}$ & $\ast$ & {$2.8$~TeV}$^{~\color{blue}\bullet\bullet}$ & $\ast$ & $2.7$~TeV$^{~\color{blue}\bullet}$ & -- \\[0.4em]
$\mathcal{C}_{\substack{lequ\\2333}}^{(3)}$ & -- & -- & \cellcolor{gray!10}$280$~TeV$^{~\bullet}$ & $100$~TeV$^{~\color{blue}\bullet}$ & $\ast$ & $77$~TeV$^{~\color{blue}\bullet}$ & $\ast$ & -- & -- \\[0.4em]
$\mathcal{C}_{\substack{ledq\\2333}}$ & $\ast$ & $\ast$ & $\ast$ & $\ast$ & {$0.9$~TeV}$^{~\color{magenta} \square}$ & $\ast$ & -- & $\ast$ & \cellcolor{gray!10} $2.2$~TeV \\[0.4em] \hline
\end{tabular}}
\caption{\small \sl Same as Table~\ref{tab:lfv-mue} for the $\tau\to \mu$ transition.
}
\label{tab:lfv-taumu} 
\end{table}

\noindent   It is important to stress that $K$- and $B$-meson decays are not competitive with the other constraints depicted in Tables~\ref{tab:lfv-mue}--\ref{tab:lfv-taumu} because of our assumption on the quark-flavor content of the operators appearing at the $\Lambda$ scale. Since the only source of quark-flavor violation that we introduce is the top Yukawa, we find that the $s\to d$, $b\to d$ and $b\to s$ decays are suppressed by $V_{ts} V_{td}^\ast$, $V_{tb} V_{td}^\ast$ and $V_{tb} V_{ts}^\ast$, respectively. For this reason, we are not able to obtain meaningful constraints from these processes. However, it is clear that these observables can be useful to probe scenarios with different flavor assumptions, with several concrete examples recently studied in the context of leptoquark models~\cite{Becirevic:2016oho,Angelescu:2018tyl,Angelescu:2021lln,Bordone:2018nbg}.

\begin{figure}[!p]
    \centering
 \hspace{-2.em}   \centerline{\includegraphics[width=0.6\linewidth]{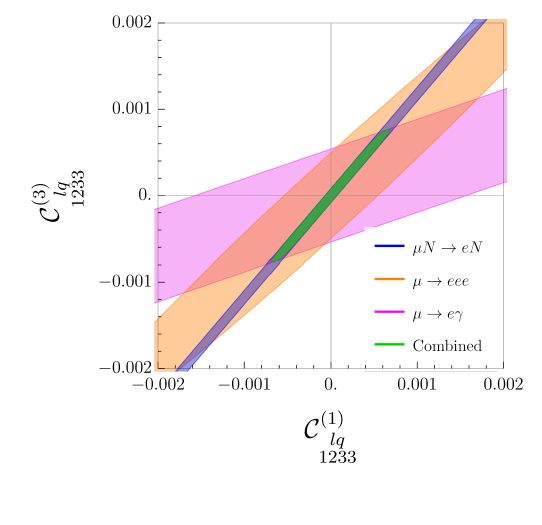}
 \hspace{-1.5em}  \includegraphics[width=0.6\linewidth]{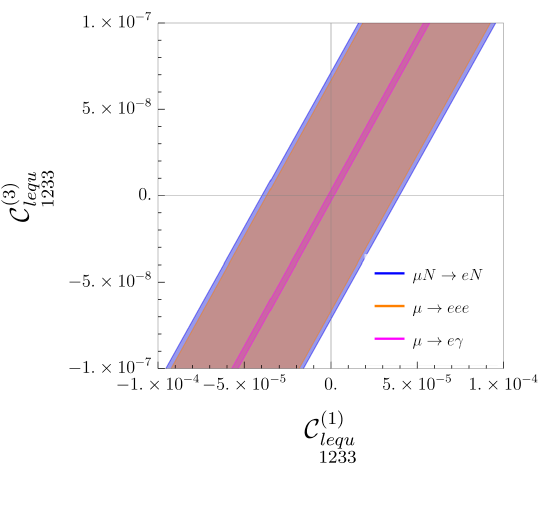}  } \\
\vspace{-1.5em}
 \hspace{-2.em}   \centerline{\includegraphics[width=0.6\linewidth]{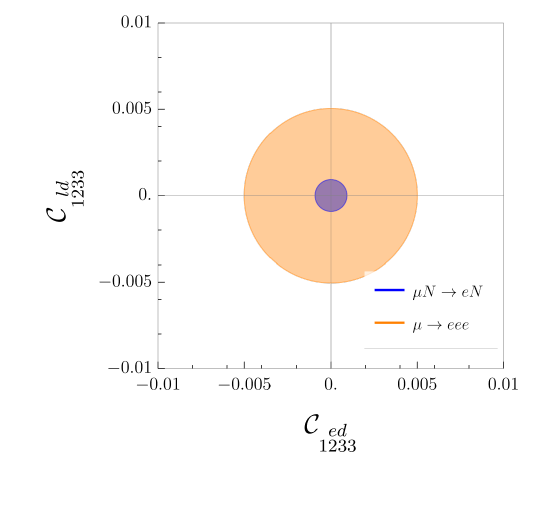}
   \hspace{-1.5em}  \includegraphics[width=0.6\linewidth]{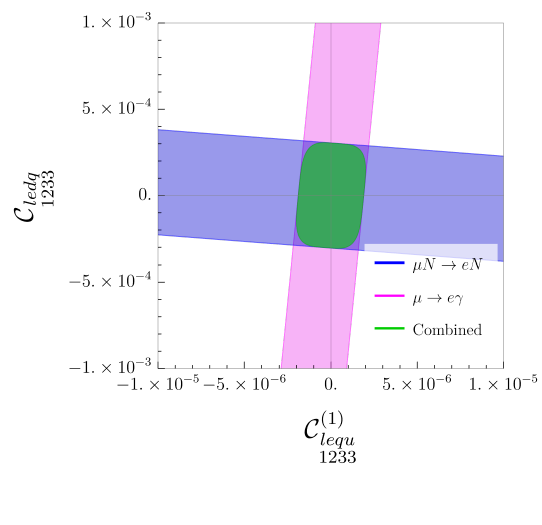}}
\vspace{-1em}
    \caption{\small \sl Constraints on selected $\mu\to e$ semileptonic Wilson coefficients, with of third generation quarks, derived to $95\%$ accuracy from $\mu N \to e N$ (blue), $\mu\to eee$ (orange) and $\mu\to e\gamma$ (magenta), cf.~Table~\ref{tab:lfv-mue}.  The combined constraint is shown in green in the upper left and bottom right panels. The EFT cutoff is fixed to $\Lambda = 1$ TeV.} 
    \label{fig:EFT_emu}
\end{figure}

\begin{figure}[p!]
    \centering
 \hspace{-2.7em}  \centerline{  \includegraphics[width=0.57\linewidth]{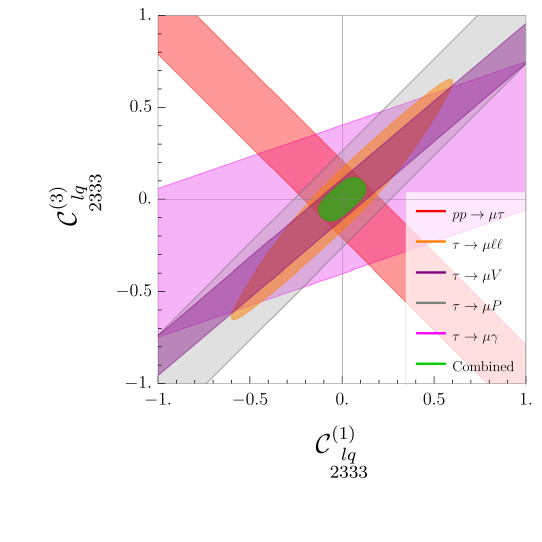}
    \includegraphics[width=0.57\linewidth]{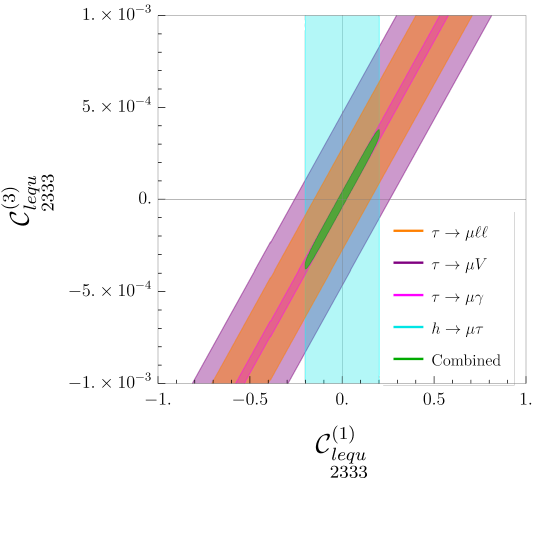}}
     \\
 \vspace{-1.5em}
 \hspace{-2.7em}   \centerline{ \includegraphics[width=0.57\linewidth]{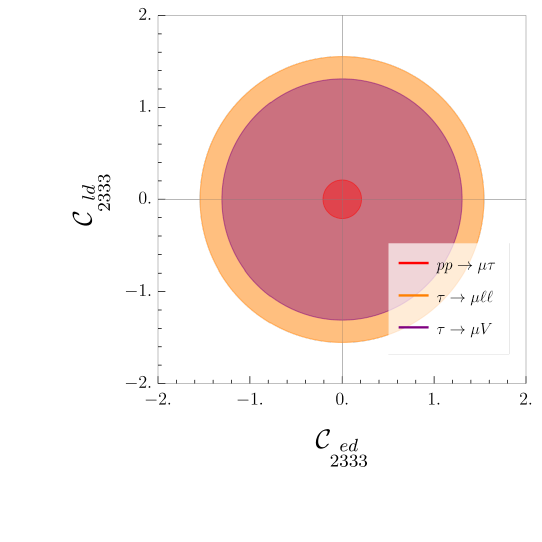} 
   \includegraphics[width=0.57\linewidth]{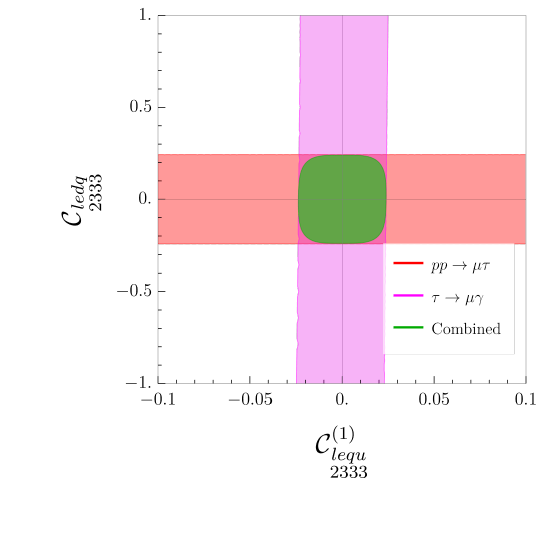}}
    \caption{\small \sl Constraints on selected $\tau\to \mu$ semileptonic Wilson coefficients, with of third generation quarks, derived to $95\%$ accuracy from $pp\to\mu \tau$ (red), $\tau\to \mu ll$ (orange), $\tau\to \mu V$ (purple), $\tau\to\mu P$ (gray) and $\tau\to \mu \gamma$ (magenta), cf.~Table~\ref{tab:lfv-taumu}.  The combined constraint is shown in green in the upper left, upper right and bottom right panels. The EFT cutoff is fixed to $\Lambda = 1$ TeV. Similar results are obtained for the $\tau\to e$ decays, which are not shown above.} 
    \label{fig:EFT_mutau}
\end{figure}

For the sake of comparison, we also quote the limits that we derive from the $\Upsilon\to \ell_i\ell_j$ decays at tree level (cf.~Table~\ref{tab:exp-lfv-had}). For instance, for the $\mathcal{C}_{ed}$ Wilson coefficients, 
\begin{equation}
    \dfrac{\Lambda}{\sqrt{|\mathcal{C}_{\substack{ed\\1233}}|}} \gtrsim 0.5 \text{ TeV}\,, \qquad\qquad     \dfrac{\Lambda}{\sqrt{|\mathcal{C}_{\substack{ed\\1333}}|}} \gtrsim 0.3 \text{ TeV}\,,   \qquad\qquad  \dfrac{\Lambda}{\sqrt{|\mathcal{C}_{\substack{ed\\2333}}|}} \gtrsim 0.3 \text{ TeV}\,,
\end{equation}
which are much weaker than the loop constraints collected in Tables~\ref{tab:lfv-mue}--\ref{tab:lfv-taumu}. Similar conclusions can be derived for the other operators that contribute to quarkonium decays at tree level. This exercise demonstrates once again the importance of accounting for the loop constraints discussed in this paper for the quark-flavor conserving operators made of $c$- and $b$-quarks, which allow us to derive far better constraints, e.g.,~via $\mu\to e$ conversion in nuclei or the various $\tau$-lepton LFV decays. These limits are also superseded by the Drell-Yan constraints, as shown in Tables~\ref{tab:lfv-mue}--\ref{tab:lfv-taumu} and as previously discussed in Ref.~\cite{Angelescu:2020uug,Descotes-Genon:2023pen}.

 The complementarity between the different types of processes is more striking when more than one Wilson coefficient is simultaneously present, as it is in fact predicted in several concrete scenarios (cf.~Sec.~\ref{ssec:concrete}).  In Fig.~\ref{fig:EFT_emu}, we plot constraints from $\mu\to e$ observables on selected pairs of effective coefficients to illustrate this complementarity. In particular, we find that it is necessary to consider more than one $\mu \to e$ observable to remove flat directions that could appear, e.g.,~in the $\mathcal{C}_{lq}^{(1)}$ vs.~$\mathcal{C}_{lq}^{(3)}$ (upper left) and $\mathcal{C}_{lequ}^{(1)}$ vs.~$\mathcal{C}_{ledq}$ (bottom right) planes. Other combinations of effective couplings are dominated by a single observable, such as $\mathcal{C}_{lequ}^{(1)}$ vs.~$\mathcal{C}_{lequ}^{(3)}$ (upper right) and $\mathcal{C}_{ed}$ vs.~$\mathcal{C}_{ld}$ (bottom left). Similar conclusions hold for the $\tau\to \mu$ observables, as shown in Fig.~\ref{fig:EFT_mutau}, where the $pp\to\mu\tau$ bounds play an important role in constraining potential flat directions. Notice that we do not show the analogous plot for the $\tau \to e$ transition since it has the same qualitative features as Fig.~\ref{fig:EFT_mutau}.

 Lastly, we have also performed a numerical comparison of our results for the observables that have been implemented in {\tt flavio}~\cite{Straub:2018kue}, by using the RGEs from the {\tt wilson} package~\cite{Aebischer:2018bkb}. These observables are the LFV decays of the $K$- and $B$-mesons, the purely leptonic processes $\ell\to \ell^\prime \gamma$ and $\ell\to 3\ell^\prime$, as well as $\mu N\to e N$ and LFV decays of the $Z$-boson. These packages are based on numerical integration of the one-loop RGEs, thus resumming the logarithms, which is a very useful cross-check of our numerical results. We find an overall good agreement between our results for the relevant Wilson coefficients, with deviations smaller than $\approx 20\%$ in most cases.~\footnote{We have verified that these small deviations are due to the running of the electroweak parameters and the SM Yukawa couplings, which for simplicity is not included in our numerical analysis.} However, there are a few disagreements that we fully understand. For instance, for the coefficient $\mathcal{C}_{ledq}$, we find more stringent constraints, e.g.,~from $\tau \to \ell P$ than {\tt flavio}, which can be traced back to the higher-dimensional gluonic operators that contribute to these processes via the (finite) one-loop effects depicted in Fig.~\ref{fig:ggtilde}. Another discrepancy that we have encountered concerns the contributions from $\mathcal{C}_{ed}$ and $\mathcal{C}_{ld}$ from $\mu\to e\gamma$, which are a factor of $\mathcal{O}(10)$ stronger for us. This disagreement can be traced back to the two-loop single-logarithm mixing of vector operators into dipoles in the low-energy EFT~\cite{Crivellin:2017rmk}, which is not implemented in {\tt flavio}.~\footnote{Note that similar discrepancies do not appear for the vector operators containing the top-quark such as $\smash{\mathcal{C}_{lq}^{(1-3)}}$, $\mathcal{C}_{eq}$, $\mathcal{C}_{eq}$ and $\mathcal{C}_{lu}$, because there are two-loop double-logarithm contributions of these coefficients in the SM EFT that are dominant over the two-loop single logarithm ones. Since {\tt wilson} integrates the RGEs numerically, the double logarithm effects are included in their results.}

\begin{table}[!t]
\renewcommand{\arraystretch}{1.8}
\centering
\resizebox{\columnwidth}{!}{\begin{tabular}{|c|c|c|cccccccccc|}
\hline 
Field  &  Spin   &  Quantum numbers  & $\mathcal{C}_{lq}^{(1)}$ &  $\mathcal{C}_{lq}^{(3)}$  & $\mathcal{C}_{lu}$  & $\mathcal{C}_{ld}$ &  $\mathcal{C}_{eq}$ &  $\mathcal{C}_{eu}$ &  $\mathcal{C}_{ed}$ &  $\mathcal{C}_{ledq}$  &  $\mathcal{C}_{lequ}^{(1)}$  &  $\mathcal{C}_{lequ}^{(3)}$ \\ \hline
$\Phi^\prime$ & 0 & $(\mathbf{1},\,\mathbf{2},\,1/2)$ &  &  &  &  &  &  &  & \checkmark & \checkmark & 
\\ \hdashline
$Z^\prime$ & 1 & $(\mathbf{1},\,\mathbf{1},\,0)$ & \checkmark & & \checkmark & \checkmark & \checkmark & \checkmark & \checkmark & & &
\\ 
$V$ & 1 & $(\mathbf{1},\,\mathbf{3},\,0)$ & & \checkmark & & & & & & & &
\\ \hdashline
$S_1$ & 0 & $(\mathbf{\bar{3}},\,\mathbf{1},\,1/3)$ & \checkmark & \checkmark & & & & \checkmark & & & \checkmark & \checkmark
\\ 
$\widetilde{S}_1$ & 0 & $(\mathbf{\bar{3}},\,\mathbf{1},\,4/3)$ & & & & & & & \checkmark & & &
\\ 
$S_3$ & 0 & $(\mathbf{\bar{3}},\,\mathbf{3},\,1/3)$ & \checkmark & \checkmark & & & & & & & &
\\ 
$R_2$ & 0 & $(\mathbf{3},\,\mathbf{2},\,7/6)$ & & & \checkmark & & \checkmark & & & & \checkmark & \checkmark
\\ 
$\widetilde{R}_2$ & 0 & $(\mathbf{3},\,\mathbf{2},\,1/6)$ & & & & \checkmark & & & & & &
\\ 
$U_1$ & 1 & $(\mathbf{3},\,\mathbf{1},\,2/3)$ &  \checkmark &  \checkmark & & & & &   \checkmark &  \checkmark & &
\\ 
$\widetilde{U}_1$ & 1 & $(\mathbf{3},\,\mathbf{1},\,5/3)$ &  &  & & & & \checkmark &   &   & &
\\ 
$V_2$ & 1 & $(\mathbf{\bar{3}},\,\mathbf{2},\,5/6)$ &  &  &  & \checkmark & \checkmark & &    & \checkmark & &
\\ 
$\widetilde{V}_2$ & 1 & $(\mathbf{\bar{3}},\,\mathbf{2},\,-1/6)$ &   &  & \checkmark &  &  & &    &   & &
\\ 
$U_3$ & 1 & $(\mathbf{3},\,\mathbf{3},\,2/3)$ &  \checkmark &  \checkmark & & & & & & & &
\\ 
 \hline
\end{tabular}}
\caption{\small \sl Bosonic mediators that can induce the semileptonic  LFV operators at tree level are classified in terms of the SM quantum numbers, $(SU(3)_c, SU(2)_L, Y)$, with $Q=Y+T_3$ as hypercharge convention. The tree-level matching between these concrete models and the SM EFT is provided in Appendix~\ref{app:mediators}.}
\label{tab:summary-mediators} 
\end{table}

\subsection{Correlations in concrete models}
\label{ssec:concrete}

To further demonstrate the relevance of our results, we consider the bosonic mediators that can induce the semileptonic operators in Table~\ref{tab:lfv-mue}--\ref{tab:lfv-taumu} at tree level and we explore the correlation between the LFV effective coefficients arising in these models~\cite{deBlas:2017xtg}. To this purpose, we assume a minimalistic flavor structure, only considering couplings to third-generation quarks and leptons with different flavors, in agreement with the assumption made in the previous Section. These mediators are collected in Table~\ref{tab:summary-mediators}. They can be a second Higgs doublet~\cite{Branco:2011iw,Crivellin:2013wna}, a singlet or triplet $SU(2)_L$ vector-boson~\cite{Pappadopulo:2014qza}, or various low-energy scalar and vector leptoquarks~\cite{Becirevic:2016zri,Dorsner:2016wpm}.~\footnote{Note, in particular, that we neglect di-quark couplings of leptoquarks since they would make the proton unstable~\cite{Dorsner:2016wpm}.} The Lagrangian of each scenario and the tree-level matching to the SM EFT Lagrangian are given in detail in Appendix~\ref{app:mediators}.

\begin{figure}[h!]
    \centering
    \vspace{1.4em}
\centerline{\includegraphics[width=0.52\linewidth]{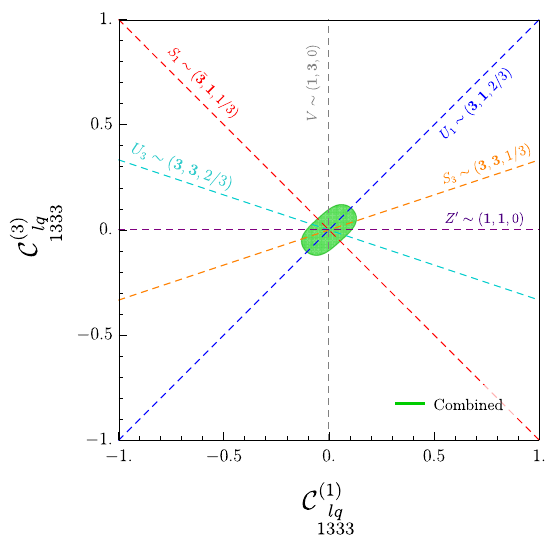} \quad \includegraphics[width=0.52\linewidth]{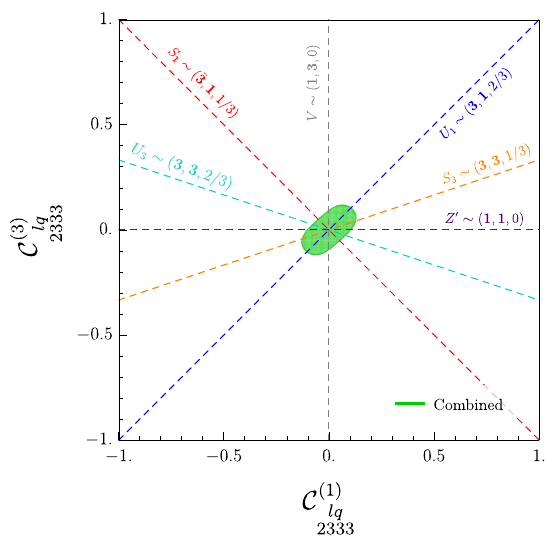}}
        \vspace{1.4em}
        \includegraphics[width=0.58\linewidth]{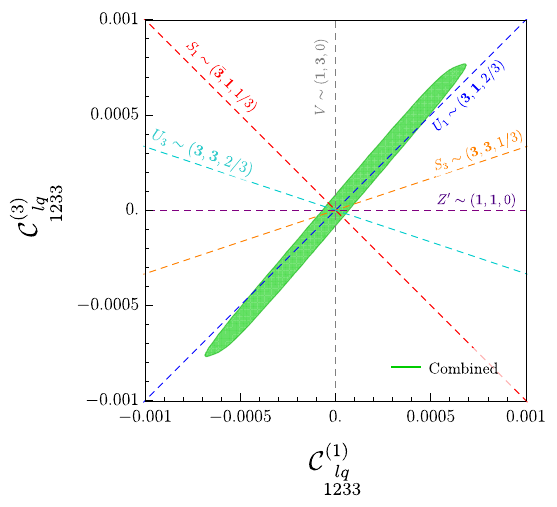}
    \caption{\small \sl Combined constraints on $\smash{\mathcal{C}_{lq}^{(1)}}$ vs.~$\smash{\mathcal{C}_{lq}^{(3)}}$, with third-generation quark flavor indices, derived from the most constraining $\tau \to e$ (top left), $\tau\to \mu$ (top right) and $\mu\to e$ (bottom) observables to $95\%$ CL.~accuracy. The correlations between these effective coefficients arising in the concrete scenarios listed in Table~\ref{tab:summary-mediators} are depicted by the dashed lines. The masses of the mediators are fixed to $\Lambda = 1$ TeV.} 
    \label{fig:mediators}
\end{figure}

In Fig.~\ref{fig:mediators}, we consider the specific example of operators coupled to left-handed fermions, which are induced by several of the concrete models listed in Table~\ref{tab:summary-mediators}, and we confront them to the global constraints from the LFV observables discussed above, by fixing the mediator masses to $\Lambda = 1$~TeV. We choose this example, since the correlations between $\mathcal{C}_{lq}^{(1)}$ and $\mathcal{C}_{lq}^{(3)}$ are fixed by the quantum numbers of the mediators in several cases, without requiring specific assumptions about the New Physics couplings. For the $\mu\to e$ transition, the least constrained scenario is $U_1\sim (\mathbf{3},\mathbf{1},2/3)$ with left-handed couplings, which lies along the diagonal $\mathcal{C}_{lq}^{(1)} = \mathcal{C}_{lq}^{(3)}$, since LFV operators containing the top-quark do not appear at tree level in this scenario~\cite{Angelescu:2018tyl,Angelescu:2021lln,Bordone:2018nbg}.~\footnote{See Ref.~\cite{Fuentes-Martin:2019ign} for detailed one-loop calculation in ultraviolet-complete models containing the $U_1$ leptoquark.} For the other scenarios, we find even more stringent constraints via the loop effects discussed above.  For the $\tau\to \ell$ transition, we obtain comparable bounds for the various models that are shown in Fig.~\ref{fig:mediators}. Finally, we note that correlations between the scalar $\smash{\mathcal{C}_{lequ}^{(1)}}$ and tensor $\smash{\mathcal{C}_{lequ}^{(3)}}$ also can arise for the scenarios with $R_2\sim (\mathbf{3},\mathbf{2},7/6)$ and $S_1\sim (\bar{\mathbf{3}},\mathbf{1},1/3)$~\cite{Feruglio:2018fxo}, cf.~Appendix~\ref{app:mediators}. In this case, the constraints on the tensor operators are far more stringent than on the scalar ones, thanks to the chirality-enhancement of the leptonic dipoles operators, as shown in the Tables~\ref{tab:lfv-mue}--\ref{tab:lfv-taumu}. These observables give the dominant bounds on the Yukawa couplings appearing in these models, see e.g.~Ref.~\cite{Mandal:2019gff}.

\section{Conclusion}
\label{sec:conclusion}

In this paper, we have studied Lepton Flavor Violation (LFV) in semileptonic observables using an Effective Field Theory (EFT) approach with a general flavor structure. Besides the tree-level effects that have been extensively studied in the literature, we have included the one-loop effects induced by the Renormalization Group Equations (RGE) below~\cite{Crivellin:2017rmk,Jenkins:2017dyc} and above~\cite{Jenkins:2013zja} the electroweak scale, as well as the relevant two-loop effects that are available in the literature~\cite{Crivellin:2017rmk}. In this way, we provided general expressions for several high-energy observables such as the decays $h \to \ell\ell^\prime$, $Z \to \ell \ell^\prime$ and $t \to c \ell \ell^\prime$, in addition to the high-energy tails of $pp \to \ell \ell^\prime$ that were studied in Ref.~\cite{Angelescu:2020uug,Descotes-Genon:2023pen}. At low energies, we have provided general expressions for the semileptonic processes $\mu N \to e N$, $M \to \ell \ell^\prime$, $M \to M^\prime \ell \ell^\prime$ and $\tau \to \ell M$, where $M^{(\prime)}$ stands for a vector or pseudoscalar meson, and the purely leptonic processes $\ell \to \ell^\prime \gamma$ and $\ell \to 3\ell^\prime$ that can constrain semileptonic operators at the loop level. In particular, we have updated the predictions for  the $M \to M^\prime \ell \ell^\prime$ decays by using the latest determination of the relevant form-factors on the lattice.

The loop effects that we consider induce important correlations between the various LFV observables and, in several instances, they allow us to derive even stronger constraints than those we would obtain at tree level. To illustrate these results, we have considered in Sec.~\ref{sec:pheno} the scenarios where only semileptonic LFV Wilson coefficients with third-generation-quarks are non-zero at the high scale $\Lambda$ and we have computed the various observables induced either at tree- or loop-level. We have obtained the lower limits on the EFT cutoff ($\Lambda$) that are given in Tables~\ref{tab:lfv-mue}--\ref{tab:lfv-taumu} by taking $\mathcal{C}_a(\Lambda)=1$, where $\mathcal{C}_a$ denotes the possible effective coefficient at the scale $\Lambda$. For the $\mu\to e$ transition, we have found that the most stringent constraints arise from the searches for $\mu \to e \gamma$, $\mu \to eee$, and $\mu N \to e N$, depending on the operator one considers, see also Ref.~\cite{Crivellin:2017rmk}. For the $\tau \to \ell$ transitions (with $\ell=e,\mu$), we highlight an interesting complementarity between the processes $\tau \to \ell \gamma$, $\tau \to \ell \ell \ell$, $\tau \to \ell P$ and $\tau \to \ell V$, which are induced at one-loop, with $pp \to \ell \tau$ that receives tree-level contributions. The complementarity between the different probes is even more evident in the two-dimensional plane, where pairs of effective coefficients are considered, as shown in Fig.~\ref{fig:EFT_emu} and \ref{fig:EFT_mutau} for the $\mu\to e$ and $\tau\to\mu$ transitions, respectively. 

It is important to emphasize that going beyond the leading logarithm solution is needed to probe several effective coefficients at the loop level (cf.~Ref.~\cite{Crivellin:2017rmk,Panico:2018hal,EliasMiro:2021jgu}). Among the effects that we have included, we accounted for the finite matching of scalar to gluonic operators, which contributes to $\mu N \to e N$ and to the $\tau \to \ell P$ decays, where $P\in \lbrace \pi^0, \eta,\eta^\prime \rbrace$~\cite{Bhattacharya:2018ryy}, as depicted in Fig.~\ref{fig:ggtilde}. We have also accounted for the two-loop single-logarithm mixing between vector operators and dipoles, which provides the most stringent constraint on several $\mu\to e$ operators, as computed in Ref.~\cite{Crivellin:2017rmk} for the low-energy EFT, cf.~Table~\ref{tab:lfv-mue}. Lastly, several two-loop double-logarithm effects are relevant to the phenomenology of $\mu\to e \gamma$ and $\tau\to\ell\gamma$. These include the mixing of a scalar into a tensor, which then mixes into dipoles, both in the low-energy EFT and within the SM EFT. Another example is the mixing of vector into tensor operators in the SM EFT, which mixes again into dipoles, and which has also been included in our phenomenological analysis. Even though these are only partial results, since the full two-loop anomalous dimension is not known yet, it shows the importance of computing these effects that can be relevant given the current precision of the experimental searches.

Finally, we have briefly discussed the implications of our results to concrete scenarios that generate semileptonic operators at tree level. We have provided the tree-level matching of these models to the SM EFT in Appendix~\ref{app:mediators}. Moreover, we have shown that these scenarios induce different correlations of the effective coefficients that can be tested experimentally, as depicted in Fig.~\ref{fig:mediators} for the effective coefficients $\mathcal{C}_{lq}^{(1)}$ and $\mathcal{C}_{lq}^{(3)}$. This example again shows the importance of accounting for loop-induced effects to constrain the size of LFV contributions to low-energy decays, with an interesting complementarity to high-energy processes.

\section{Acknowledgments}
\label{sec:acknowledgment}

We thank Peter Stangl for useful exchanges on {\tt wilson}~\cite{Aebischer:2018bkb} and {\tt flavio}~\cite{Straub:2018kue}, Lukas Allwicher for useful discussions, and Sébastien Descotes-Genon for comments on the manuscript. This project has received support from the European Union’s Horizon 2020 research and innovation programme under the Marie Skłodowska-Curie grant agreement N$^\circ$~860881-HIDDeN. I.P.~receives funding from
``P2IO LabEx (ANR-10-LABX-0038)" in the framework ``Investissements d’Avenir" (ANR-11-IDEX-0003-01) managed by the Agence Nationale de la Recherche (ANR), France.

\appendix

\section{Conventions}
\label{app:conventions}

We consider the same notation of Ref.~\cite{Jenkins:2013zja} for the operators in the Warsaw basis~\cite{Grzadkowski:2010es}. Quark and lepton weak-doublets are denoted by $q$ and $l$, while up- and down-type quarks, and lepton singlets are denoted by $u$, $d$, and $e$, respectively. The covariant derivative acting on a quark-doublet reads
\begin{equation}
D_\mu=\partial_\mu + i g^\prime \, Y B_\mu + i g\, \frac{\tau^I}{2} W^I_\mu + i g_3\, T^A G_\mu^A \,,
\end{equation}
where $T^A=\lambda^A/2$ are the $SU(3)_c$ generators, $\tau^I$ are the Pauli matrices and $Y$ denotes the hypercharge. The Yukawa interactions are defined in the flavor basis as follows,
\begin{equation}
-\mathcal{L}_{\mathrm{yuk}} = H^\dagger\, \bar{d}\, y_d\, q + \widetilde{H}^\dagger \,\bar{u}\, y_u\, q + H^\dagger \, \bar{e}\, y_e\, l+\mathrm{h.c.}\,,
\end{equation}

\noindent where $y_f$ (with $f\in \lbrace u,d,e \rbrace$) stand for the Yukawa matrices and flavor indices have been omitted. The SM Higgs doublet is denoted by $H$, with the conjugate field defined as $\widetilde H\equiv \epsilon H^\ast$ and the $SU(2)$ anti-symmetric tensor is  $\epsilon \equiv i\tau^2$.
We opt to work in the basis where $y_\ell$ and $y_d$ are diagonal matrices, while 
$y_u$ depends on the CKM matrix, $V \equiv V_{\mathrm{CKM}}$ (i.e., $y_u = \hat{y}_u V$).

\section{LEFT basis for $b\to s \ell_i \ell_j$}
\label{app:left-basis}

For the convenience of the reader, we provide the matching of our operator basis for the $b\to s \ell_i \ell_j$ transition to the one that is usually considered in the literature on $B$-meson LFV decays~\cite{Becirevic:2016zri,Becirevic:2016oho},
\begin{align}
\mathcal{H}_\mathrm{eff} =  -\frac{4
    G_F}{\sqrt{2}}V_{tb}V_{ts}^\ast &\sum_{a=9,10,S,P}
  \Big{[} C^{ij}_a(\mu)\mathcal{O}^{ij}_a(\mu) + C^{ij}_{a^\prime}(\mu) \mathcal{O}^{ij}_{a^\prime}(\mu)\Big{]}
+\mathrm{h.c.}\,,
\end{align}
where
\begin{align}
\label{eq:C_LFV}
\mathcal{O}_{9}^{ij}
  &=\frac{e^2}{(4\pi)^2}(\bar{s}\gamma_\mu P_{L}
    b)(\bar{\ell}_i\gamma^\mu\ell_j)\,, \qquad\qquad\hspace*{0.4cm}
\mathcal{O}_{S}^{ij} =
  \frac{e^2}{(4\pi)^2}(\bar{s} P_{R} b)(\bar{\ell}_i \ell_j)\,,\\
\mathcal{O}_{10}^{ij} &=
    \frac{e^2}{(4\pi)^2}(\bar{s}\gamma_\mu P_{L}
    b)(\bar{\ell}_i\gamma^\mu\gamma^5\ell_j)\,,\qquad\qquad
\mathcal{O}_{P}^{ij} =
  \frac{e^2}{(4\pi)^2}(\bar{s} P_{R} b)(\bar{\ell}_i \gamma^5 \ell_j)\,. \nonumber
\end{align}
The two operator bases are related via
\begin{align}
     C_{9}^{\ell_i \ell_j} &= \frac{\pi}{\lambda_t\alpha_{em}} \Big{(}  C^{(d)}_{\substack{V_{RL}\\ ij23}} +C^{(d)}_{\substack{V_{LL}\\ ij23}} \Big{)} \,, & C_{9^\prime}^{\ell_i \ell_j} &= \frac{\pi}{\lambda_t\alpha_{em}} \Big{(}  C^{(d)}_{\substack{V_{RR}\\{ij23}}} + C^{(d)}_{\substack{V_{LR}\\ ij23}} \Big{)}\,,\\[0.4em]
     C_{10}^{\ell_i \ell_j} &= \frac{\pi}{\lambda_t\alpha_{em}} \Big{(} C^{(d)}_{\substack{V_{RL}\\ ij23}} - C^{(d)}_{\substack{V_{LL}\\ij23}} \Big{)} \,, & C_{10^\prime}^{\ell_i \ell_j} &= \frac{\pi}{\lambda_t\alpha_{em}} \Big{(} C^{(d)}_{\substack{V_{RR}\\ij23}} - C^{(d)}_{\substack{V_{LR}\\ij23}} \Big{)} \,,\nonumber\\[0.4em]
     C_{S}^{\ell_i \ell_j} &= \frac{\pi}{\lambda_t\alpha_{em}} \Big{(} C^{(d)}_{\substack{S_{RR}\\ij23}} + C^{(d)}_{\substack{S_{LR}\\ij23}}  \Big{)}\,, &      C_{S^\prime}^{\ell_i \ell_j} &= \frac{\pi}{\lambda_t\alpha_{em}} \Big{(} C^{(d)}_{\substack{S_{RL}\\ij23}} + C^{(d)}_{\substack{S_{LL}\\ij23}}  \Big{)} \,,\nonumber\\[0.4em]
     C_{P}^{\ell_i \ell_j} &= \frac{\pi}{\lambda_t\alpha_{em}} \Big{(} C^{(d)}_{\substack{S_{RR}\\ij23}} - C^{(d)}_{\substack{S_{LR}\\ij23}} \Big{)} \,, &
     C_{P^\prime}^{\ell_i \ell_j} &= \frac{\pi}{\lambda_t\alpha_{em}} \Big{(} C^{(d)}_{\substack{S_{RL}\\ij23}} - C^{(d)}_{\substack{S_{LL}\\ij23}} \Big{)} \,, \nonumber
\end{align}

\noindent with $\lambda_t=V_{tb} V_{ts}^\ast$\,. The results for the $b\to d\ell_i \ell_j$ transition can be obtained by a trivial replacement of the flavor indices.

\section{Matching to the SM EFT}
\label{app:SMEFT-matching}

The matching between LEFT coefficients defined above with the SM EFT operators introduced in Sec.~\ref{ssec:SMEFT-LFV} reads:

\subsection*{Vector Operators}
\begin{align}
C_{\substack{V_{LL}\\ijkl}}^{(u)} &= \dfrac{v^2}{\Lambda^2} \mathcal{C}_{\substack{lq\\ijkl}}^{\prime (1-3)} +2 g^Z_{u_L} \delta_{kl} \dfrac{v^2}{\Lambda^2}\,\mathcal{C}_{\substack{Hl\\ij}}^{(1+3)}\,, \quad &
    C_{\substack{V_{RR}\\ijkl}}^{(u)} &= \dfrac{v^2}{\Lambda^2} \mathcal{C}_{\substack{eu\\ijkl}}+2g^Z_{u_R} \delta_{kl} \dfrac{v^2}{\Lambda^2}\,\mathcal{C}_{\substack{He\\ij}}\,, &\\[0.35em]
    C_{\substack{V_{LR}\\ijkl}}^{(u)}&= \dfrac{v^2}{\Lambda^2} \mathcal{C}_{\substack{lu\\ijkl}}+2 g^Z_{u_R} \delta_{kl} \dfrac{v^2}{\Lambda^2}\,\mathcal{C}_{\substack{Hl\\ij}}^{(1+3)}\,,&
    C_{\substack{V_{RL}\\ ijkl}}^{(u)} &= \dfrac{v^2}{\Lambda^2}  \mathcal{C}^\prime_{\substack{eq\\ijkl}}+ 2g^Z_{u_L} \delta_{kl} \dfrac{v^2}{\Lambda^2}\,\mathcal{C}_{\substack{He\\ij}}\,, &
    \nonumber\\[0.35em]  
   C_{\substack{V_{LL}\\ ijkl}}^{(d)} &= \dfrac{v^2}{\Lambda^2} \mathcal{C}_{\substack{lq\\ ijkl}}^{(1+3)}+2 g^Z_{d_L} \delta_{kl} \dfrac{v^2}{\Lambda^2} \,\mathcal{C}_{\substack{Hl\\ ij}}^{(1+3)}\,, & C_{\substack{V_{RR}\\ijkl}}^{(d)} &= \dfrac{v^2}{\Lambda^2}\mathcal{C}_{\substack{ed\\ijkl}}+2 g^Z_{d_R} \delta_{kl} \dfrac{v^2}{\Lambda^2}\,\mathcal{C}_{\substack{He\\ij}}\,, &\nonumber\\[0.4em]
    C_{\substack{V_{LR}\\ ijkl}}^{(d)} &= \dfrac{v^2}{\Lambda^2} \mathcal{C}_{\substack{ld\\ijkl}}+2 g^Z_{d_R} \delta_{kl} \dfrac{v^2}{\Lambda^2}\,\mathcal{C}_{\substack{Hl\\ij}}^{(1+3)}\,,&
    C_{\substack{V_{RL}\\ijkl}}^{(d)} &=  \dfrac{v^2}{\Lambda^2} \mathcal{C}_{\substack{eq\\ijkl}} + 2g^Z_{d_L} \delta_{kl} \dfrac{v^2}{\Lambda^2}\,\mathcal{C}_{\substack{He\\ij}}\,, &\nonumber\\[0.35em]
    C_{\substack{V_{LL}\\ijkk}}^{(\ell)} &= \frac{\upsilon^2}{\Lambda^2} \mathcal{C}_{\substack{ll\\ ijkk}} +2g^Z_{e_L}  \dfrac{v^2}{\Lambda^2}\,\mathcal{C}_{\substack{Hl\\ ij}}^{(1+3)}  \,, &
     C_{\substack{V_{RR}\\ ijkk}}^{(\ell)} &= \frac{\upsilon^2}{\Lambda^2}  \mathcal{C}_{\substack{ee\\ijkk}} +2g^Z_{e_R}  \dfrac{v^2}{\Lambda^2}\,
     \mathcal{C}_{\substack{He\\ij}}   \,, &\nonumber\\[0.35em]
    C_{\substack{V_{LR}\\ ijkk}}^{(\ell)} &= \frac{\upsilon^2}{\Lambda^2}  \mathcal{C}_{\substack{le\\ijkk}} ++2g^Z_{e_R}  \dfrac{v^2}{\Lambda^2} \, \mathcal{C}_{\substack{Hl\\ ij}}^{(1+3)} \,, &
    C_{\substack{V_{RL}\\ ijkk}}^{(\ell)} &= \frac{\upsilon^2}{\Lambda^2}\mathcal{C}_{\substack{le\\ kkij}} +2g^Z_{e_L}  \dfrac{v^2}{\Lambda^2}\, \mathcal{C}_{\substack{He\\ij}} \,, &\nonumber
\end{align}

\noindent where the couplings of the $Z$-boson to the SM fermions are defined as $\smash{g_{f_L}^Z = T_3^f-Q_f\sin^2 \theta_W}$ and $\smash{g_{f_R}^Z = -Q_f\sin^2 \theta_W}$, where $T_3^f$ stands for the third-component of the weak isospin and $Q_f$ is the electric charge of the fermion $f$. The primed coefficients are defined in Eq.~\eqref{eq:WC2} and we use again the shorthand notation $\mathcal{C}_{lq}^{(1\pm 3)} = \mathcal{C}_{lq}^{(1)} \pm \mathcal{C}_{lq}^{(3)}$ for the left-handed operators.

\subsection*{Scalar Operators}

\begin{align}
     C^{(u)}_{\substack{S_{RL}\\ ijkl}} &= -\delta_{kl}\frac{ \upsilon^3 m_{u_k}}{\sqrt{2} m_h^2\Lambda^2}\, \mathcal{C}_{\substack{eH\\ij}} \,, \qquad & \qquad
    C_{\substack{S_{LR}\\ ijkl}}^{(u)} &= -\delta_{kl} \frac{v^2}{\Lambda^2} \frac{v m_{u_k}}{\sqrt{2}m_h^2}\,
    \mathcal{C}^\ast_{\substack{eH\\ji}}   \,,&\\[0.4em]
    C^{(u)}_{\substack{S_{RR}\\ ijkl}} &= -\frac{\upsilon^2}{\Lambda^2} \,\mathcal{C}_{\substack{lequ\\ijkl}}^{\prime (1)} - \frac{v^2}{\Lambda^2} \frac{v m_{u_k}}{\sqrt{2}m_h^2}\,\mathcal{C}_{\substack{eH\\ij}} \,,&
    C_{\substack{S_{LL}\\ ijkl}}^{(u)} &= -\frac{\upsilon^2}{\Lambda^2} \mathcal{C}_{\substack{lequ\\jilk}}^{\prime (1)\ast} -\delta_{kl} \frac{v^2}{\Lambda^2} \frac{v m_{u_k}}{\sqrt{2}m_h^2}\,\mathcal{C}^\ast_{\substack{eH\\ji}}   \,,&\nonumber\\[0.4em]  
   C^{(d)}_{\substack{S_{RL}\\ ijkl}} &= \frac{\upsilon^2}{\Lambda^2} \,\mathcal{C}_{\substack{ledq\\ijkl}} -\delta_{kl} \frac{v^2}{\Lambda^2} \frac{v m_{d_k}}{\sqrt{2}m_h^2}\, \mathcal{C}_{\substack{eH\\ij}}
     \,, &
     C^{(d)}_{\substack{S_{LR}\\ijkl}} &= \frac{\upsilon^2}{\Lambda^2} \,\mathcal{C}_{\substack{ledq\\jilk}}^\ast -\delta_{kl} \frac{v^2}{\Lambda^2} \frac{v m_{d_k}}{\sqrt{2}m_h^2}\,\mathcal{C}^\ast_{\substack{eH\\ji}}  \,,\nonumber\\[0.4em]
     C^{(d)}_{\substack{S_{RR}\\ ijkl}} &= -\delta_{kl}\frac{ \upsilon^3 m_{d_k}}{\sqrt{2}m_h^2\Lambda^2} \mathcal{C}_{\substack{eH\\ij}} 
     \,, &
     C^{(d)}_{\substack{S_{LL}\\ijkl}} &= -\delta_{kl}\frac{ \upsilon^3 m_{d_k}}{\sqrt{2} m_h^2\Lambda^2}\, \mathcal{C}^\ast_{\substack{eH\\ji}}  \,,\nonumber\\[0.4em]
    C^{(\ell)}_{\substack{S_{RL}\\ ijkk}} &= - \frac{v^2}{\Lambda^2} \frac{v m_{\ell_k}}{\sqrt{2}m_h^2}\,\mathcal{C}_{\substack{eH\\ij}} \,,&
    C_{\substack{S_{LR}\\ ijkk}}^{(\ell)} &= -\frac{v^2}{\Lambda^2} \frac{v m_{\ell_k}}{\sqrt{2}m_h^2}\, \mathcal{C}^\ast_{\substack{eH\\ji}}   \,,&\nonumber\\[0.4em]
    C^{(\ell)}_{\substack{S_{RR}\\ ijkk}} &= -\frac{v^2}{\Lambda^2} \frac{v m_{\ell_k}}{\sqrt{2}m_h^2}\, \mathcal{C}_{\substack{eH\\ij}} \,,&
    C_{\substack{S_{LL}\\ ijkk}}^{(\ell)} &=  -\frac{v^2}{\Lambda^2} \frac{v m_{\ell_k}}{\sqrt{2}m_h^2}\, \mathcal{C}^\ast_{\substack{eH\\ji}}   \,.&\nonumber
\end{align}

\subsection*{Tensor Operators}
\begin{align}
    C^{(u)}_{\substack{T_R\\ijkl}} &=  -\frac{\upsilon^2}{\Lambda^2}  \mathcal{C}_{\substack{lequ\\ijkl}}^{\prime (3)}  \,,&
    C_{\substack{T_L\\ijkl}}^{(u)} &= -\frac{\upsilon^2}{\Lambda^2} [\mathcal{C}_{\substack{lequ\\jilk}}^{\prime (3)}]^\ast  \,,&\nonumber\\[0.4em]
    C_{\substack{T_L\\ ijkl}}^{(d)} &= C_{\substack{T_R\\ ijkl}}^{(d)}=0\,,\\[0.4em] 
    C_{\substack{T_L\\ijkk}}^{(\ell)} &= C_{\substack{T_R\\ ijkk}}^{(\ell)}=0\,.\nonumber
\end{align}

\subsection*{Leptonic Dipole Operators}

\begin{align}
    C_{\substack{D_R\\ij}}^{(\ell)} &= \frac{\upsilon^3}{\sqrt{2}m_{\ell} \Lambda^2} \Big{(}-\sin{\theta_W} \,\mathcal{C}_{\substack{eW \\ ij}} + \cos {\theta_W} \,\mathcal{C}_{\substack{eB \\ ij}}  \Big{)} \,,  \\[0.4em]
    C_{\substack{D_L\\ij}}^{(\ell)} &= \frac{\upsilon^3}{\sqrt{2}m_{\ell} \Lambda^2} \Big{(}-\sin{\theta_W} \,\mathcal{C}_{\substack{eW\\ ji}}^\ast  + \cos \theta_W \, \mathcal{C}_{\substack{eB\\ ji}}^\ast  \Big{)} \,. \nonumber
\end{align}

\section{Semileptonic differential formulas}
\label{app:diff-semileptonic}

\subsection{$P\to P^\prime \ell_i^- \ell_j^+$}

The expression for the differential distribution of $P\to P^\prime \ell_i^- \ell_j^+$ can be expressed in full generality as 
\begin{align}
    \dfrac{\mathrm{d}\Gamma}{\mathrm{d}q^2\, \mathrm{d}\cos\theta_\ell} = a(q^2) + b(q^2) \cos\theta_\ell + c(q^2)\, \cos^2 \theta_\ell\,,
\end{align}

\noindent where $\theta_\ell$ stands for the angle between $\ell_i^-$ and the $P^\prime$ meson line-of-flight, in the rest frame of the lepton-pair rest frame, and $q^2$ is the dilepton invariant mass. The kinematical coefficients $a(q^2)$, $b(q^2)$ and $c(q^2)$ can be expressed in terms of effective coefficients defined in Eq.~\eqref{eq:left-redef-1}--\eqref{eq:left-redef-4} as follows,
\begin{align}
a(q^2) &= \mathcal{N}_P \frac{\lambda_{\ell}^{1/2}}{4q^2} \bigg{[} |h_0^V(q^2)|^2 + |h_0^A(q^2)|^2  + \frac{(m_{\ell_i} - m_{\ell_j})^2}{q^2} |h_t^{VS}(q^2)|^2 + \frac{(m_{\ell_i} + m_{\ell_j})^2}{q^2} |h_t^{AP}(q^2)|^2 \bigg{]} \,,\nonumber\\[0.4em]
b(q^2) &= \mathcal{N}_P \frac{\lambda_{\ell}^{1/2}}{2q^2}(m_{\ell_i}^2 - m_{\ell_j}^2)  \, \text{Re}\Big{[}h_0^V(q^2) \big{(}h_t^{VS}(q^2)\big{)}^\ast+h_0^A(q^2) h_t^{AP}(q^2)^\ast\Big{]}   \,,\nonumber\\[0.4em]
c(q^2) &= - \mathcal{N}_P \frac{\lambda_{\ell}}{4q^4} \Big{(} |h_0^V(q^2)|^2 + |h_0^A(q^2)|^2 \Big{)} \,,
\end{align}
where the normalization is defined by, 
\begin{align}
\mathcal{N}_P = \dfrac{\tau_P G_F^2 \lambda_P^{1/2} \lambda_\ell^{1/2}}{512 \pi^3 m_P^3} \,,
\end{align}
and we define $\lambda_P = \lambda(m_P^2,m_{P^\prime}^2,q^2)$ and $\lambda_\ell =\lambda(m_{\ell_i}^2,m_{\ell_j}^2,q^2)$, with $\lambda(a^2,b^2,c^2)=(a^2-(b-c)^2)(a^2-(b+c)^2)$. The transversity amplitudes for a decay based on the $d_k \to d_l \ell_i^- \ell_j^+$ transition is given by the following equations,
\begin{align}
     h_0^V(q^2) &= \frac{C^{(q)}_{VV}}{\sqrt{q^2}} f_+(q^2) \lambda_P^{1/2} \,, \\  h_0^A(q^2) &= \frac{C^{(q)}_{AV}}{\sqrt{q^2}} f_+(q^2) \lambda_P^{1/2} \,, \\ 
     h_t^{VS}(q^2) &= \Big{(} C^{(q)}_{VV} + \frac{C^{(q)}_{SS}}{m_{\ell_i} - m_{\ell_j}} \frac{q^2}{m_{q_k}-m_{q_l}} \Big{)} \frac{m_P^2-m_{P^\prime}^2}{\sqrt{q^2}} f_0(q^2) \,, \\
     h_t^{AP}(q^2) &= \Big{(} C^{(q)}_{AV} + \frac{C^{(q)}_{PS}}{m_{\ell_i} + m_{\ell_j}} \frac{q^2}{m_{q_k}-m_{q_l}} \Big{)} \frac{m_P^2-m_{P^\prime}^2}{\sqrt{q^2}} f_0(q^2) \,.
\end{align}
where flavor indices should be replaced following Eq.~\eqref{eq:wc-semilep}. The form-factors in the above equations are defined as usual,
\begin{align}
    \langle P^\prime(k)\vert \bar{q}_l\gamma^\mu q_k \vert P(p) \rangle = \Big{[}(p+k)^\mu-\dfrac{(m_P^2-m_{P^\prime}^2)}{q^2}q^\mu\Big{]} f_+(q^2) + \dfrac{m_P^2-m_{P^\prime}^2}{q^2}q^\mu f_0(q^2)
\end{align}
\noindent where $f_+$ ($f_0$) stands for the vector (scalar) form-factors. Note, in particular, that the above expressions are in agreement with the ones obtained for the charged current decays $d\to u \ell \bar{\nu}$ in the limit where $m_{\ell}\to 0$, as provided e.g.~in Ref.~\cite{Becirevic:2020rzi}.~\footnote{Notice that we have neglected tensor operators in the above formulas since they represent sub-leading corrections to the $d_k\to d_l \ell_i^-\ell_j^+$ transition in the SM EFT. See Ref.~\cite{Gratrex:2015hna} for a calculation of the branching fractions including these operators.}

\subsection{$P\to V\ell_i^- \ell_j^+$}
\label{app:diff-semileptonic-V}

The expressions for the angular distributions of $P\to V\ell_i^- \ell_j^+$ have been derived in Ref.~\cite{Becirevic:2016zri}. For completeness, we collect these expressions in this Appendix,
\begin{align}
    \dfrac{\mathrm{d}\Gamma }{\mathrm{d}q^2 \mathrm{d}\cos \theta_\ell \mathrm{d}\cos \theta_K \mathrm{d}\phi} = \dfrac{9}{32\pi} I(q^2,\theta_\ell,\theta_K,\phi)\,,
\end{align}

\noindent with
\begin{align}
    I(q^2,\theta_\ell,\theta_K,\phi) &\equiv I_1^s(q^2)\sin^2\theta_K + I_1^c(q^2) \cos^2\theta_K+[I_2^s(q^2) \sin^2\theta_K+I_2^c(q^2)\,\cos^2\theta_K]\cos 2\theta_\ell\nonumber\\[.4em] 
&+I_3(q^2)\sin^2\theta_K \sin^2\theta_\ell \cos 2\phi+I_4(q^2)\sin 2\theta_K \sin 2\theta_\ell \cos \phi \nonumber \\[.4em] 
&+ I_5(q^2) \sin 2\theta_K\sin \theta_\ell\cos\phi+[I_6^s(q^2)\sin^2\theta_K+I_6^c(q^2)\,\cos^2\theta_K]\cos \theta_\ell \nonumber \\[.4em] 
&+I_7(q^2)\sin 2\theta_K \sin \theta_\ell \sin \phi + I_8(q^2)\sin 2\theta_K \sin 2\theta_\ell \sin\phi \nonumber\\[.4em] 
&+I_9(q^2) \sin^2\theta_K \sin^2\theta_\ell \sin 2 \phi\,,
\end{align}

\noindent where we adopt the same angular conventions used for $\overline{B}\to \overline{K}^{\ast 0}(\to K \pi) \ell_i^- \ell_j^+$ in Ref.~\cite{Becirevic:2016zri}. These angular coefficients can be expressed in terms of effective coefficients defined in Eq.~\eqref{eq:left-redef-1}--\eqref{eq:left-redef-4}~\cite{Becirevic:2016zri},
\begin{align}
I_1^s(q^2) &=\Big{[}|A_{\perp}^L|^2+|A_{\parallel}^L|^2+ (L\to R) \Big{]}\frac{\lambda_\ell +2 [q^4-(m_{\ell_i}^2-m_{\ell_j}^2)^2]}{4 q^4}\nonumber\\
&\hspace{3cm}+\frac{4 m_{\ell_i} m_{\ell_j}}{q^2}\,\mathrm{Re}\left(A_{\parallel}^L A_{\parallel}^{R\ast}+A_{\perp}^L A_{\perp}^{R\ast}\right)\,,  \\[0.4em]
I_1^c(q^2) &= \bigl[|A_0^L|^2+|A_0^R|^2 \bigr]\frac{q^4-(m_{\ell_i}^2-m_{\ell_j}^2)^2}{q^4}+\frac{8 m_{\ell_i} m_{\ell_j}}{q^2} \mathrm{Re}(A_0^L A_0^{R\ast}-A_t^L A_t^{R\ast}) \nonumber\\[0.4em]
&\hspace{3cm}-2\frac{(m_{\ell_i}^2-m_{\ell_j}^2)^2-q^2 (m_{\ell_i}^2+m_{\ell_j}^2)}{q^4}\bigl(|A_t^L|^2+|A_t^R|^2\bigr)\,,\\[0.4em]
I_2^s(q^2) &= \frac{\lambda_\ell}{4 q^4} \,[|A_\perp^L|^2+|A_\parallel^L|^2+(L\to R)]\,,  \\[0.4em]
I_2^c(q^2) &= - \frac{\lambda_\ell}{q^4}\,(|A_0^L|^2+|A_0^R|^2)\,,  \\[0.4em]
I_3(q^2) &= \frac{\lambda_\ell}{2 q^4}\, [|A_\perp^L|^2-|A_\parallel^L|^2+(L\to R)]\,, \\[0.4em]
I_4(q^2) &= - \frac{\lambda_\ell}{\sqrt{2} q^4}\, [\mathrm{Re}(A_\parallel^L A_0^{L\ast})+(L\to R)]\,, \\[0.4em]
I_5(q^2) &= \frac{\sqrt{2}\lambda_\ell^{1/2}}{q^2} \Big{[} \mathrm{Re}(A_0^L A_\perp^{L\ast}-(L\to R)) -\frac{m_{\ell_i}^2-m_{\ell_j}^2}{q^2} \mathrm{Re}(A_t^L A_\parallel^{L\ast}+(L\to R))\Big{]}\,, \\[0.4em]
I_6^s(q^2) &=- \frac{2 \lambda_\ell^{1/2}}{q^2}[\mathrm{Re}(A_\parallel^L A_\perp^{L\ast}-(L\to R))]\,, \\[0.4em]
I_6^c(q^2) &= - \frac{4\lambda_\ell^{1/2}}{q^2}\frac{m_{\ell_i}^2-m_{\ell_j}^2}{q^2} \mathrm{Re}(A_0^L A_t^{L\ast}+(L\to R))\,,\\[0.4em]
I_7(q^2) &= - \frac{\sqrt{2}\lambda_\ell^{1/2}}{q^2} \Big{[} \mathrm{Im}(A_0^L A_\parallel^{L\ast}-(L\to R))+\frac{m_{\ell_i}^2-m_{\ell_j}^2}{q^2} \, \mathrm{Im}(A_\perp^{L}A_t^{L\ast} +(L\to R))\Big{]}\,,  \\[0.4em]
I_8(q^2) &= \frac{\lambda_\ell}{\sqrt{2}q^4}\,\mathrm{Im}(A_0^{L}A_\perp^{L\ast} +(L\to R)),  \\[0.4em]
I_9(q^2) &=- \frac{\lambda_\ell}{q^4}\,\mathrm{Im}(A_\perp^L A_\parallel^{L\ast} +A_\perp^R A_\parallel^{R\ast} )\,,
\end{align}
where the transversity amplitudes are defined by
\begin{align}
A_{\perp}^{L(R)} &= \mathcal{N}_V  \sqrt{2} \lambda_V^{1/2}  \dfrac{V(q^2)}{m_P+m_V} \big{(}C_{VV}^{(q)} \mp C_{VA}^{(q)} \big{)}\,, \\[0.4em]
A_{\parallel}^{L(R)} &= \mathcal{N}_V \sqrt{2} (m_P+m_V) A_1(q^2) \big{(}C_{AV}^{(q)} \mp C_{AA}^{(q)} \big{)}\,, \\[0.4em]
A_0^{L(R)} &= \dfrac{\mathcal{N}_V (m_P+m_V)}{2 m_V \sqrt{q^2}} \bigg{[}(m_P^2-m_V^2-q^2)A_1(q^2)-\dfrac{\lambda_V A_2(q^2)}{m_P+m_V}\bigg{]}
\big{(}C_{AV}^{(q)} \mp C_{AA}^{(q)} \big{)}\,, \\[0.4em]
A_t^{L(R)} &=\mathcal{N}_V \dfrac{\lambda_P^{1/2}}{\sqrt{q^2}} A_0(q^2) \bigg{[}C_{AV}^{(q)}\mp C_{AA}^{(q)} - \dfrac{q^2}{m_{q_k}+m_{q_l}}\bigg{(}\dfrac{C_{PS}^{(q)}}{m_{\ell_i}-m_{\ell_j}}\mp \dfrac{C_{PP}^{(q)}}{m_{\ell_i}+m_{\ell_j}}\bigg{)}\bigg{]}\,, 
\end{align}

\noindent where, for simplicity, flavor indices are omitted. The square of the normalization reads
\begin{align}
\mathcal{N}_V^{\,2} \equiv \dfrac{\tau_P\, G_F^2 \lambda_V^{1/2}\lambda_\ell^{1/2}}{3 \times 2^{10}\pi^3 m_P^3}\,,
\end{align}
where $\lambda_V = \lambda(m_P^2,m_{V}^2,q^2)$ and $\lambda_\ell =\lambda(m_{\ell_i}^2,m_{\ell_j}^2,q^2)$, and the $P\to V$ form-factors are defined as follows~\cite{Altmannshofer:2008dz},
\begin{align}
    \langle V(k)\vert \bar{q}_l\gamma^\mu (1-\gamma_5)q_k &\vert P(p) \rangle = \varepsilon_{\mu\nu\rho\sigma} \varepsilon^{\ast\nu}p^\rho k^\sigma \dfrac{2 V(q^2)}{m_P+m_V}-i \varepsilon_\mu^\ast (m_P+m_V)A_1(q^2)\\
    &+i(p+k)_\mu (\varepsilon^\ast\cdot q) \dfrac{A_2(q^2)}{m_P+m_V}+iq_\mu (\varepsilon^\ast\cdot q)\dfrac{2 m_V}{q^2} [A_3(q^2)-A_0(q^2)]\,,\nonumber
\end{align}
where $\varepsilon^\mu$ is the polarization vector of the $V$-meson, and the form-factor $A_3(q^2)$ is related to $A_{1,2}(q^2)$ via $2m_VA_3(q^2)=(m_P+m_V) A_1(q^2)-(m_P-m_V)A_2(q^2)$. Notice, once again, that we have not included the contributions from tensor operators in the above expressions.

\section{SM Yukawa running}
\label{app:yuk-rge}

In this Appendix, we briefly discuss the impact of the SM Yukawa renormalization, which can impact semileptonic observables through loop-induced non-diagonal elements that are not necessarily present in the ultraviolet. More specifically, the Yukawa RGEs read~\cite{Machacek:1983tz}
\begin{align}
16 \pi^2 \dfrac{\mathrm{d}\,y_d}{\mathrm{d}\log\mu} \simeq \dfrac{3}{2}\big{(}y_d\, y_d^\dagger\, y_d -y_d\, y_u^\dagger\, y_u \big{)}+3\,\mathrm{Tr}\big{[}y_u^\dagger\, y_d + y_d^\dagger\, y_d\big{]}y_d-8\, g_3^2 y_d\,,
\end{align}
where the up- and down-type quark Yukawas are denoted by $y_u$ and $y_d$, respectively, and we have neglected the lepton Yukawas and the electroweak gauge couplings, cf.~Appendix~\ref{app:conventions} for our notation. Keeping only the top-quark Yukawa contribution, it is straightforward to show that off-diagonal contributions are induced via the Yukawa running,
\begin{align}
\label{eq:Yuk-rge}
\qquad y_{\substack{d\\ij}}(\mu=m_t) \simeq \dfrac{3 V_{ti} V_{tj}^\ast \,y_b y_t^2}{32\pi^2}\log \bigg{(}\dfrac{\Lambda}{m_t}\bigg{)}\,,\qquad\qquad\quad (i\neq j)\,,
\end{align}

\noindent where $y_b$ and $y_t$ are the physical top- and bottom-quark Yukawas to first approximation. The contributions in Eq.~\eqref{eq:Yuk-rge} must be reabsorbed when diagonalizing the SM Yukawas at $\mu \simeq \mu_\mathrm{ew}$ via the following rotation of the left-handed down-type quarks,
\begin{equation}
U_{d_L} \simeq {\mathbb{1}_{3\times 3}} +
\frac{3y_t^2}{32\pi^2}\log \bigg{(}\frac{\Lambda}{m_t}\bigg{)}\begin{pmatrix}
   \begin{array}{ccc}
  0  & 0 & V_{td}^\ast V_{tb}  \\
  0  & 0 & V_{ts}^\ast V_{tb}  \\
   -V_{td} V_{tb}^\ast  & -V_{ts} V_{tb}^\ast  & 0 \\
  \end{array} 
  \end{pmatrix}
 \;,  
\end{equation}

\noindent whereas the other fermions are not affected by the running to first approximation. These effects are included in our analysis, but we find that they are sub-leading for the effective scenarios and the observables that we consider in Sec.~\ref{sec:pheno}. Explicit examples where these effects could be larger have been discussed e.g.~in Ref.~\cite{Coy:2019rfr}.

\section{Tree-level mediators}
\label{app:mediators}

In this Appendix, we collect the Lagrangian of each tree-level mediator considered in Sec.~\ref{ssec:concrete}, as well as the matching to the SM EFT semileptonic Wilson coefficients at tree level. For simplicity, we omit $SU(3)_c$ and $SU(2)_L$ indices in the expressions below.

\begin{itemize}
    \item[$\bullet$] $\Phi^\prime\sim (\mathbf{1},\,\mathbf{2},\,\frac{1}{2})$\,:~\footnote{For simplicity, we assume that $\Phi^\prime$ does not acquire a vacuum expectation value and does not mix with the SM Higgs.}  
    \begin{itemize}
        \item[-] UV Lagrangian:
        \begin{align}
        \mathcal{L}_{\Phi^\prime} \supset - (y_{\Phi^\prime}^d)_{ij} \,\Phi^{\prime \dagger}
     \Bar{d}_iq_j - (y_{\Phi^\prime}^u)_{ij} \,\widetilde{\Phi}^{\prime\dagger}\bar{u}_i 
     q_j- (y_{\Phi^\prime}^e)_{ij}\, \Phi^{\prime\dagger}
     \Bar{e}_il_j + \mathrm{h.c.} 
        \end{align}
        \item[-] SM EFT coefficients:
        \begin{align}
        \frac{1}{\Lambda^2} \mathcal{C}^{(1)}_{\substack{lequ\\ijkl}} &= \frac{(y_{\Phi}^u)_{kl} (y_{\Phi}^e)_{ji}^{\ast}}{M_{\Phi}^2}\,, \qquad\quad
     \frac{1}{\Lambda^2} \mathcal{C}_{\substack{ledq\\ijkl}} = \frac{(y_{\Phi}^d)_{kl} (y_{\Phi}^e)_{ji}^{\ast}}{M_{\Phi}^2}\,.
        \end{align}
        \end{itemize}
    
    \item[$\bullet$] $Z^\prime\sim (\mathbf{1},\,\mathbf{1},\,0)$\,:
    \begin{itemize}
        \item[-] UV Lagrangian:
        \begin{align}
            \mathcal{L}_{Z^\prime} \supset \sum_{\psi = l, q, e, u, d} (g_{Z^\prime}^{\psi})_{ij}\,\Bar{\psi}_i &  \gamma_{\mu}\psi_j \,Z^{\prime \mu} \,.
        \end{align}
        \item[-] SM EFT coefficients:
        \begin{align}
            \frac{1}{ \Lambda^2 } \mathcal{C}^{(1)}_{\substack{lq\\ijkl}}  &= -\frac{ (g_{Z^\prime}^l)_{ij} (g_{Z^\prime}^q)_{kl}^\ast}{M_{Z^\prime}^2} \,, 
            &
            \frac{1}{ \Lambda^2 } \mathcal{C}_{\substack{eq\\klij}} &= -\frac{ (g_{Z^\prime}^e)_{ij} (g_{Z^\prime}^q)_{kl}^\ast}{M_{Z^\prime}^2} \,, \nonumber\\[0.5em]
            \frac{1}{ \Lambda^2 }\mathcal{C}_{\substack{lu\\ijkl}} &= -\frac{ (g_{Z^\prime}^l)_{ij} (g_{Z^\prime}^u)_{kl}^\ast}{M_{Z^\prime}^2} \,,
            &
            \frac{1}{ \Lambda^2 }\mathcal{C}_{\substack{eu\\ijkl}} &= -\frac{ (g_{Z^\prime}^e)_{ij} (g_{Z^\prime}^u)_{kl}^\ast}{M_{Z^\prime}^2} \,, \\[0.5em]
            \frac{1}{ \Lambda^2 }\mathcal{C}_{\substack{ld\\ijkl}} &= -\frac{ (g_{Z^\prime}^l)_{ij} (g_{Z^\prime}^d)_{kl}^\ast}{M_{Z^\prime}^2} \,,
            &
            \frac{1}{ \Lambda^2 } \mathcal{C}_{\substack{ed\\ijkl}} &= -\frac{ (g_{Z^\prime}^e)_{ij} (g_{Z^\prime}^d)_{kl}^\ast}{M_{Z^\prime}^2} \,. \nonumber
        \end{align}
    \end{itemize}
    
    \item[$\bullet$] $V \sim (\mathbf{1},\,\mathbf{3},\,0)$\,: 
    \begin{itemize}
        \item[-] UV Lagrangian:
        \begin{align}
            \mathcal{L}_{V} \supset (g_{V}^q)_{ij} \,  (\bar{q}_i \tau^I \gamma^{\mu} q_j)\,V^{ I}_{\mu}  +(g_{V}^l)_{ij}\,  (\bar{l}_i \tau^I \gamma^{\mu} \l_j)\,V^{ I}_{\mu} \,.
        \end{align}
        \item[-] SM EFT coefficients:
        \begin{align}
            \frac{1}{\Lambda^2} \mathcal{C}^{(3)}_{\substack{lq\\ijkl}} & = - \frac{(g_{V}^l)_{ij}(g_{V}^q)_{kl}}{M_{V}^2} \;.
        \end{align}
    \end{itemize}

    \item[$\bullet$] $S_1\sim (\mathbf{\bar{3}},\,\mathbf{1},\,\frac{1}{3})$\,: 
    \begin{itemize}
    \item[-] UV Lagrangian: 
    \begin{align}
        \mathcal{L}_{S_1} \supset (y^L_{S_1})_{ij} \,\Bar{q}_i^c i\tau_2 l_j S_1 + (y^R_{S_1})_{ij}\, \Bar{u}_i^c e_j S_1 + \text{h.c.} 
        \end{align}
        \item[-] SM EFT coefficients: 
        \begin{align}
        \frac{1}{\Lambda^2} \mathcal{C}^{(1)}_{\substack{lq\\ijkl}} &= \frac{(y^L_{S_1})_{lj} (y^L_{S_1})^\ast_{ki}}{4M_{S_1}^2} \;,
        & 
        \frac{1}{\Lambda^2} \mathcal{C}^{(3)}_{\substack{lq\\ijkl}} &= -\frac{(y^L_{S_1})_{lj} (y^L_{S_1})^\ast_{ki}}{4M_{S_1}^2} \;, \nonumber \\[0.5em]
        \frac{1}{\Lambda^2} \mathcal{C}^{(1)}_{\substack{lequ\\ijkl}} &= \frac{(y^R_{S_1})_{lj} (y^L_{S_1})^\ast_{ki}}{2M_{S_1}^2} \;,
        &
        \frac{1}{\Lambda^2} \mathcal{C}^{(3)}_{\substack{lequ\\ijkl}} &= -\frac{(y^R_{S_1})_{lj} (y^L_{S_1})^\ast_{ki}}{8M_{S_1}^2} \;, \\[0.5em]
     \frac{1}{\Lambda^2} \mathcal{C}_{\substack{eu\\ijkl}} &= \frac{(y^R_{S_1})_{lj}  (y^R_{S_1})^\ast_{ki}}{2M_{S_1}^2} \;. \nonumber
 \end{align}
    \end{itemize}

    \item[$\bullet$] $\widetilde{S}_1\sim (\mathbf{\bar{3}},\,\mathbf{1},\,\frac{4}{3})$\,: 
    \begin{itemize}
    \item[-] UV Lagrangian: 
    \begin{align}
        \mathcal{L}_{\widetilde{S}_1} \supset (y_{\widetilde{S}_1}^{R})_{ij} \,\Bar{d_i^c} e_j \,\widetilde{S}_1 + \text{h.c.} 
        \end{align}
        \item[-] SM EFT coefficients: 
        \begin{align}
        \frac{1}{\Lambda^2} \mathcal{C}_{\substack{ed\\ijkl}} = \frac{ (y_{\widetilde{S}_1}^{R})_{lj} (y_{\widetilde{S}_1}^{R})^\ast_{ki}}{2M_{\widetilde{S}_1}^2} \;.
    \end{align}

    \end{itemize}

    \item[$\bullet$] $S_3\sim (\mathbf{\bar{3}},\,\mathbf{3},\,\frac{1}{3})$\,:
    \begin{itemize}
      \item[-] UV Lagrangian:   
    
    \begin{align}
        \mathcal{L}_{S_3} \supset & (y^L_{S_3})_{ij}\, \big{(}\Bar{q}_i^c i\tau_2 \tau^I l_j \big{)} \, S_3^I+ \text{h.c.} 
        \end{align}
        \item[-] SM EFT coefficients:
        \begin{align}
        \frac{1}{\Lambda^2} \mathcal{C}^{(1)}_{\substack{lq\\ijkl}} &= \frac{3(y^L_{S_3})_{lj} (y^L_{S_3})^\ast_{ki}}{4M_{S_3}^2} \;,
        & \frac{1}{\Lambda^2} \mathcal{C}^{(3)}_{\substack{lq\\ijkl}} = \frac{(y^L_{S_3})_{lj} (y^L_{S_3})^\ast_{ki}}{4M_{S_3}^2} \;.
    \end{align}
    \end{itemize}
    
    \item[$\bullet$] $R_2\sim (\mathbf{3},\,\mathbf{2},\,\frac{7}{6})$\,:
    \begin{itemize}
    \item[-] UV Lagrangian: 
    \begin{align}
        \mathcal{L}_{R_2} \supset - (y_{R_2}^{L})_{ij}\,& \bar{u}_i R_2 i\tau_2 l_j - (y_{R_2}^{R})_{ij} \, \bar{q}_i e_j R_2 + \text{h.c.}  
        \end{align}
        \item[-] SM EFT coefficients:
        \begin{align}
        \frac{1}{\Lambda^2} \mathcal{C}_{\substack{lu\\ijkl}} &= -\frac{ (y_{R_2}^{L})_{kj}(y_{R_2}^{L})_{li}^\ast}{2M_{R_2}^2} \;, 
        &
        \frac{1}{\Lambda^2} \mathcal{C}_{\substack{eq\\ijkl}} &= -\frac{ (y_{R_2}^{R})_{kj}(y_{R_2}^{R})_{li}^\ast}{2M_{R_2}^2} \;, \nonumber \\[0.5em]
        \frac{1}{\Lambda^2} \mathcal{C}^{(1)}_{\substack{lequ\\ijkl}} &= \frac{(y_{R_2}^{R})_{kj}(y_{R_2}^{L})_{li}^\ast}{2M_{R_2}^2} \;,  
        &
        \frac{1}{\Lambda^2} \mathcal{C}^{(3)}_{\substack{lequ\\ijkl}} &= \frac{(y_{R_2}^{R})_{kj}(y_{R_2}^{L})_{li}^\ast}{8M_{R_2}^2} \;.
    \end{align}
    \end{itemize}

        \item[$\bullet$] $\widetilde{R}_2\sim (\mathbf{3},\,\mathbf{2},\,\frac{1}{6})$\,:
    \begin{itemize}
    \item[-] UV Lagrangian: 
    \begin{align}
        \mathcal{L}_{\widetilde{R}_2} \supset -(y^{L}_{\widetilde{R}_2})_{ij} \,\Bar{d_i} \widetilde{R}_2 i\tau_2  l_j  + \text{h.c.}  
        \end{align}
        \item[-] SM EFT coefficients:
        \begin{align}
        \frac{1}{\Lambda^2} \mathcal{C}_{\substack{ld\\ijkl}} = -\frac{(y^{L}_{\widetilde{R}_2})_{kj}(y^{L}_{\widetilde{R}_2})^\ast_{li}}{2 M^2_{\widetilde{R}_2}} \;.
    \end{align}
    \end{itemize}
    
    \item[$\bullet$] $U_1\sim (\mathbf{3},\,\mathbf{1},\,\frac{2}{3})$\,: 
    \begin{itemize}
    \item[-] UV Lagrangian: 
    \begin{align}
        \mathcal{L}_{U_1} \supset (x_{U_1}^{L})_{ij} \, \Bar{q}_i\gamma^{\mu} l_j\,U_{1\mu} + (x_{U_1}^{R})_{ij}\, \bar{d}_i\gamma^{\mu} e_j\, U_{1\mu} + \text{h.c.}
        \end{align}
        \item[-] SM EFT coefficients: 
        \begin{align}
        \frac{1}{\Lambda^2} \mathcal{C}^{(1)}_{\substack{lq\\ijkl}} &= -\frac{ (x_{U_1}^{L})_{kj} (x_{U_1}^{L})_{li}^\ast}{2M_{U_1}^2} \;,
        &
        \frac{1}{\Lambda^2} \mathcal{C}^{(3)}_{\substack{lq\\ijkl}} &= -\frac{ (x_{U_1}^{L})_{kj} (x_{U_1}^{L})_{li}^\ast}{2M_{U_1}^2} \;, \nonumber \\[0.5em]
        \frac{1}{\Lambda^2} \mathcal{C}_{\substack{ed\\ijkl}} &= -\frac{ (x_{U_1}^{R})_{kj} (x_{U_1}^{R})_{li}^\ast}{M_{U_1}^2} \;, 
        &
        \frac{1}{\Lambda^2} \mathcal{C}_{\substack{ledq\\ijkl}} &= \frac{2 (x_{U_1}^{R})_{kj} (x_{U_1}^{L})_{li}^\ast}{M_{U_1}^2} \;.
    \end{align}
    \end{itemize}
    
    \item[$\bullet$] $\widetilde{U}_1 \sim (\mathbf{3},\, \mathbf{1},\,\frac{5}{3})$\,:
    \begin{itemize}
        \item[-] UV Lagrangian:
        \begin{align}
            \mathcal{L}_{\widetilde{U}_1} \supset (x_{\widetilde{U}_1}^{R
        })_{ij} \,\bar{u}_i \gamma_{\mu}  e_j \, \widetilde{U}^{\mu}_1+ \text{h.c.}
        \end{align}
        \item[-] SM EFT coefficients:
        \begin{align}
            \frac{1}{\Lambda^2} \mathcal{C}_{\substack{eu\\ijkl}} = -\frac{(x_{\widetilde{U}_1}^{R})_{kj} (x_{\widetilde{U}_1}^{R})^\ast_{li} }{M_{\widetilde{U}_1}^2} \;.
        \end{align}
        
    \end{itemize}

\item[$\bullet$] $V_2 \sim (\mathbf{\bar{3}},\, \mathbf{2},\,\frac{5}{6})$\,:
    \begin{itemize}
        \item[-] UV Lagrangian:
        \begin{align}
            \mathcal{L}_{V_2} \supset (x_{V_2}^{L})_{ij} \,\bar{d}^c_i \gamma_{\mu} V^{\mu}_2 i \tau_2 l_j + (x_{V_2}^{R})_{ij} \,\bar{q}^{c}_i \gamma_{\mu} V^{\mu}_2 i \tau_2 e_j + \text{h.c.}
        \end{align}
        \item[-] SM EFT coefficients:
        \begin{align}
            \frac{1}{\Lambda^2} \mathcal{C}_{\substack{eq\\ijkl}} &= \frac{(x_{V_2}^{R})_{lj} (x_{V_2}^{R})^\ast_{ki} }{M_{V_2}^2} \;,
            &
            \frac{1}{\Lambda^2} \mathcal{C}_{\substack{ld\\ijkl}} &= \frac{(x_{V_2}^{L})_{lj} (x_{V_2}^{L})^\ast_{ki} }{M_{V_2}^2} \;, \nonumber \\[0.5em]
            \frac{1}{\Lambda^2} \mathcal{C}_{\substack{ledq\\ijkl}} &=  - \frac{2(x_{V_2}^{R})_{lj} (x_{V_2}^{L})^\ast_{ki} }{M_{V_2}^2} \;.
        \end{align}
    \end{itemize}    

\item[$\bullet$] $\widetilde{V}_2 \sim (\mathbf{\bar{3}},\, \mathbf{2},\,-\frac{1}{6})$\,:
    \begin{itemize}
        \item[-] UV Lagrangian:
        \begin{align}
            \mathcal{L}_{\widetilde{V}_2} \supset (x_{\widetilde{V}_2}^{L})_{ij}\, \bar{u}^c_i \gamma_{\mu} \widetilde{V}^{\mu}_2 i\tau_2 l_j + \text{h.c.}
        \end{align}
        \item[-] SM EFT coefficients:
        \begin{align}
            \frac{1}{\Lambda^2} \mathcal{C}_{\substack{lu\\ijkl}} = \frac{(x_{\widetilde{V}_2}^{L})_{lj} (x_{\widetilde{V}_2}^{L})^\ast_{ki} }{M_{\widetilde{V}_2}^2} \;.
        \end{align}
        
    \end{itemize}

    \item[$\bullet$] $U_3\sim (\mathbf{3},\,\mathbf{3},\,\frac{2}{3})$\,: 
    \begin{itemize}
    \item[-] UV Lagrangian: 
    \begin{align}
        \mathcal{L}_{U_3} \supset & (x_{U_3})_{ij} \, \big{(}\bar{q}_i \gamma^{\mu} \tau^I l_j\big{)}  U_{3\mu}^{I}  + \text{h.c.}
        \end{align}
        \item[-] SM EFT coefficients: 
        \begin{align}
        \frac{1}{\Lambda^2} \mathcal{C}^{(1)}_{\substack{lq\\ijkl}} &= -\frac{3 (x_{U_3})_{kj}(x_{U_3})_{li}^\ast}{2M_{U_3}^2} \;, 
        &
        \frac{1}{\Lambda^2} \mathcal{C}^{(3)}_{\substack{lq\\ijkl}} &= \frac{ (x_{U_3})_{kj}(x_{U_3})_{li}^\ast}{2M_{U_3}^2} \;.
    \end{align}
    \end{itemize}
\end{itemize}

\


\end{document}